\documentstyle[12pt]{article}

\input tcilatex
\begin{document}

\author{Mario A. Castagnino \\
Instituto de Astronom\'{i}a y F\'{i}sica del Espacio.\\
Casilla de Correos 67,Sucursal 28.\\
1428 Buenos Aires, Argentina. \and Edgerd Gunzig \\
Instituts Internationaux de Physique et de Chimie, 1050 Bruxelles, Belgium.
\and Mario Castagnino \\
Instituto de Astronom\'ia y F\'isica del Espacio \\
C.C. 67, Suc. 28, 1428 Buenos Aires, Argentina}
\title{Dynamics, Thermodynamics, and Time-Asymmetry. }
\date{July 16,1995. }
\maketitle

\begin{abstract}
There are two schools, or lines of thought, that try to unify the apparently
divergent laws of dynamics and thermodynamics and to explain the observed
time-asymmetry of the universe, and most of its sub-systems, in spite of the
fact that these systems are driven by time-symmetric evolution equations.
They will be called the coarse-graining and the fine-graining schools (even
if these names describe only a part of their philosophy). Coarse-graining
school obtains time-asymmetry via a projection of the state space on a space
of ''relevant'' states. The corresponding projection of the primitive
reversible evolution laws yields effective irreversible evolution laws for
the relevant states. Fine-graining always use the same primitive reversible
evolution laws. But these laws (in adequate extensions of the usual spaces
where these laws are formulated) have a set of solutions $S$ that can be
decompose in two subsets $S_{+\text{ }}$ and $S_{-}$ of time asymmetric
solutions. Choosing one of these two sets, as the arena to formulate the
theory, time asymmetry is established. The aim of these lectures is to
explain, in the simplest- self-contained, unbiased, and, honest way, the
main characteristics of both schools and to point out the advantages and
disadvantages of both formalism, in such a way that, the polemic between the
schools, turns out to be explicit and organized in the mind of the reader
(who will be considered the supreme judge to give the final verdict).

Some cosmological features of the theory will be also considered, mainly the
problem of the low entropy initial state of the universe
\end{abstract}

\title{Dynamics, thermodynamics, and time-asymmetry}

\begin{itemize}
\item  \vspace{2in}

\item  PACS Nrs. 05.20-y, 03.65, BZ, 05.30-d
\end{itemize}

\section{Introduction.}

In these lectures we will study and try to solve two, long standing,
problems of theoretical physics.

\subsection{The problem of time asymmetry.}

The problem of the existence of the arrow of time or, what is the same
thing, the problem of time asymmetry of the universe, can be stated in two
questions:

{\it i.How can it be that the universe is time-asymmetric if all the
relevant physical laws are time symmetric?}

{\it ii. Why all arrows of time point in the same direction?}

In fact, universe has several time asymmetries, namely the various arrows of
time: thermodynamical, electromagnetical, psicological, etc., while the main
laws of nature are time-symmetric (because,as usual, we will neglect the
laws of weak interaction, since it is very difficult to imagine a mechanism
that explain the time asymmetry of the universe based in these laws [1]).

In these lectures we would like to answer these questions given an adequate
mathematical formalism to the problem and using several, old and new, well
known ideas ([2],[3],[4]). In doing so we must first precise two important
words: {\it conventional }and {\it substantial} ([1],[5]). In mathematics we
use to work with {\it identical} objects, like points, the two directions of
an axis, the two semi cones of a null cone, etc. In physics there are also
identical objects: like identical particles, spin directions, etc.-Among
identical objects there is always a mathematical transformation that
exchange these object leaving the system unmodified. If we are forced to
call identical objects with different names we will say that be are
establishing a {\it conventional difference }among these objects, e. g. when
we call + and - the two directions of an axis or''past'' and ''future'' the
two semi cones of a null cone. If some objects are different we will say
that there is a{\it \ substantial difference} among them. The problem of the
arrow of time is that past and future are only conventionally different, in
usual physical theories, while we have the filling that past and future are
substantially difference, in facts in the past events had happened, while in
the future event could only happen.

In theories endowed with time-symmetric evolution equations, as those we
will deal with, it is quite impossible to find time substantial asymmetry
using rigorous mathematical manipulations. But normally we can find in these
theories, as we shall see, two identical mathematical structures, one
related with the past and one related with the future, e. g. two subspaces
of the space of solutions of the theory. Nevertheless these structures are
only conventionally different, because they are related with a time
inversion. But within these structures past is substantially different than
future. To chose one of these structures,or the other, is physically
irrelevant, since time inversion exchange one structure with the other,
leaving the universe unchanged. Therefore to create an arrow of time we just
conventionally chose one of the structures. This choice is irrelevant, as
irrelevant as to chose one face of a dice if all faces are marked with the
same number. But when we have chose one of the structures a substantial
difference is also created, between past and future, within these
structure,and an arrows of time appears. This is the method we will use to
create all the arrows of time, both in coarse-graining and fine-graining
cases (see section 6).

To show that all the arrows of time point in the same direction we will
consider that the master arrow of time is the cosmological one. We will show
that the universe expansion creates a thermodynamical instability in the
universe, in such a way that the thermodynamical arrow of time must
necessarily point in the same direction. We will refer to the literature for
the problem of the coincidence of the other arrows of time with the
cosmological-master arrow.

\subsection{The problem of the unification of dynamics and thermodynamics.}

A particular. but very important, case of the first problem, is the problem
of the unification of the time-symmetric dynamical laws with the
time-asymmetric thermodynamical laws. In fact, it is reasonable to think
that thermodynamical laws could be demonstrated using the classical or
quantum dynamical laws. But, it seems that this is not the case for the
second law of thermodynamics, that says that entropy increases, in
irreversible evolutions, leading the system to a state of thermodynamical
equilibrium or maximal entropy. This problem can be state as follows:

{\it i.-Liouville equation is the time-symmetric evolution equation for
classical distribution functions or quantum density matrices }$\rho .$

{\it ii.-This equation prevent the definition of any function of }$\rho
:F(\rho )$ ({\it constructed only with }$\rho $ {\it and mathematical
elements of the Liouville-phase space) such that }$\stackrel{\bullet }{%
F(\rho )}>0$ , {\it namely it is impossible, as a consequence of Liouville
theorem, to define a Lyapunov variable, i.e. a growing function of }$\rho $, 
{\it e. g.: the volume or the support of a characteristic distribution
function }$\rho $ {\it is time constant, Gibbs and conditional entropies are
time constant [6], etc. etc.}

{\it iii.-Nevertheless we actually see that the evolution leads the system
to a thermodynamical equilibrium with a maximal entropy stationary state }$%
\rho _{*}$ {\it .}

Therefore the problem is to combine Liouville theorem with the obvious fact
that usual physical systems have a tendency to go to a thermodynamical
equilibrium. The solution of the problem is based in a theorem by Mackey and
Lasota [6] (Theorem 4.3.1 below):

{\bf Theorem}{\it : Let } $S(t)$ {\it be an ergodic transformation, with
stationary equilibrium density } $\rho _{*}$ ({\it of the associated
Frobenius-Perron operator } $P(t)$ {\it in a phase space of finite } $\rho
_{*}-${\it measure}). {\it Then if } $S(t)$ {\it is } $\rho _{*}$-{\it %
mixing if and only if } $P(t)\rho $ {\it is weakly convergent to } $\rho
_{*} $ {\it i.e.:
\[
\stackunder{t\rightarrow \infty }{\lim (P(t)\rho |}g)=(\rho
_{*}|g)................(1.2.1) 
\]
for all bounded measurable functions }$g$.

i.e. if the time evolution is phase space is $S(t)$ and the corresponding
time evolution of the distribution functions is $\rho (t)=P(t)\rho (0),$ and
this evolution is mixing, a chaotic property of evolutions that we shall
define below, and if there is an equilibrium density such that $P(t)\rho
_{*}=\rho _{*},$ then eq. (1.2.1) can be proved.

But:
\[
\stackunder{t\rightarrow \infty }{\lim }P(t)\rho \neq \rho
_{*}..................................(1.2.2) 
\]
in fact: as we shall see in many cases this limit do not even exist.
Therefore we have a weak limit but we have not a strong limit (i.e. a limit
in the norm).

Nevertheless we never see or measure $\rho $. What we see and measure are
mean values of physical quantities $O$ such that:
\[
<O>_\rho =<\rho |O>...........................(1.2.3) 
\]
Thus what we actually see is that:
\[
\stackunder{t\rightarrow \infty }{\lim }<O>_\rho =<O>_{\rho
_{*}}......................(1.2.4) 
\]
In fact, all the mean values of the physical quantities go to their
equilibrium mean values if the evolution of the system is $\rho _{*}$%
-mixing. So the solution of the problem is quite easy:

{\it i.-Liouville theorem is embodied in eq. (1.2.2): the system do not
go-(strongly) toward the equilibrium states.}

{\it ii.-Tendency toward equilibrium is embodied in eq. (1.2.4): the mean
values of all the physical quantities goes to their equilibrium values.}

Clearly these facts are not contradictory. We will call to this solution the 
{\it non-graining} solution.

As chaotic-mixing systems are very frequent in the universe the problem is
essentially solved. What it is left to be studied are the different technics
to deal with the detail calculations. These technics try to find some
logical modification of the theory in order to solve the missing limit
(1.2.2), which, even if unnecessary from the mathematical point of view, it
is the way physics used to think, (or love to think) at least up to now. In
fact there are two technics:

\subsubsection{Coarse-Graining.}

Let us define an arbitrary, but time independent, projector:
\[
P=|g)(g|,.......(g|g)=1...............(1.2.1.1) 
\]
and let us define a coarse-graining density function as:
\[
\stackrel{\sim }{\rho }=P\rho =|g)(g|\rho
)............................(1.2.1.2) 
\]
From eq. (1.2.1) we have:
\[
\stackunder{t\rightarrow \infty }{\lim }|g)(g|P(t)\rho )=|g)(g|\rho
_{*})............(1.2.1.3) 
\]
and therefore:
\[
\stackunder{t\rightarrow \infty }{\lim }\stackrel{\sim }{\rho (t)}=\stackrel{%
\sim }{\rho }_{*}..............................................(1.2.1.4) 
\]
that would be the coarsed-graining version of eq. (1.2.2) and the main
equation of the first technic (of course the same thing happens with the
general projector $\Pi =\sum |g_i)(g_i|,$ $(g_i|g_j)=\delta _{ij})$. It is
easy to demonstrate that (1.2.1.4 is a limit in norm. It is also evident
that eq. (1.2.1.4) can be obtained with a quite arbitrary state $|g)$ and
that all the philosophy, typical of the coarse-graining technic, namely the
definition and consideration of macroscopic and microscopic states [8] is
just an intuitive justification to give a physical meaning to the limit
(1.2.1.4). But as this justification is really unnecessary, since the
relevant and important limit is (1.2.4), the physical explanation of all the
philosophy of the coarse-graining technic can be philosophically criticized
[9]. This is the main problem with coarse-graining. It is an arbitrary
method. It works perfectly well but it is difficult to justify based on
physical-philosophical (metaphysical?)-arguments.

In fact, coarse-graining contains the miss-leading statement:{\it \ we
cannot see microscopic states (i.e. }$\rho $) {\it but we can see
macroscopic states (i.e. }$\stackrel{\sim }{\rho }$). This statement leads
to the problem of finding an unique an reasonable definition for these
macrostates. This problem is unsolved and, in our opinion, it will remain
unsolved since $|g)$ is essentially arbitrary. Also, if we arbitrary chose
some definition of macrostates, we are introducing a physical element that
really it is alien to the system itself, and therefore this definition, even
if natural in particular examples, will be suspicious from a general point
of view.

The correct ''no-graining'' statement is: {\it we cannot directly measure
microscopic states (i.e. $\rho $}){\it , we can only measure mean values of
physical quantities or observables (}among them the projector $P=|g)(g|$ and
therefore the arbitrarily defined macroscopic states). This statement is
completely true at the classical and quantum levels [10] and refers to {\it %
all physical observables.} Then we can rigorously say, e.g. that the two
thermodynamical variables $<p>$ and $<v>$ define the thermodynamical
macrostate of a perfect gas. etc.

\subsubsection{ Fine Graining.}

Let ${\cal L}${\cal \ }is the Hilbert-Liouville space of the physical states 
$\rho $ and ${\cal L^{\times }=L}$ the space of the linear operator on $%
{\cal L}${\cal .}. We may think that not all $O\in {\cal L=L^{\times }}$ is
a physical admissible observable. In fact, observables are measured by real
physical devices, that very likely are free of sophisticated mathematical
behaviors: e.g. are related with continuous and derivable functions and not
with discontinuous non derivable functions, even if square-integrable. So it
is reasonable that $O(x)$ would be ,e.g. a Schwarz function (we will precise
this point in section 5). So let us call $\Phi $ to the space of physically
admissible observables such that:
\[
\Phi \subset {\cal L=L^{\times }.............................}(1.2.2.1) 
\]
If we consider the dual $\Phi ^{\times }$ of $\Phi $ we have a Gel'fand
triplet (cf. Appendix 4.A.):
\[
\Phi \subset {\cal L=L^{\times }}\subset \Phi ^{\times
}...................(1.2.2.2) 
\]
(as we shall see in section 5 if we give to the functions of $\Phi $ some
analyticity properties we can consider also the time-asymmetry problem
within this framework). We will work with states that belong to $\Phi
^{\times }$, e.g.: $\rho _{*}$ normally belongs to this space. As it is well
know we define the functional $A$ addition of the functionals $B$ and $C$ :$%
A=B+C$ as the functional $A[g]=(A|g)$ defined by:
\[
A[g]=(A|g)=B[g]+C[g]=(B|g)+(C|g) 
\]
for all $g\in \Phi $. The same method is used to define the product of a
functional by a number. Analogously, if we have a sequence of functionals: $%
A_1,A_2,...$the limit $A=\stackunder{i\rightarrow \infty }{\lim }A_i$ is
defined as the functional such that:
\[
A[g]=(A|g)=\stackunder{i\rightarrow \infty }{\lim }A_i[g]=\stackunder{%
i\rightarrow \infty }{\lim }(A_i|g) 
\]
for all $g\in \Phi .$

Then, as $\rho $ and $\rho _{*}$ can be considered as functional on $\Phi $
eq. (1.2.1) reads:
\[
\stackunder{t\rightarrow \infty }{\lim }P(t)\rho =\rho
_{*}.........................(1.2.2.3) 
\]
and we have found a rigorous ''strong'' limit corresponding to eq. (1.2.1)
[11]. Perhaps the main problem with the fine-graining technic is that it is
usual to consider the states of $\Phi ^{\times }\setminus {\cal L}$ as
unphysical states or just effective states, where some characteristic of
real physical states have been neglected (as Zeno and Khalfin effects).
Nevertheless we can also say that every state that can be used to measure
the mean values of all observables of $\Phi $ is a physical states, and this
is the case with all the states of $\Phi ^{\times }$. But this point it is
not completely clear today.

So neither technics is completely sinless. Nevertheless as the physical real
problem is solved by the Mackey and Lasota theorem, we can say that all this
sins are venial sins. On the other hand both technics have some advantages:
e. g.:

i.-Coarse graining works just with one physical space, ${\cal L.}$ Also
coarse-graining is unavoidable to calculate global thermodynamical variables
like temperature or pressure, but

ii.-The time evolution of $\rho (t)$ can be computed easier using the
fine-graining technic, since we have the vectors of space $\Phi ^{\times }$
that can be used to find new spectral expansion for the observables of the
problem. Once we know $\rho (t)$ we can compute $\stackrel{\sim }{\rho }%
=P\rho (t),$ while the direct computation of $\stackrel{\sim }{\rho }(t)$
,using coarse-graining technics directly, can be more difficult [12].

These lectures will be almost devoted to study the new ideas of no-graining
and fine-graining, since the coarse graining technic is well known (cf.
[11],[12],[13])

The lectures are organized as follows:

In section two we will describe the dynamics, both classical and quantum,
and define the notions of time-symmetry and reversibility. This section is
based in paper [15].

In section three we will deal with thermodynamics and we will give different
definitions of entropy. This, and the two next section are based in paper
[6], but we have added the new mathematical and physical structures,
recently appeared and studied.

In section four we will introduce the classical evolution equations and we
will study the ergodic, mixing, and, exact transformations.

In section five will see the quantum evolution equations. We will study the
no-graning and fine graining ideas, both in models with discrete and
continuous spectra, and we will consider the Friedrichs model, for pure and
mixed states.

In section six we will study the coarse-graining projectors and the
fine-graining traces. We will also study the problem of time asymmetry.

In section seven we will review the main equation of thermodynamics in
curved space-time.

In section eight we will consider the coordination of the arrows of time.
This section is mostly based in references [2], [4], and, [16].

In section nine we will drown our conclusions.

\section{Dynamics.}

In this section we will review the formalism that we will use in this work
and we will see how the notions of reversibility and time-asymmetry are
introduced.

\subsection{Classical formalism.}

A classical system with $N$ degrees of freedom is characterized by its
Hamiltonian

\[
H=H(x)=H(q_i,p_i),............(2.1.1) 
\]

\noindent a function of $x,$ the generic point in the 2N-dimensional phase
space $X$ or a function of the configuration variables $q_i$ and the
momentum variables $p_i$ ($i=1,....,N$). The system is solved if we compute
the functions

\[
q_i=q_i(t), 
\]
\[
p_i=p_i(t),........or 
\]
\[
x=x(t).......................(2.1.2) 
\]

\noindent solutions of the Hamilton equations

\[
{\frac{dq_i}{{dt}}}=\partial _{p_i}H, 
\]
\[
{\frac{dp_i}{{dt}}}=-\partial _{q_i}H,..........(2.1.3) 
\]

\noindent satisfying, at time $t=0$, the conditions

\[
q_i(0)=q_i^0, 
\]
\[
p_i(0)=p_i^0...........(2.1.4) 
\]
The solution of the system of differential equation s (2.1.3) is the map $%
S(t):X\rightarrow X$, defined by:
\[
S(t)[x(0)]=x(t)........(2.1.4^{\prime }) 
\]
We also call $S_t=S(t)$ and these $S_t$ form a group. If $A\subset X$ is a
subset of the phase space we can compute the image of $A$, namely $S_t(A)=B.$
Then if $\mu _L$ is the Lebesgue measure on $X$ we can formulate the
Liouville

{\bf Theorem 2.1.1.}If $S(t)$ is the map obtained solving the classical
dynamical evolution and $A$ a $\mu _L$-measurable set of $X$ then:
\[
\mu _L(S(t)A)=\mu _L(A)............(2.1.4") 
\]
i.e.: Classically the evolution preserve the ''volumes'' of phase space.

Let us now define the notion of reversibility. Experimentally it is
impossible to change the direction of time. The best we can do in order to
simulate a time inversion, is to film the motion under study and project the
film backward. Then, if $q_i=q_i(t)$ and $p_i=p_i(t)$ gives the real motion,
the law of the fictitious motion obtained playing backward the film will be $%
q_i=q_i(-t)$, $p_i=-p_i(-t)$, where to change $t$ by $-t$ is simply an easy
way to avoid to define new initial data (the final ones of the reversed
motion). We can deduce that the time reversal operator $T$ acts on the
configuration variables and the momentum variables as [1],[17]

\[
T(q_i,p_i)=(Tq_i,Tp_i)=(q_i,-p_i)...........(2.1.5) 
\]

We can now consider the data (2.1.4) (that we have called ``conditions at
zero time'' and not ``initial conditions'' in order to avoid any reference
to time, even though we shall follow the common use in other sections) and
compute the reversed data

\[
q^{rev}_i(0)=q_i(0),
\]

\[
p_i^{rev}(0)=-p_i(0)...........(2.1.6) 
\]

\noindent With these conditions, ``at zero time'' we can calculate, using
Eq. (2.1.3), a new real motion that we will call $q_i^{rev}(t)$, $%
p_i^{rev}(t)$. We will say that the motion is reversible if

\[
q^{rev}_i(t)=q_i(-t),
\]

\[
p_i^{rev}(t)=-p_i(-t),..........(2.1.7) 
\]

\noindent namely, if the motion in the backward film agrees with a real
motion with reversed conditions at zero time (we see that the initial
conditions of one motion are the final ones of the other).

Usually $H$ (cf. Eq. (2.1.1)) is quadratic in the $p_i$, so that

\[
TH(q_i,p_i)=H(Tq_i,Tp_i)=H(q_i,-p_i)=H(q_i,p_i)...........(2.1.8) 
\]
In this case we will say that the hamiltonian is time-symmetric.

\noindent Then, if we make a $T$ transformation (2.1.5) on Eq. (2.1.3), we
find

\[
{\frac{dq_i}{{dt}}}=-\partial _{(-p_i)}H, 
\]
\[
-{\frac{d(-p_i)}{{dt}}}=-\partial _{q_i^{rev}}H,..........(2.1.9) 
\]

\noindent and if we now change $t$ by $-t$ we find again Eqs. (3.1.3) as

\[
{\frac{dq_i}{{d(-t)}}}=\partial _{(-p_i)}H, 
\]

\[
{\frac{d(-)p_i}{{d(-t)}}}=-\partial _{q_i}H...........(2.1.10) 
\]

\noindent From this equation and eq. (2.1.3) a motion $(q_i^{rev},p_i^{rev})$
with data (2.1.6) must satisfy (2.1.7). Therefore

{\bf Theorem 2.1.2.}, A usual Hamiltonian, quadratic in the $p_i^{\prime }s$%
, yields a reversible motion.

The only condition to obtain a reversible motion is eq. (2.1.8), namely that
the hamiltonian would be time-symmetric.

Then reversible motion form a group. But irreversible motions do not form a
group, since the inverse of such a motions do not even exist, because they
are not real motions.

We will further say that the initial conditions are time-symmetric, with
respect to $t=0$ if $p_i(0)=0$ or:
\[
q_i(0)=q_i(0) 
\]
\[
p_i(0)=-p_i(0)............(2.1.11) 
\]
Then, if the motion is reversible, we will have:
\[
q_i(t)=q_i(-t) 
\]
\[
p_i(t)=-p_i(-t).............(2.1.12) 
\]
We call this motion time -symmetric with respect to $t=0,$ since the curves $%
q_i(t)$ are symmetric with respect to the vertical axis and the curves $%
p_i(t)$ are symmetric with respect to the origin of the coordinate system,
as in fig. 0. Therefore:

{\bf Theorem 2.1.3.} If the motion is reversible and the condition at $t=0$
is time-symmetric, the motion is time-symmetric with respect to $t=0.$

If all the motion would be time-symmetric with respect to $t=0$ it would be
impossible to define any arrow of time at $t=0$, since past and future would
look like exactly the same from this instant of time.

\subsection{Quantum formalism.}

The quantum wave function for the same system treated in Sect. 2.1. reads

\[
\Phi (q_i,t)=\langle q_i|\Phi (t)\rangle ..........(2.2.1) 
\]
This function belongs to a Hilbert space ${\cal H}=L^2$. Namely if we
introduce the inner product:
\[
(\Phi ,\Psi )=\int \Phi ^{*}\Psi d^Nq......(2.2.1^{\prime }) 
\]
it is $(\Phi ,\Phi )<\infty $, and usually is normalized as $(\Phi ,\Phi )=1$

\noindent and satisfies the Schr\"odinger equation

\[
i{\frac{{{\partial }\Phi (q_i,t)}}{{\partial t}}}={H\Phi (q_{i,}t)}%
,..........(2.2.2) 
\]

\noindent from which we can find the time evolution of the wave function $%
\Phi (q_i,t)$ by imposing conditions at zero time

\[
\Phi (q_i,0)=\Phi ^0(q_i)...........(2.2.3) 
\]
Then:
\[
\Phi (q_i,t)=e^{-iHt}\Phi (q_i,0)=u(t)\Phi
^0(q_i).................(2.3.3^{\prime }) 
\]

Since we are now working in the configuration representation, in which the
position and momentum operators are

\[
\hat q_i=q_i, 
\]
\[
\hat p_i=-i\partial _{q_i},..........(2.2.4) 
\]

\noindent the quantum version of Eq. (2.1.5) is

\[
T\Phi (q_i,t)=\Phi ^{*}(q_i,t)...........(2.2.5) 
\]

\noindent For, if

\[
\langle \hat p_i\rangle _\Phi =\int \Phi ^{*}(q)\big(-i\partial _{q_i}\big)%
\Phi (q)dq,<\widehat{q_i>_\Phi =\int \Phi ^{*}(q)q_i\Phi (q)dq_i} 
\]

\noindent then

\[
\langle \hat p_i\rangle _{\Phi ^{*}}=\int \Phi (q)\big(-i\partial _{q_i}\big)%
\Phi ^{*}(q)dq=-\langle \hat p_i\rangle _\Phi ,<\widehat{q}_i>_{\Phi ^{*}}=< 
\widehat{q}_i>_\Phi 
\]

\noindent (for more details see [17]). Then, the wave function of the
inverted motion will have as zero time data

\[
\Phi _{rev}(q_i,0)=\Phi ^{*}(q_i,0)=\Phi ^{0*}(q_i),..........(2.2.6) 
\]

\noindent and the motion will be reversible if

\[
\Phi _{rev}(q_i,t)=\Phi ^{*}(q_i,-t),..........(2.2.7) 
\]

\noindent which is the quantum version of eqs. (2.1.7).

If $H$ is a hamiltonian quadratic in $p$, it is easy to see that $H$ is real
(or time-symmetric) namely:

\[
H=H^{*}...........(2.2.8) 
\]
Then we can formulate the

{\bf Theorem 2.2.1.} If the hamiltonian is real the corresponding evolution
is reversible.

Proof.:

\noindent From eqs. (2.2.6) and (2.2.8) we can obtain (2.2.7), since

\[
\Phi _{rev}(t)=e^{-iHt}\Phi _{rev}(0)=e^{-iHt}\Phi ^{*}(0)=(e^{iHt}\Phi
(0))^{*}=\Phi ^{*}(-t),..........(2.2.9) 
\]

\noindent where we have omitted the variables $q_i$. Then, as in the
classical case, a usual Hamiltonian yields a reversible motion.$\Box $

We can also show directly that Eq. (2.2.2) is t-invariant, but we prefer the
proof above because the role played by the condition at zero time can be
seen explicitly.

As in the classical, case reversible motion form a group, since $%
u^{-1}(t)=u(-t)$ is a real motion, which is not the case for irreversible
motion where this motion is not a real one. If $u(t_1)u(t_2)=u(t_1+t_2)$ for 
$t_1,t_2\geq 0$ only, we will say that these motions form a semigroup. This
is the case for irreversible motions.

Let us now repeat all this formalism, that we have so far introduced in the
configuration representation, in an abstract way. The state of the system is
defined by the ket $|\Phi (t)\rangle $ that belongs to the Hilbert space of
states ${\cal H}$, and satisfies the Schr\"odinger equation

\[
i{\frac d{{dt}}}|\Phi (t)\rangle =H|\Phi (t)\rangle ,..........(2.2.10) 
\]
The inner product is symbolized as $<\Phi |\Psi >=(\Phi ,\Psi ),$ and the
normalization is $<\Phi |\Phi >=1.$

\noindent $|\Phi (t)>$ can be found solving eq. (2.2.10) with the condition
at zero time

\[
|\Phi (0)\rangle =|\Phi ^0\rangle ,..........(2.2.11) 
\]
Namely:
\[
|\Phi (t)>=e^{-iHt}|\Phi (0)>=u(t)|\Phi (0)>......(2.2.11^{\prime }) 
\]

\noindent Then, the $T$ transformation can be defined as [17],[18]

\[
T|\Phi (t)\rangle =K|\Phi (t)\rangle =|\Phi ^{*}(t)\rangle
,..........(2.2.12) 
\]

\noindent which means that we must conjugate the wave function in the
configuration representation and then go to the generic representation. $K$
is known as the Wigner operator. More precisely, let ($\Phi (q_i,t))=|\Phi
(q_i,t))$ be the coordinates of the state vector in the configuration
representation (wave function) and $|\Phi (,t)>$ the coordinates of the same
vector in a generic representation; then

\[
|\Phi (t)>=U|\Phi (q_i,t)),\ \ \ UU^{\dagger }=1...........(2.2.13) 
\]

\noindent Let $K_0$ be the conjugation operator in the configuration
representation

\[
K_0|\Phi (q_i,t))=|\Phi ^{*}(q_i,t));..........(2.2.14) 
\]

\noindent then 
\[
K|\Phi (t)\rangle =KU|\Phi (q_i,t))=UK_0|\Phi (q_i,t))=UK_0U^{\dagger }|\Phi
(t)\rangle ...........(2.2.15) 
\]

\noindent Namely, if $K_0$ is the conjugation in the configuration
representation, the Wigner operator $K$ in a generic representation reads

\[
K=UK_0U^{\dagger }...........(2.2.16) 
\]

It is easy to show that in the configuration representation we have $K_0$
has the following properties:

\noindent a) $K_0$ is an antilinear, antiunitary operator, namely [17]:

a1) $K_0(\alpha |1\rangle + \beta |2\rangle) = \alpha ^* K_0 |1\rangle +
\beta ^* K_0 |2\rangle;$

a2) if $|\widehat{2}\rangle =K_0|2\rangle $, \ \ $\langle \widehat{1}%
|=\langle 1|K_0^{\dagger }$, and $\hat A=K_0AK_0^{\dagger }$, then ~~$%
\langle \hat 1|\hat A|\hat 2\rangle =\langle 1|A|2\rangle ^{*};$

a3) $(\langle1|K_0)|2\rangle = \langle1|(K_0|2\rangle)^*,$

\noindent i.e., parentheses cannot be omitted.

\noindent b) $K_0^2=1$ (at least for spin zero fields [17]);

c)
\[
K_0 \hat q_i K_0^{\dagger}=\hat q_i;
\]

\[
K_0 \hat p_i K_0^{\dagger}=-\hat p_i;
\]

d)
\[
K_0cK_0^{\dagger }=c^{*}\ \ {\rm if}\ \ c\in {\cal C}. 
\]
.

\noindent Therefore, $K_0 K_0^{\dagger}=1$ and $K_0 i K_0^{\dagger}=-i$.

From (2.2.16) it is also easy to show that $K$ has the same properties.

As an exercise we can repeat formulae (2.2.6) to (2.2.9) in a generic
representation. The time reversal is given by Eq. (2.2.12). The reversed
initial condition is

\[
|\Phi (0)_{rev}\rangle =K|\Phi (0)\rangle ,..........(2.2.17) 
\]

\noindent and the condition of reversible motion reads

\[
|\Phi (t)_{rev}\rangle =K|\Phi (-t)\rangle ...........(2.2.18) 
\]

We will say that $H$ is real if

\[
H=KHK^{\dagger },..........(2.2.19) 
\]

\noindent and usually $H$ is endowed with this property, because Eq. (2.2.8)
is satisfied in the configuration basis. Then, from Eqs. (2.2.17) and
(2.2.19) we can deduce again theorem 2.2.1, now in a generic coordinate
system:

\[
\vert \Phi(t)_{rev}\rangle=e^{-iHt}\vert \Phi(0)_{rev}\rangle=e^{-iHt}K\vert
\Phi(0)\rangle=
\]

\[
=K\big(K^{\dagger }e^{-iHt}K|\Phi (0)\rangle \big)=Ke^{iHt}|\Phi (0)\rangle
=K|\Phi (-t)\rangle ...........(2.2.20) 
\]

\noindent Then, as in the classical case, a usual real Hamiltonian yields a
reversible motion.

In general, we will call a ket $\vert 1\rangle$ (bra $\langle1\vert$) real if

\[
K|1\rangle =|1\rangle ,....................or 
\]

\[
\langle 1|K^{\dagger }=\langle 1|,..........(2.2.21) 
\]

\noindent and an operator $A$ real if 
\[
KAK^{\dagger }=A...........(2.2.22) 
\]

From (2.2.19) we see that a usual Hamiltonian is a real operator.

A basis $\{|i\rangle \}$ will be a real basis if all its kets are real:

\[
K|i\rangle =|i\rangle ...........(2.2.23) 
\]

In a real basis, $K$ is just the conjugation of the coordinates of the
vectors, or of the coordinates of the operators:

\[
K\vert \phi\rangle=K\sum_ic_i\vert i\rangle=\sum_ic_i^*\vert i\rangle,
\]

\[
KAK^{\dagger }=K\big(\sum_{ij}c_{ij}|i\rangle \langle j|\big)K^{\dagger
}=\sum_{ij}c_{ij}^{*}|i\rangle \langle j|...........(2.2.24) 
\]

\noindent Therefore, the configuration basis is $\{|x>\}$ real.

We will say that the conditions at $t=0$, are time-symmetric if:
\[
|\Phi (0)>=K|\Phi (0)>...............(2.2.25) 
\]
namely $|\Phi (0)>$ is real. Then, if the evolution is reversible we have:
\[
|\Phi (t)>=K|\Phi (-t)>..............(2.2.26) 
\]
and we will say that the evolution is time-symmetric with respect to $t=0.$
So we have the

{\bf Theorem 2.2.2.} If the evolution is reversible and the initial
condition is time-symmetric, the evolution is time-symmetric.

Then we can repeat what we have said in the classical case. If all the
quantum evolutions would be time-symmetric with respect to $t=0$ it would be
impossible to define a quantum arrow of time at $t=0.$

\subsection{Statistical formalism.}

We shall treat simultaneously the classical and quantum cases in order to
establish an analogy or unified formalism that we shall use below.
Nevertheless, it must be clear that there is a great difference between the
classical and quantum cases.

We will call the classical {\it distribution function or density} (resp.,
the quantum {\it density matrix}) a function (resp., matrix) endowed with
the following properties:

\[
\rho (q_i,p_i)\geq 0;..or...\rho (x)\geq 0 
\]

\[
\parallel \rho \parallel =\int_X\rho (q_i,p_i)dq_idp_i=\int_X\rho
(x)dx=1,..........(2.3.1) 
\]

\noindent where $X$ is the phase space. Distribution functions $\rho $
belong to a $L^1 $ Hilbert space called the classical Liouville space.

(resp., in the quantum mechanical formalism:

\[
\rho=\rho^{\dagger};
\]

\[
tr(\rho)=1;
\]

\[
\rho _{\alpha \alpha }\geq 0)...........(2.3.2) 
\]
Density matrices $\rho $ belong to a space ${\cal L=H\times H}$ called the
quantum Liouville space.)

\noindent $\rho$ satisfies the Liouville equation

\[
i\partial _t\rho =L\rho ,..........(2.3.3) 
\]

\noindent where 
\[
L=i\{H,..\}_{PB}..........(2.3.5) 
\]

\noindent (resp., 
\[
L=[H,..]=H\times 1-1\times H...........(2.3.5) 
\]
(cf. eq. (2.A.24) for the definition of $\times $)).

Therefore the time evolution, in both classical and quantum cases, is:
\[
\rho (t)=e^{-iLt}\rho (0)=U(t)\rho (0)..........(2.3.5^{\prime }) 
\]

The $T$ transformation of a density function is

\[
T\rho (q_i,p_i)=\rho ^{\prime }(q_i,p_i)=\rho (q_i,-p_i)..........(2.3.6) 
\]

\noindent (resp., the $T$ transformation of a density matrix is

\[
T\rho =\rho ^{\prime }=K\rho K^{\dagger }={\cal K}\rho
.........................(2.3.6^{\prime }). 
\]
where ${\cal K}=K\times K^{\dagger }).$

From Eq. (2.3.3), if the Hamiltonian is a usual time-symmetric one, we have
classically

\[
T L \rho(q_i,p_i)=T i \{H,\rho\}_{PB}=
\]

\[
Ti\sum_i\partial _{q_i}H\partial _{p_i}\rho -\partial _{p_i}H\partial
_{q_i}\rho = 
\]

\[
i \sum_i\partial_{q_i}H\partial_{-p_i}T\rho-\partial_{-p_i}H
\partial_{q_i}T\rho=
\]

\[
-i\{H,T\rho \}_{PB}=-LT\rho ;..........(2.3.7) 
\]

\noindent therefore, if we $T$-transform classically Eq. (2.3.2), we obtain

\[
i\partial _tT\rho =-LT\rho ..........(2.3.8) 
\]

\noindent (resp., if we $T$-transform the quantum Liouville equation
(2.3.2), we obtain, if the Hamiltonian is a real usual one,

\[
KiK^{\dagger }\partial _tK\rho K^{\dagger }=(K\times K^{\dagger
})L(K^{\dagger }\times K)K\rho K^{\dagger };..........(2.3.8^{\prime }) 
\]

\noindent but $K i K^{\dagger}=-i$, and \ \ $(K\times
K^{\dagger})L(K^{\dagger}\times K)=KHK^{\dagger} \times 1 - 1 \times K H
K^{\dagger}= L$, so

\[
-i\partial _tK\rho K^{\dagger }=LK\rho K^{\dagger
},........................(2.3.8") 
\]

\noindent i.e., the same equation as the classical one (2.3.8)).

In both cases a minus sign appears. In the reverted solution we must change $%
t$ by $-t$, namely:
\[
T(t)=t^{\prime }=-t................................(2.3.9) 
\]
So we have proved the

{\bf Theorem 2.3.1.}The Liouville equation remains invariant under $T$
transformations for a usual time-symmetric Hamiltonian.

Thus we have shown the complete isomorphism of the classical and quantum
formalisms. From now on we will mainly use the quantum formalism, since it
is the one that is better known by physicists. Let us therefore review the
main properties of the usual Hamiltonian, in a real basis to simplify the
treatment. From the equations of Sect. 2.2. we have

\[
H=H^{\dagger},
\]

\[
H=H^*,
\]

\[
H=H^T,..........(2.3.10) 
\]

\noindent namely, the Hamiltonian is

I- self-adjoint, because it is an observable;

II- real, because for the usual Hamiltonian the motion is reversible;

III- as a consequence, it is also symmetric.

\noindent $\rho $ belongs to a set that, endowed with the inner product
(2.A.1) of the appendix, becomes the Liouville-Hilbert space ${\cal L}$.
From Eqs. (2.A.29) and (2.3.10)) we can prove that the Liouvillian has the
following properties in real basis:

\[
L=L^{\dagger},
\]

\[
L=L^*,
\]

\[
L=-L^a,
\]

\[
L=-L^T...........(2.3.11) 
\]

Then, from (2.3.11) and (2.A.18) we have

\[
(iL)=(iL)^a..........(2.3.10) 
\]

This property is important since, from it, we can deduce that the matrix $%
\rho $ remains Hermitian under the evolution satisfying to the Liouville
equation (2.3.2). In fact, it follows from Eq. (2.A.18) that a product of
self-associated commuting operators is also self-associated. Then, $%
(iL)=(iL)^a$ implies $(e^{-iLt})^a=e^{-iLt}$ and $\rho (0)=\rho (0)^{\dagger
}$ implies $e^{-iLt}\rho (0)=[e^{-iLt}\rho (0)]^{\dagger }$, namely, $\rho
(t)=\rho (t)^{\dagger }$.

Finally, let we prove once more that if the Liouvillian is real the
evolution is reversible. Based on Eqs. (2.1.7) and (2.2.7) we define a
reversible motion, in a real basis, as

\[
\rho _{rev}(t)=\rho ^{*}(-t),..........(2.3.13) 
\]

\noindent where $\rho_{rev}(t)$ is the motion with reversed condition at
zero time:

\[
\rho _{rev}(0)=\rho ^{*}(0)...........(2.3.14) 
\]
Now we can prove the

{\bf Theorem2.3.2. }If the liouvillian is real the evolution is reversible.

Proof.:

Then, with the same reasoning as for eqs. (2.2.9) we have:

\[
\rho _{rev}(t)=e^{-iLt}\rho _{rev}(0)=e^{-iLt}\rho ^{*}(0)=[e^{iLt}\rho
(0)]^{*}=\rho ^{*}(-t),..........(2.3.15) 
\]

\noindent which shows that a motion with a real Liouvillian is reversible $%
\Box $

.In a generic basis eqs. (2,3,11) t0 (2.3.15) read as follows:

The liouvillian is real or time symmetric if:.
\[
{\cal K}L{\cal K}^{\dagger
}=L,.......................................(2.3.16) 
\]
The evolution is time-symmetric if:
\[
\rho _{rev}(t)=K\rho (-t)K^{\dagger }={\cal K}\rho (-t).............(2.3.17) 
\]
The conditions at time $t=0$ are time-symmetric if:
\[
\rho _{rev}(0)=K\rho (0)K^{\dagger }={\cal K}\rho (0)...............(2.3.18) 
\]
A real liouvillian and time-symmetric conditions at $t=0$ yields a
time-symmetric evolution since:
\[
\rho _{rev}(t)=e^{-iLt}\rho _{rev}(0)=e^{-iLt}K\rho (0)K^{\dagger
}=K[e^{iLt}\rho (0)]K^{\dagger }=K\rho (-t)K^{\dagger }......(2.3.19) 
\]
The condition at $t=0$ will be called time-symmetric if:
\[
\rho (0)=K\rho (0)K^{\dagger }={\cal K}\rho (0)...................(2.3.20) 
\]
then if the evolution is irreversible we have:
\[
\rho (-t)={\cal K}\rho (t)....................................(2.3.21) 
\]
and we will say that the whole evolution is time symmetric and we can repeat
what we have said in the previous cases. So we have the

{\bf Theorem 2.3.3.} If the evolution is reversible and the condition at $%
t=0 $ is time-symmetric the evolution is time-symmetric with respect to $%
t=0. $

Proof.:

If the liouvillian satisfies eq. (2.3.16) and condition at $t=0$ satisfies
eq. (2.3.20) all the evolution is time-symmetric, since:
\[
{\cal K}\rho (t)={\cal K}(e^{-iLt}\rho (0))=e^{i{\cal K}L{\cal K}^{\dagger
}t}{\cal K}\rho (0)=e^{iLt}\rho (0)=\rho (-t)..........(2.3.22) 
\]
Therefore the motion is time-symmetric if $L$ is real and the condition at
time $t=0$ is time symmetric.$\Box $

\subsection{Appendix 2 A. Mathematical theory of superspace and
superoperators [19].}

Let us make a small mathematical interlude, to define the notions of {\it %
superspace} and {\it superoperators}.

\subsubsection{The quantum case}

Let us consider a Hilbert space ${\cal H}$ and the space ${\cal L=H\times H}$
of matrices on ${\cal H}$, i.e., the Liouville-Hilbert space. Matrices will
be symbolized by greek lower case letters $\alpha $, $\beta $,....,$\rho $,
with coordinate $\alpha _{ij}$, $\beta _{ij}$,....,$\rho _{ij}$. We will
call the linear space of matrices the {\it superspace} ${\cal L}$ and the
matrices {\it supervectors}. Let us define an inner product in the
superspace ${\cal L}$:

\[
\alpha \cdot \beta =(\alpha |\beta )=tr(\alpha ^{\dagger }\beta
)=\sum_{ij}\alpha _{ij}^{*}\beta _{ij}...........(2.A.1) 
\]
Using this inner product {\cal L} becomes a $L^2$ Hilbert space.

The norm of a supervector is thus

\[
||\alpha ||=\alpha \cdot \alpha =\sum_{ij}|\alpha _{ij}|^2\geq
0...........(2.A.2) 
\]

Let us consider the linear operators in superspace, that we shall call {\it %
superoperators}, and that we shall represent by capital Latin letters $A$, $%
B $, ...., $L$, with coordinates $A_{ij,kl}$, $B_{ij,kl}$. Superoperators
act on matrices as

\[
A\alpha = \beta,
\]

\[
\alpha A=\beta ...........(2.A.3) 
\]

\noindent We will use, for these two equations, the following rule for
indices

\[
\sum_{kl}A_{ij,kl} \alpha_{kl}=\beta_{ij},
\]

\[
\sum_{kl}\alpha _{lk}^TA_{lk,ji}=\beta _{ji}^T...........(2.A.4) 
\]

In the first equation we have used the usual multiplication ''row by
column'' and $\alpha $ and $\beta $ are considered as column vectors. In the
second one we have transposed $\alpha $ and $\beta $ since, in these case,
they are considered as row vectors.

Since the superoperators have four indices we can define more operations
defining transposed and adjoints than for ordinary two-indices matrices. So,
we define, for a superoperator A, its

$a$- {\it Transposed} $A^T$ as the superoperator such that

\[
A\alpha =\alpha A^T..........(2.A.5) 
\]

\noindent for all $\alpha \in {\cal L}$.

Then, 
\[
A_{ij,kl}\alpha _{kl}=\alpha _{kl}A_{lk,ji}^T,..........(2.A.6) 
\]

\noindent so 
\[
A_{ij,kl}=A_{lk,ji}^T...........(2.A.7) 
\]

\noindent Of course,

\[
(A^T)^T=A,
\]

\[
(A_1A_2)^T=A_2^TA_1^{,}..........(2.A.8) 
\]

\noindent and $A$ is symmetric (antisymmetric) if

\[
A=A^T,\ \ (A=-A^T)...........(2.A.9) 
\]

$b$- {\it Adjoint} $A^{\dagger}$ as the superoperator such that

\[
A\alpha =(\alpha ^{\dagger }A^{\dagger })^{\dagger }..........(2.A.10) 
\]

\noindent for all $\alpha \in {\cal L}$. Then,

\[
A_{ij,kl}\alpha _{kl}=(\alpha _{lk}^{*}A_{lk,ji}^{\dagger })^{\dagger
}=\alpha _{lk}(A_{lk,ij}^{\dagger })^{*},..........(2.A.11) 
\]

\noindent so 
\[
A_{ij,kl}^{*}=A_{kl,ij}^{\dagger }...........(2.A.12) 
\]

\noindent Of course, 
\[
(A^{\dagger})^{\dagger}=A,
\]

\[
(A_1A_2)^{\dagger }=(A_2^{\dagger }A_1^{\dagger }),..........(2.A.13) 
\]

\noindent and $A$ is Hermitian (anti-Hermitian) if

\[
A=A^{\dagger },\ \ (A=-A^{\dagger })...........(2.A.14) 
\]

$c$- {\it Associated} $A^a$ as the superoperator such that

\[
A\alpha =(A^a\alpha ^{\dagger })^{\dagger }..........(2.A.15) 
\]

\noindent for all $\alpha \in {\cal L}$. Then,

\[
A_{ij,kl}\alpha _{kl}=(A_{ij,kl}^a\alpha _{kl}^{\dagger })^{\dagger
}=(A_{ij,kl}^a\alpha _{lk}^{*})^{\dagger }=(A^a*_{ji,kl}\alpha
_{lk}),..........(2.A.16) 
\]

\noindent so 
\[
A_{ij,kl}^{*}=A_{ji,lk}^a...........(2.A.17) 
\]

\noindent Of course,

\[
(A^a)^a=A,
\]

\[
(A_1A_2)^a=A_1^aA_2^a,..........(2.A.18) 
\]

\noindent and an operator is {\it adjoint-symmetric} (or self-associated) if

\[
A=A^a...........(2.A.19) 
\]

An adjoint-symmetric operator acting on a Hermitian matrix gives another
Hermitian matrix. For, if

\[
\alpha =\alpha ^{\dagger },\ \ A=A^a,..........(2.A.20) 
\]

\noindent then from Eq. (2.A.15) we have

\[
A\alpha =(A\alpha )^{\dagger }...........(2.A.21) 
\]

\noindent Putting all together, we have

\[
A_{ij,kl}=A_{lk,ji}^T=(A_{kl,ij}^{\dagger
})^{*}=(A_{ji,lk}^a)^{*},..........(2.A.22) 
\]

\noindent and therefore

\[
A^{aT}=A^{\dagger }...........(2.A.23) 
\]

Let us now define a superoperator as a product of two operators, $%
A=\alpha\times \beta$, in the following way:

\[
A\gamma =\alpha \gamma \beta ,\ \ \forall \gamma ,..........(2.A.24) 
\]

\noindent or, equivalently,

\[
\sum_{kl} A_{ij,kl} \gamma_{kl}=\sum_{kl} \alpha_{ik} \gamma_{kl} \beta_{lj},
\]

\noindent that is,

\[
A_{ij,kl}=\alpha _{ik}\beta _{lj}...........(2.A.25) 
\]

Then,

\[
\sum_{kl}\gamma _{kl}A_{lk.ji}=\sum_{kl}\gamma _{kl}\alpha _{lj}\beta
_{ik}=\sum_{kl}\beta _{ik}\gamma _{kl}\alpha _{lj}..........(2.A.26) 
\]

\noindent and, from (2.A.4), 
\[
\gamma A=\beta \gamma \alpha ...........(2.A.27) 
\]

\noindent Therefore, we have from (2.A.24) and (2.A.27),

\[
(\alpha \times \beta) \gamma=\alpha \gamma \beta,
\]

\[
\gamma (\alpha \times \beta )=\beta \gamma \alpha ...........(2.A.28) 
\]

The choice of the index position,in eq. (2.A.4), was made in order to obtain
these simple multiplication rules. It is easy to prove that

\[
(\alpha \times \beta)^T = \beta \times \alpha;
\]

\[
(\alpha \times \beta)^{\dagger}=\alpha^{\dagger} \times \beta^{\dagger};
\]

\[
(\alpha \times \beta )^a=\beta ^{\dagger }\times \alpha ^{\dagger
}...........(2.A.29) 
\]

The product $\times $ can be used to define the time inversion of matrices,
since a time inverted matrix is (Eq. (2.2.22))):

\[
T\rho =K\rho K^{\dagger }=(K\times K^{\dagger })\rho ={\cal K}\rho
...........(2.A.30) 
\]

\noindent From this equation we can deduce the time inversion rule of
superoperators, namely,

\[
TA=(K\times K^{\dagger })A(K\times K^{\dagger })^{\dagger }={\cal K}A{\cal K}%
^{\dagger };..........(2.A.31) 
\]

\noindent since (cf. Eq. (2.A.29))

\[
(K \times K^{\dagger})^{\dagger}=K^{\dagger} \times K,
\]

\noindent we have the alternative expression

\[
TA=(K\times K^{\dagger })A(K^{\dagger }\times K)...........(2.A.32) 
\]

We can also compute $(\alpha \times \beta)(\gamma \times \delta)$:

\[
\sum_{kl}(\alpha \times \beta)_{ij,kl} (\gamma \times \delta)_{kl,nm}=
\]

\[
\sum_{kl}\alpha_{ik}\beta_{lj}\gamma_{kn}\delta_{ml}=\sum_{kl}\alpha_{ik}%
\gamma_{kn} \delta_{ml}\beta_{lj}=
\]

\[
(\alpha \gamma \times \delta \beta )_{ij,nm},..........(2.A.33) 
\]

\noindent namely, 
\[
(\alpha \times \beta )(\gamma \times \delta )=(\alpha \gamma \times \delta
\beta )...........(2.A.34) 
\]

\subsubsection{Classical case.}

As we have seen the quantum Liouville space is transformed in a $L^2$
Hilbert space by the inner product (2.A.1). In the same way it is convenient
to define an inner product in the classical Liouville space ${\cal L}$
,namely:
\[
(\rho |\sigma )=\int_X\rho ^{*}(x)\sigma (x)dx.............(2.A.35) 
\]
Using this inner product and Wigner functions (App. 6.A) the classical $%
{\cal L}$ becomes also a $L^2$ Hilbert space, and the classical analogs of
the quantum equation of the previous subsection can be found. Also we can
use the Wigner function integral of appendix 6. A to make the analogy
explicit.

\section{Thermodynamics.}

\subsection{Classification of the different types of second laws.}

The first law of thermodynamics is just the conservation of the energy.
There is no conflict between dynamics and thermodynamics about this law. The
problem is to derive the second law of thermodynamics based in dynamical
considerations. The second law is expressed in many forms by the different
authors, therefore we will begin our research by a classification of these
forms.

Let $S(t)$ denote the thermodynamical entropy of a closed system:

i.-We will call a first-order second law to the statement:
\[
S(t)\geq S(t^{\prime })........................(3.1.1) 
\]
if $t\geq t^{\prime }$, thus according to this form the entropy cannot
decrease.

ii.-An stronger assertion would be a second-order second law: Eq. (3.1.1) is
satisfied and also:
\[
\stackunder{t\rightarrow +\infty }{\lim }%
S(t)=S_{*}.....................(3.1.2) 
\]
in this case we assert that the system entropy converges to a steady-state
value $S_{*}$, which may not be unique, e.g. it can be the entropy of a
metastable state. Different preparations of the system could yield different
final metastable states.

iii.-The final and stronger form, or third-order form, of the second law is:
Eqs. (3.1.1) and (3.1.2) are satisfy but also the limit (3.1.2) is unique.
In this case the entropy of the systems evolve to a unique maximum,
irrespective of the way the system was prepared.

We will find these different forms of the second law below.

\subsection{Dynamics and densities.}

We will consider more generic systems than the ones of section 2, in order
to be as general as possible. This is not done just by the sake of
mathematical generality but because we will be forced to consider such a
systems below, if we want to solve our problems.So let us consider a system
operating in a phase space $X$, with a evolution law $S_t$ more general than
(2.1.3), i. e. a mapping $S_t:X\rightarrow X$, that change the point $x$ of $%
X$ as $t$ changes. $X$- may have finite dimension $d$ or infinite dimension,
and $t$ can be discrete or continuous. We will consider only ''autonomous''
processes: i.e. such that $S_t(S_{t^{\prime }}(x))=S_{t+t^{\prime }}(x),$ $%
S_0(x)=x$. Thus the mapping $S$ can form either a group of transformation
when $t,t^{\prime }\in R,($or $Z)$ ( e. g. the evolutions with
time.symmetric hamiltonian or liouvillian of sec. 2) or a semigroup if $%
t,t^{\prime }\in R^{+}($or $N$). In the two last cases ($R^{+},N)$ an
equation like (2.1.7) do not exists and the evolution is necessarily
irreversible.

For every point $x_0$ the successive point $S_t(x_0)$ are a system
trajectory. To study an infinite number of initial point or an infinite
number of trajectories we introduce the density functions $\rho (x)\in
L^1(X) $ ,namely:
\[
\int_X|\rho (x)|dx<\infty ,...........(3.2.1) 
\]
and such that:
\[
\rho (x)\geq 0,\parallel \rho (x)\parallel =1,.......(3.2.2) 
\]
where:
\[
\parallel \rho (x)\parallel =\int_X|\rho (x)|dx.........(3.2.1) 
\]
is the $L^1$-norm of $\rho $. We use to postulate that a thermodynamics
system is a system that has , at a given time, states distributed throughout
the phase space $X$ and the distribution of these states is characterized by
the density function $\rho (x)$.

We will call the $\rho $-measured $\mu _\rho (A)$ of the set $A\subset X$ to:
\[
\mu _\rho (A)=\int_A\rho (x)dx...........(3.2.4) 
\]
The Lebesgue (non-normalized usual) measure of a set $A$-will be denoted $%
\stackrel{-}{\mu _L}(A)$. The uniform density will be 
\[
\rho _L(x)=\frac 1{\stackrel{-}{\mu _L}(X)}.....................(3.2.5) 
\]
and therefore Lebesgue normalized measure is $\mu _{\rho _L}(X)=\mu _L(X)=1.$%
We always write $\stackrel{-}{\mu _L}(dx)=dx.$

Finally $X$ can be either Gibbs phase space $\Gamma $ or Boltzmann phase
space $\mu $ [6].

\subsection{Gibbs entropy.}

This entropy is defined as:
\[
H(\rho )=-\int_X\rho (x)\log \rho (x)dx.....(3.3.1) 
\]
It is and additive quantity, namely the Gibbs-entropy of a system form by
two subsystem is the sum of the two corresponding entropies. Then it is
called an extensive quantity. Gibbs-entropy can be written as:
\[
H(f)=\int_X\eta (\rho (x))dx................(3.3.2) 
\]
where the $\eta (\rho )$ function is defined as:
\[
\eta (\rho )=-\rho \log \rho ..for..\rho >0..and..\eta (0)=0....(3.3.3) 
\]
and it is endowed with the property:
\[
\eta (\rho )\leq (\rho -\sigma )\eta ^{\prime }(\sigma )+\eta (\sigma
)........(3.3.4) 
\]
combining these last two formulae we can prove the Gibbs inequality:
\[
\rho -\rho \log \rho \leq \sigma -\rho \log \sigma ,..for..\rho ,\sigma
>0..........(3.3.5) 
\]
If $\rho $ and $\sigma $ are two normalized density function, integrating
the last equation we have:
\[
-\int_X\rho (x)\log \rho (x)dx\leq -\int_X\rho (x)\log \sigma (x)dx..(3.3.6) 
\]
Only when $\rho =\sigma $ does the equality hold in eqs.(3.3.4), (3.3.5),
and (3.3.6)

\subsection{Microcanonical and canonical ensembles.}

Let us consider an space $X$ with a finite Lebesgue measure: $\stackrel{-}{%
\mu _L}(X)<\infty $.Then the only density that will make Gibbs-entropy
maximal is the uniform density of eq. (3.2.5). Precisely:

{\bf Theorem 3.4.1.}When $\stackrel{-}{\mu _L}(X)<\infty $ the density that
maximized the Gibbs-entropy is the uniform density, $\rho _L(x)$ (cf. eq.
(3.2.5). For any other density $\rho \neq \rho _{L,}H(\rho )<H(\rho _L).$

Proof.: Chose an arbitrary density $\rho $ ,thus from eq. (3.3.6) we have:
\[
H(\rho )\leq -\int_X\rho (x)\log \sigma (x)dx.......................(3.4.1) 
\]
However, if $\sigma (x)=1/\stackrel{-}{\mu }_L(X)$ the integrated Gibbs
inequality (3.3.6) gives:
\[
H(\rho )\leq -\log \left[ \frac 1{\stackrel{-}{\mu }_L(X)}\right]
................................(3.4.2) 
\]
since $\rho $ is normalized to one. The equality holds if $\rho =\rho _L$ ,
but the entropy corresponding to $\rho _L$ is:
\[
H(\rho _L)=-\log \left[ \frac 1{\stackrel{-}{\mu }_L(X)}\right]
...............................(3.4.3) 
\]
therefore $H(\rho )\leq H(\rho _{L)}$ for any density $\rho $ and $H(\rho
)<H(\rho _L)$ for $\rho \neq \rho _{L.}$ Clearly if $X$ is normalized so $%
\stackrel{-}{\mu }_L(X)=1,$ then $H(\rho )\leq 0.$ $\Box $

The uniform density it is also called the density of a microcanonical
ensemble, and, as we can see, to define it we do not need to use any
particular property of the thermodynamical system under consideration.

Another,even more interesting theorem is the following:

{\bf Theorem 3.4.2.}Assume that a non-negative measurable function $\alpha
(x)$ is given as well as an average or expectation mean value $<\alpha
>_\rho $ of that function over the entire $X$, weighted by the density $\rho 
$:
\[
<\alpha >_\rho =\int_X\alpha (x)\rho (x)dx........(3.4.4) 
\]
Then the maximum of the Gibbs-entropy $H(\rho )$ subject to the constraint $%
<\alpha >_\rho =const.$ occurs for the density:
\[
\rho _{*}(x)=Z^{-1}e^{-\nu \alpha (x)}................(3.4.5) 
\]
where:
\[
Z=\int_Xe^{-\nu \alpha (x)}dx................(3.4.6) 
\]
and $\nu $-is implicitly defined by the normalization condition:
\[
<\alpha >_\rho =Z^{-1}\int_X\alpha (x)e^{-\nu \alpha (x)}dx......(3.4.7) 
\]
Proof: The proof again uses the integrated Gibbs equality (3.3.6) so:
\[
\begin{array}{c}
H(\rho )\leq -\int_X\rho (x)\log \rho _{*}(x)dx=-\int_X\rho (x)[-\log Z-\nu
\alpha (x)]dx= \\ 
=\log Z+\nu \int_X\rho (x)\alpha (x)dx=\log Z+\nu <\alpha >_\rho
\end{array}
(3.4.8) 
\]
However, it is equally easy to show that:
\[
H(\rho _{*})=\log Z+\nu <\alpha >_\rho ..................(3.4.9) 
\]
and therefore $H(\rho )\leq H(\rho _{*})$, with the equality holding if and
only if $\rho =\rho _{*},$ $\Box $

If $\alpha (x)$ is the energy of the system $\rho _{*}$ is density of the
Gibbs canonical ensemble at temperature $T=\nu ^{-1}$. (With many
constraints $<\alpha _i>_\rho $ we would define the density of a grand
canonical ensemble).

We use to postulate also that there is a one-to-one correspondence between
thermodynamical equilibrium states and the states of maximum entropy. Then,
from the last theorems it would be natural to postulate also that the
thermodynamical entropy $S$ coincide with Gibbs entropy $H(\rho )$. In fact
with this postulate we can obtain usual equilibrium thermodynamical. But, as
we shall see below, this identification is not what we need to base a
non-equilibrium thermodynamics, since Gibbs-entropy has not the wright
properties, in this case.

\subsection{Reversible and irreversible systems.}

In section 2 the properties of the hamiltonian force the motion to be either
reversible or irreversible.. But in this section we are studying more
general motions so we are force to repeat these definition for this more
general cases. Nevertheless, in order to prove some theorems, the motions
cannot be completely general so we will restrict ourselves to motion
produced by Markov operators.

Any linear operator: $P_t:L^1\rightarrow L^1$ such that:
\[
(a)..P_t\rho \geq 0,.......(b)..\parallel P_t\rho \parallel =\parallel \rho
\parallel .........(3.5.1) 
\]
for all $t\in R,\rho \geq 0,\rho \in L^1$ is a Markov operator, i.e. an
operator that acting on a density gives a density. Markov operator have a
number of useful properties. The most important is that if $\rho \in L^1$
and it is not restricted to $\rho \geq 0$, then:
\[
\parallel P_t\rho \parallel \leq \parallel \rho \parallel
...........................................(3.5.2) 
\]
which is known as contractive property.

A Markov operator is reversible (or time-symmetric) if:
\[
(a)..P_0\rho =\rho ,..(b)..P_t(P_{t^{\prime }}\rho )=P_{t+t^{\prime }}\rho
.....(3.5.3) 
\]
for all $t,t^{\prime }\in R$ (or $Z$ ), namely reversible Markov operators
form a group. Evolution operator $U(t)=e^{iLt}$ of eq. (2.3.5') is an
example of reversible Markov operator. It is so because it is generated by a
time-symmetric or real liouvillian $L$.

However, if in the last definition we substitute $R$ and $Z$ by $R^{+}$ and $%
N$ we have the definition of an irreversible Markov operator. Irreversible
Markov operators form a semigroup.

Gibbs entropy cannot be use in non-equilibrium theory since it may decrease
under the action of some Markov operators (cf. [6]), therefore we cannot use
this entropy to formulate a second law of thermodynamic, even in the
first-order form. Nevertheless Gibbs-entropy is completely successful in
equilibrium situations, so the entropy we will chose, for non-equilibrium
situations, must coincide with Gibbs entropy at equilibrium.

\subsection{Conditional entropy.}

If $\rho $ and $\sigma $ are two densities such that supp$\rho \subset $supp$%
\sigma $, then the conditional entropy of density $\rho $, with respect to
density $\sigma ,$ is:(3.6.2)
\[
H_C(\rho |\sigma )=-\int_X\rho (x)\log \frac{\rho (x)}{\sigma (x)}%
dx..............(3.6.1) 
\]
The conditional entropy is always definite, i. e.: it is finite or equal to -%
$\infty $. As it is evident the conditional entropy measured the deviation
of density $\rho $ from density $\sigma $.

Conditional entropy has two very important properties:

i.-Since $\rho $ and $\sigma $ are both densities the integrated Gibbs
inequality (3.3.6) implies that $H_C(\rho |\sigma )\leq 0$ . It is only when 
$\rho =\sigma $ that the equality hold.

ii.-If $\rho _L$ is the constant density of the microcanonical ensemble
throughout the phase space $X$ then $H_C(\rho |\rho _L)=H(\rho )-\log 
\stackrel{-}{\mu _L}(X)$. Therefore conditional entropy is a generalization
of Gibbs entropy in this case.

As $H_C(\rho |\rho _{*})=0$ when $\rho =\rho _{*}$ it is reasonable to
postulate that:
\[
S-S_{*}=H_C(\rho |\rho _{*})................(3.6.2) 
\]
e.g. when $\rho _{*}$ is the density of the canonical ensemble. We will see
that this definition is completely satisfactory and that using equation
(3.6.2) we can formulate the second law of thermodynamics in its second and
third-order forms. The first result, along these lines, is a weak,
first-order form, of the law of thermodynamics, namely that the conditional
entropy is never decreasing, as it is proved by the

{\bf Theorem 3.6.1.}[20] Let $P_t$ be a Markov operator. Then:
\[
H_C(P_t\rho |P_t\sigma )\geq H_C(\rho |\sigma ).........(3.6.3) 
\]
for all densities $\rho $ and $\sigma $.

A second result is the following: if $\sigma =\rho _{*}$ is stationary,
namely $P_t\rho _{*}=\rho _{*}$ , then:
\[
H_C(P_t\rho |\rho _{*})\geq H_C(\rho |\rho _{*})...........(3.6.4) 
\]
Thus this conditional entropy is always a non decreasing function bounded
above and $H_{\max }=H_C(\rho _{*}|\rho _{*})=0$. Therefore this conditional
entropy converge as $t\rightarrow \infty $ ,though more information about
the evolution is required to find the limiting value. Furthermore if the
stationary density is uniform, namely the one of the microcanonical ensemble
we have:
\[
H(P_t\rho )\geq H(\rho ).........................(3.6.5) 
\]
for all non-negative $\rho $. Now $H_{\max }=-\log [1/\stackrel{-}{\mu _L}%
(X)]$ and as, in the general case we have convergency when $t\rightarrow
\infty $.

Therefore eq. (3.6.2) seems a reasonable assumption. But when the Markov
operator is reversible all these nice inequalities become equalities and the
problem of the thermodynamical entropy reappears. In fact:

{\bf Theorem 3.6.2. }If $P_t$ is a reversible Markov operator, then the
conditional entropy is absolutely constant for all times $t$ and equal to
the value determinated by the choice of the initial densities $\rho $ and $%
\sigma $. That is:
\[
H_C(P_t\rho |P_t\sigma )=H_C(\rho |\sigma )............(3.6.6) 
\]
for all $t.$

Proof.: Since $P_t$ is reversible, by the previous theorem it follows that:
\[
\begin{array}{c}
H_C(P_{t+t^{\prime }}\rho |P_{t+t^{\prime }}\sigma )=H_C(P_tP_{t^{\prime
}}\rho |P_tP_{t^{\prime }}\sigma ) \\ 
\geq H_C(P_t\rho |P_t\sigma )\geq H_C(\rho |\sigma )...(3.6.7)
\end{array}
\]
for all $t,t^{\prime }$ since $P_t$ is reversible$.$ So let us chose $%
t^{\prime }=-t,$ then for all times we have:
\[
H_C(\rho |\sigma )\geq H_C(P_t\rho |P_t\sigma )\geq H_C(\rho |\sigma
).....(3.6.8) 
\]
and therefore:
\[
H_C(P_t\rho |P_t\sigma )=H_C(\rho |\sigma )......................(3.6.9) 
\]
for all $t.$ $\Box $

So in this case the conditional entropy is, for ever, fixed and determined
by the method of preparation of the system. So we have gain nothing if the
Markov operator is reversible.

\subsection{Appendix 3.A : The physical interpretation of non-equilibrium
Gibbs-entropy.}

Gibbs-entropy is completely successful in equilibrium cases. Therefore, even
if not the correct definition of entropy in non-equilibrium cases, it must
have some physical meaning in these last cases. In fact, to create an
unstable state, with a decreasing of entropy, it is necessary to provide
some energy to the system, on the other hand, if a systems evolves from an
unstable state toward equilibrium it release energy and the entropy grows.
Therefore there we must find some relation like
\[
\Delta S\approx -C\Delta E+C^{\prime }.....,C>0,..........(3.A.1) 
\]
This is the case since from eq. (3.6.2) we can obtain:
\[
\Delta S=\Delta H_C=-\int \rho _2\log \frac{\rho _2}{\rho _{*}}dx+\int \rho
_1\frac{\rho _1}{\rho _{*}}dx........(3.A.2) 
\]
where $\rho _1$ and $\rho _2$ are the initial and final distribution
functions. Now let us suppose that the final state is a canonical ensemble
equilibrium state at temperature $T$ (cf.eq.(3.4.7). Then we obtain:
\[
\Delta S=\Delta H-\frac 1T\int (\rho _2-\rho _1)\omega dx=\Delta H-\frac{%
\Delta E}T.........(3.A.3) 
\]
so we have obtained the equation we were looking for: in fact, there is a
relation like (3.A.1) where the coefficient is $C=T^{-1}$ and $C^{\prime
}=\Delta H..$ We have also obtained the interpretation of the variation of
the Gibbs entropy in irreversible evolutions: it is the difference between $%
\Delta S$ and the linear term $T^{-1}\Delta E,$ in such a way that if $%
\Delta H=0,$ all the energy is used to produce entropy or all the entropy is
used to produce energy. Therefore the difference of Gibbs entropy measure,
some how, the efficient of the system to produce entropy or energy. On the
other hand, if the system is isolated, so $\Delta E=0,$ and if the
equilibrium state corresponds to the one of a canonical ensemble, Gibbs
entropy coincide with thermodynamic and conditional entropy.

\section{The Classical Evolution.}

In this section we will study ''classical evolutions'', in the sense that
these evolution are not quantum ones. Nevertheless the evolutions will be as
general as the one of the previous section: i. e., not necessarily those of
section 2.

\subsection{The Frobenius-Perron operator.}

A transformation $S_t$ is called a measurable transformation if $\mu _{*}($$%
S_t^{-1}(A))$ is well defined for all subsets $A\subset X,$ where $%
S_t^{-1}(A)=B$ is the counterimage of $A$-namely: $S_t(B)=A$. Let us remark
that even if a unique $S_t^{-1}(x)$ may not exist (as ion the case of
irreversible evolutions) the counterimage do exists since it is the set of
all the points $x\in B$ that will go to $A$ under the action of $S_{t\text{.}%
}$

The transformation is non-singular if $\mu
_{*}(S_t^{-1}(A))=0\Longleftrightarrow \mu _{*}(A)=0$.

If $S_t$ is a non-singular transformation, then the unique operator: $%
P_t:L^1\rightarrow L^1$ defined by:
\[
\int_AP_t\rho (x)dx=\int_{S_t^{-1}(A)}\rho (x)dx..................(4.1.1) 
\]
is called the Frobenius-Perron operator corresponding to $S_t$. For each $%
S_t $ the Frobenius-Perron operator is unique. If $\rho \geq 0$ then $%
P_t\rho \geq 0.$ As $S_t^{-1}(X)=X$ then $\parallel P_t\rho \parallel
=\parallel \rho \parallel $ and these operators are Markov operators.
Operator $U(t)$ of eq. (2.3.5') is a Frobenius-Perron operator.

$S_t$ is $\rho $-measure preserving if:
\[
\mu _\rho (S_t^{-1}(A))=\mu _\rho (A) 
\]

for all sets $A$. Measure-preserving transformation are necessarily non
singular, we will also say that the measure $\mu _\rho $ is invariant under
the transformation. Liouville theorem shows that transformation $U(t)$ of
eq. (2.3.5') is Lebesgue-measure preserving.

We will call a state $\rho $ steady if $P_t\rho =\rho ,$ for all $t.$ We
will call it also a state of ''thermodynamical equilibrium'', and we will
symbolize it by $\rho _{*}.$

The relation between invariant measures and Frobenius-Perron operator is
stated by the

{\bf Theorem 4.1.1.-[}6{\bf ].} Let $S_t$ be a non singular transformation
and $P_t$ its Frobenius-Perron operator. Then there exist an state of
thermodynamic equilibrium whose density $\rho _{*}$ is an stationary state
of $P_t$ if and only if the measure $\mu _{*}$:
\[
\mu _{*}(A)=\int_A\rho _{*}(x)dx....................(4.1.2) 
\]
is invariant.

Therefore transformation $U(t)$ that preserve Lebesgue measure, has
necessarily a equilibrium steady state, e. g. the uniform state of the
microcanonical-ensemble. But theorem 4.1.1.says nothing about the uniqueness
of the equilibrium state. We shall discuss this problem in the next section.

A point $x\in A\subset X$ it is called a recurrent point if there is some
time $t>0$ such that $S_t(x)\in A$. An important result is Poincar\'e
recurrence

{\bf Theorem 4.1.2.-}Let $S_t$ be a transformation with an invariant measure 
$\mu _{*}$ operating in a finite space $X,$ $\mu _{*}(X)<\infty $, and let $%
A $ be a subset of $X$ with positive $\rho _{*}$-measure. Then there exists
a point $x$ in $A$ that is recurrent.

Proof.: Assume the contrary, i. e. that there are no recurrent point in $A.$
This then implies that $S_t^{-1}(A)\cap A=\emptyset $ for all times $t>0$,
and thus that $S_{t^{\prime }}^{-1}(A)\cap S_t^{-1}(A)=\emptyset $ for all
positive times $t\neq t^{\prime }$. However, since $S_t$ is measuring
preserving, this implies that $\mu _{*}(S_t^{-1}(A))=\mu _{*}(S_{t^{\prime
}}^{-1}(A))$ and this, coupled with the pairwise disjoint nature of the sets 
$S_t^{-1}(A)$ and $S_{t^{\prime }}^{-1}(A)$ leads to:
\[
\sum_{t=0}^\infty \mu _{*}(A)=\sum_{t=0}^\infty \mu _{*}(S_t^{-1}(A))=\mu
_{*}\left[ \bigcup_{t=0}^\infty S_t^{-1}(A)\right] \leq \mu _{*}(X)<\infty
.......(4.1.3) 
\]
The only way in which this inequality can be satisfied is for $\mu _{*}(A)$
to be zero, which is a contradiction. Thus we conclude that $A$ contains
recurrent points. $\Box $

Therefore, in ordinary mechanical motion of finite systems almost any point
is recurrent, since the sets of non recurrent point have measure zero. This
fact seems to prevent the existence of irreversible evolutions, namely it is
impossible to reach a final equilibrium state, since the system will came
back as closed to its initial condition as we wish, if we wait enough. The
time we must wait it is called the Poincar\'e recurrence time. There are two
ways to avoid this problem:

i.-The practical way is to compute the recurrence time. It turns out that in
usual system (say with a number of molecules of the order of Avogadro
number) the time if much bigger than the age of the universe, so the
returning to the initial conditions is practically unobservable.

ii.-The theoretical way is to consider that irreversibility is not a notion
of classical mechanics, but a notion that can only be defined in statistical
mechanics, where we deals with statistical ensembles of identical system.
Then the recurrent time of the ensemble, namely the time such that we
reobtain the initial condition in{\it \ each one of the infinite identical
systems} is, of course, infinite and the problem is theoretically solved.

In the following subsections we shall study some properties of dynamical
system ordered by their increasing chaotic behavior.

\subsection{Ergodicity.}

It would be interesting to know if the equilibrium state of theorem 4.1.1 is
unique or not. To answer this question we must introduce some new concepts.

i.-A set $A$ such that $S_t^{-1}(A)=A$ is called an invariant set.

ii.-Any invariant set $A$ such that $\mu _{*}(A)=0$ or $\mu _{*}(X\setminus
A)=0$ is called trivial.

iii.-A non-singular transformation $S_t$ is called $\rho _{*}$-ergodic if
every invariant set $A$ is a trivial subset of the phase space $X$. I. e.,
either $\mu _{*}(A)=0$ or $\mu _{*}(X\setminus A)=0.$ This means that, if we
consider a generic non-singular subset $A,$ the time evolved counter-image
of this subset, $S_t^{-1}(A),$ will wonder around all $X$ since $A$ cannot
be invariant.

iv.-If $\rho _{*}$ is the uniform density of the microcanonical ensemble we
will say that $S_t$ is uniformly ergodic.

The motion within almost all tori of integrable classical mechanical systems
is ergodic [21],[22].Ergodicity is therefore a very usual property of the
mechanical systems of section 2.

The connection on the uniqueness of the equilibrium state and the properties
of the operators is stated in the following

{\bf Theorem 4.2.1.}[6].-Let $S_t$ be a non-singular transformation and $P_t$
the corresponding Frobenius-Perron operator. $S_t$ is $\rho _{*}$-ergodic if
and only if $P_t$ has a unique state of thermodynamic equilibrium with
associated stationary density $\rho _{*},$ namely a density such that $%
P_t\rho _{*}=\rho _{*}$.

Hence ergodicity is the necessary and sufficient condition for the
uniqueness of thermodynamical equilibrium, and allows us to formulate a
third-order form second law. But this is of course half the picture, because
we must also understands why the system evolves to this equilibrium state.

Let us sate an important

{\bf Theorem 4.2.2.}[6]. Let $S_t$ be a non singular transformation and $P_t$
the corresponding Frobenius-Perron operator with stationary density $\rho
_{*}>0$ for all points in phase space $X$. Then $S_t$ is $\rho _{*}$-ergodic
if and only if $\{P_t\rho \}$ is C\'esaro convergent to $\rho _{*}$ for all
densities $\rho $. i.e., if and only if
\[
\stackunder{t\rightarrow \infty }{\lim \frac 1t\sum_{k=0}^{t-1}(P_k\rho
,\sigma )=(\rho _{*},\sigma )}...............(4.2.1) 
\]
in the discrete time case, or if and only if:
\[
\stackunder{T\rightarrow \infty }{\lim }\frac 1T\int_0^T(P_t\rho ,\sigma
)dt=(\rho _{*},\sigma )..........(4.2.2) 
\]
in the continuous time case, for all bounded measurable functions $\sigma $
and where ($\rho ,\sigma )=\int_X\rho (x)\sigma (x)\mu (x)dx$ (in this case $%
\mu (x)$ is an arbitrary measure) is a generalization of inner product
(2.A.2.1)

\subsection{Mixing.}

This will be the main property of dynamical system that we will study, it
serve to guarantee the approach of the system to an equilibrium state.

Let $S_t$ be a $\rho _{*}$-measure preserving transformation operating in a
normalized space $X$ ($\mu _{*}(X)=1$). Then $S_t$ is called $\rho _{*}$%
-mixing if:
\[
\stackunder{t\rightarrow \infty }{\lim }\mu _{*}(A\cap S_t^{-1}(B))=\mu
_{*}(A)\mu _{*}(B).............(4.3.1) 
\]
for all sets $A$ and $B$. If $\rho _{*}$ is the uniform density of the
microcanonical ensemble then we will say that $S_t$ is uniformly mixing.

Some tori of mechanical non-integrable system are broken, thus a chaotic
motion in phase space takes place. Chaos, most likely with mixing
properties, is very frequent in mechanical systems. [21],[22]

A very important and popular example of uniformly mixing transformation is
the, so called, baker transformation that operates in the phase space $%
X=[0,1]\times [0,1]$ and it is defined by the following procedure:

i.-squeeze the $1\times 1$ square to a $2\times \frac 12$ rectangle, and,

ii.-cut the rectangle vertically into 2 rectangles and pile them up to form
another $1\times 1$ rectangle.

in doing so the point of the square will move as:
\[
(x,y)\rightarrow S(x,y)=\left\{ 
\begin{array}{cc}
(2x,\frac 12y), & if:..0\leq x\leq \frac 12 \\ 
(2x-1,\frac 12+\frac 12y), & if:..\frac 12\leq x\leq 1
\end{array}
\right. ......(4.3.2) 
\]
The transformation is shown in fig. 1, where in the first square, the one
corresponding to $t=0$, the lower half is shadowed and corresponds to a
subset B. It is easy to see that this transformation is reversible. The fate
of this area B, evolving to the future, is shown in the right side of the
figure,(it is transformed in a great number of horizontal strips with area
1/2) and evolving to the past, in the left side (the strips are now
vertical). An smaller subset A is also shown. It is then easy to verified
that condition (4.3.1.) is fulfilled (the final measure of $S_t(B)\cap A$
will be $\frac 12\mu _L(A),$ being the initial measure of $B$ just $\frac 12$
so eq.(4.3.1) is satisfy). We will study this transformation in all detail
in the subsection 4.5 using the fine-graining technic.

Much more complicated mixing evolutions than (4.3.2) can be invented. In
fact, baker's transformation is the simplest of all, it is the simplest
model of the famous Gibbs ink drop. Gibbs try to explain the essence of
irreversibility with the ink drop model. If a drop of blue ink is introduced
in a glass of water, even if the volume of the ink drop remains constant (as
the volume of any subset of mechanical phase space, according to Liouville
theorem) we will have, after a while, an homogeneous mixture of bluish
water, What it is happens is that the motion of the water is mixing and
therefore the ink drop is deformed (even if its volume is constant) in such
a way that it is transformed in a set of very thin filaments that are
present in every portion of the water giving the sensation that the water
has become bluish. The growing of this filaments-like structure gives an
arrow of time and it is for Gibbs the essence of irreversibility. This
phenomenon is modeled by the baker's transformation. In fact, let us
consider a small rectangle $a\times b$ within the square $1\times 1$, let
say a small task of lower quality flower within the bread mass. The height
of the task will successively became: $\frac 12b,\frac 14b,...,\frac 1tb,...$%
while the base of the task will became: $2a,4a,...ta,....$in such a way that
the area is conserved. Eventually a time will arrive such that $ta>1$ and
then the task will be cutted in two, and then in four, eight, etc., in such
a way that it will become a set of horizontal filaments of decreasing
height, namely a ''cubistic'' picture of the ink drop, so bakers
transformation is just a model of the ink drop phenomenon.

If now we consider the much more complicated evolution of the ink drop, and
if the volume of the ink drop is the 1\% of the volume of the water, it is
clear that the motion of usual water is mixing according to definition
(4.3.1), as bakers transformation. In fact, if the motion is mixing, when $%
t\rightarrow \infty $ every subset $A\subset X$ will have a 1\% of ink and,
therefore, the distribution of ink will become homogeneous. As this is the
case with the real ink drop we can conclude that the real motion is mixing.

It is a straightforward consequence of the definition that $\rho _{*}$%
-mixing implies $\rho _{*}$-ergodicity. In fact, if $B$ is an invariant set
eq. (4.3.1) reads:
\[
\mu _{*}(A\cap B)=\mu _{*}(A)\mu _{*}(B) 
\]
for all set $A$. Now if we take $B=A$ we obtain $\mu _{*}(B)=[\mu _{*}(B)]^2$
and therefore either $\mu _{*}(B)=0,$ or $\mu _{*}(B)=1,$ so $\mu
_{*}(X-B)=0.$ So the evolution is $\rho _{*}$-ergodic.

Now we have arrived to our most important

{\bf Theorem 4.3.1.}[6].- Let $S_t$ be an ergodic transformation, with
stationary density $\rho _{*}$ of the associated Frobenius-Perron operator,
operating in a phase space of finite $\rho _{*}$-measure. Then $S_t$ is
mixing if and only if $\{P_t\rho \}$ is weakly convergent to $\rho _{*}$,
i.e.,
\[
\stackunder{t\rightarrow \infty }{\lim }(P_t\rho ,\sigma )=(\rho _{*},\sigma
)..................(4.3.3) 
\]
where $\sigma $ is a bounded measurable function.

If a sequence is weakly convergent it is also C\'esaro convergent, so we can
see again that mixing evolutions are ergodic.

So the mixing property assures a weak convergence of $\{P_t\rho \}$ to $\rho
_{*}$. But, e. g., in the example of the baker transformation the strong
limits toward the far past or the far future do not exit. In fact, the
support of any distribution function (if it has a measure $<1)$ will be a
set of infinity horizontal toward the future or vertical lines toward the
past. These set can not be the support of any regular distribution function.
Nevertheless the weak limit (4.3.3) do exists with $\rho _{*}=1$.

The physical meaning of theorem 4.3.1 is very clear: Let us consider a
(non-viscous) fluid in motion in a cubic box. As energy is conserved the
motion will never stops, and therefore, according to the laws of mechanics,
equilibrium will never be attain, and $P_t\rho $ will have no limit. This
will be the case if the motion is oscillatory, namely a pressure wave that
oscillate back and forth between two parallel walls of the box. But if the
motion is mixing it is so complicated that there are portions of the fluid
moving in every direction near every point of the box. In this case if we
take the inner product $(P_t\rho |\sigma )$ we are making an average that
goes to an equilibrium average $(\rho _{*}|\sigma )$ when $t\rightarrow
\infty .$ Therefore, even if there is always motion, the motion average
gives an image of equilibrium. This is the profound meaning of theorem
4.3.1. and the way to obtain a synthesis of the apparent contradiction of
dynamics and thermodynamics:

-even if the dynamics says that the energy is conserved and the motion will
never stops

-there is a thermodynamical equilibrium in average, because the motion is
mixing.

From this point on fine-graining and coarse graining follow different paths,
as we have explained in the introduction and we will discuss below. But let
us remember that the problem is not completely solved, since all the nice
inequalities of subsection 3.6, that are necessary to explain the second law
, are equalities for reversible system and all system in nature are
considered to be, al least microscopically, reversible.

There are system endowed with properties more chaotic than mixing, they are:

i.- Kolmogorov systems [6] that necessarily are mixing [23].

ii.-Anosov systems-[6],[22],[24].

iii.-Bernoulli systems, the most chaotics of all [25]. Baker transformation
is, in fact, a Bernoulli system [26]..

\subsection{Exactness.}

We will now introduce a property that (apparently) will solve all our
problems

If $S_t$ is a $\rho _{*}$-measure preserving transformation operating in a
phase space $X$, then $S_t$ is said to be $\rho _{*}$-exact if:
\[
\stackunder{t\rightarrow \infty }{\lim }\mu
_{*}(S_t(A))=1................................(4.4.1) 
\]
for all sets $A$ of non zero measure. This is possible even if $S_t$ is $%
\rho _{*}$-measure preserving since a evolution is measure preserving if eq.
(4.1.2) is satisfied and this equation is not equivalent to $\mu
_{*}(S_t(A))=\mu _{*}(A),$ if the evolution is not reversible. Renyi map is
a good example.

Let us consider a dyadic Renyi map:
\[
R:[0,1)\rightarrow [0,1),...x\rightarrow Rx=2x(mod1).........(4.4.2) 
\]

As the length of any subset $A$ is multiplied by two in each transformation,
this map is exact since it satisfies eq. (4.4.1). Anyhow it is also measure
preserving. In fact, let as consider, e. g. the subset $A=[0,\frac 12)$, $%
R^{-1}(A)$ is $[0,\frac 12)\cup [\frac 12,\frac 34)$ and, therefore, both
subsets have measure $\frac 12.$

If $\rho _{*}$ is the uniform density of the microcanonical ensemble we say
that $S_t$ is uniformly exact.

The essential think to understand is that reversible system cannot be exact.
In fact, for reversible $\rho _{*}$-measure preserving transformation we
have:
\[
\mu _{*}(S_t(A))=\mu _{*}[S_t^{-1}(S_t(A))]=\mu _{*}(A).....(4.4.3) 
\]
thus the definition of exactness is violated. Since usually classical
dynamical system are measure preserving, by the Liouville theorem, and
reversible they are not exact. Nevertheless, as we shall see exactness is
really the property we are looking for. Precisely:

{\bf Theorem 4.4.1.}[6] If $S_t$ is a $\rho _{*}$-measure preserving
transformation operating on a finite normalizable phase space $X$ and $P_t$
is the associated Frobenius-Perron operator corresponding to $S_t$, then $%
S_t $ is $\rho _{*}$-exact if and only if:
\[
\stackunder{t\rightarrow \infty }{\lim }\parallel P_t\rho -\rho
_{*}\parallel =0.................................(4.4.3) 
\]
Therefore: Ergodicity corresponds to C\'esaro convergence, Mixing
corresponds to weak convergence, and, exactness corresponds to strong
convergence (i.e. convergence in the norm). A strongly convergence sequence
is also weakly convergent, thus we can deduce that exact evolution era also
mixing evolution and therefore ergodic evolutions. Moreover, since we are
looking for a strong limit we see that working with ordinary distribution
functions we will find this limit only if the transformation is exact, but
ordinary classical (microscopical) system are not exact since they are
reversible and measure preserving. As an example we have shown that the
reversible baker transformation has not strong limits toward the past and
the future, in fact baker transformation, being reversible cannot be exact.
Thus our problem is now clearly stated: if we want a strong limit our
evolutions must be exact. but exact evolutions are not reversible and all
microscopical transformation are reversible, therefore we can not have a
strong limit. Furthermore we have also the

{\bf Theorem 4.4.2.}[6].Let $P_t$ be a Markov operator operating in phase
space $X$. Then the conditional entropy of $P_t\rho $ with respect to
density $\rho _{*}$ goes to a maximum value of zero as $t\rightarrow \infty
,i.e.,$%
\[
\stackunder{t\rightarrow \infty }{\lim }H_C(P_t\rho |\rho
_{*})=0...................................(4.4.5) 
\]
if and only if $P_t$ is $\rho _{*}$-exact.

This theorem tell us the necessary and sufficient criteria to be able to
state the second law of thermodynamics in third-order form, namely, for the
entropy of the system to converge to its maximum value regardless of the way
in which the system was prepared. This condition is that the system must
evolve according to an exact transformation. But such systems do not exist
in nature. So dynamics cannot be related, at least trivially, with
thermodynamics. Therefore our theory must be modified one way or the other.

\subsection{Mixing studied by the fine-graining technic.}

Mixing evolution are studied by the fine-graining technic in papers [26],
[27], and [28], using a perturbative method, that can be implemented in any
example. For didactical reason we will present the most important mixing
evolutions and we refer to the papers above for the general perturbative
method.

\subsubsection{The Renyi maps.}

The $\beta -$adic Renyi map $R$ on the interval $[0,1)$ is the
multiplication, modulo 1, by the integer $\beta \geq 2:$%
\[
R:[0.1)\rightarrow [0,1):....x\rightarrow Rx=\beta
x..(mod..1),..........(4.5.1.1) 
\]
The forward iteration of the Renyi map $n$ times, define a ''cascade'' or
time evolution with time $t=n\in Z$. This evolution preserve only the
Lebesgue measure, as we have shown after eq. (4.4.3). The density functions $%
\rho (x)$ evolve according the Frobenius-Perron operator $U$ :
\[
U\rho (x)=\frac 1\beta \sum_{r=0}^{\beta -1}\rho \left( \frac{x+r}\beta
\right) ............................................(4.5.1.2) 
\]
Gel'fand-Maurin theorem 4.A.1. tell us that we can found an spectral
expansion in the eigenvectors of this operator in an adequate rigged Hilbert
space. In fact, using the perturbative methods of papers [26],[27],and [28],
the spectral decomposition of $U$ can be found and reads:
\[
U=\sum_{n=0}^\infty \frac 1{\beta ^n}|B_n)(\widetilde{B}%
_n|.............................................(4.5.1.3) 
\]
where $:$%
\[
|B_n(x))=|x^n+\sum_{m=0}^{n-1}x^m\frac{n!}{m!(n-m)!}%
B_{n-m}).............(4.5.1.4) 
\]
where the $B_n(x)$ is the n-degree Bernoulli polynomial defined by the
generating function:
\[
\frac{ze^{zx}}{e^z-1}=\sum_{n=0}^\infty \frac{B_n(x)}{n!}%
z^n..........................................(4.5.1.4) 
\]
and:
\[
(\widetilde{B}_n|= 
\begin{array}{c}
(1|.............................................................n=0 \\ 
(\frac{(-1)^{(n-1)}}{n!}\{\delta ^{(n-1)}(x-1)-\delta
^{(n-1)}(x)\}|......n=1,2,...
\end{array}
(4.5.1.5) 
\]
where $(1|$ is the constant distribution function. From eq. (4.5.1.5) we can
see that the elements of the spectral decomposition (4.5.1.3) do not belong
to ${\cal L}$ but to a larger space where the Dirac $\delta $ must have a
precise mathematical meaning. This space is,in fact, a rigged Hilbert space
that we shall define below, in full agreement with the Gel'fand-Maurin
theorem. The system $\{|B_n),(\widetilde{B}_n|\}$ is bi-orthonormal and
complete, namely:
\[
(\widetilde{B}_n|B_m)=\delta
_{nm}.................................................(4.5.1.6) 
\]
\[
\sum_{n=0}^\infty |B_n)(\widetilde{B}%
_n|=1...............................................(4.5.1.7) 
\]
The spectral decomposition (4.5.1.3) acquires a precise mathematical meaning
if we define, as test space $\Phi $ the space of polynomials ${\cal P}$
.This space is dense in ${\cal L}$=$L^2$ , nuclear (in fact, it is the union
of an infinite and discrete set of finite dimensional spaces), complete,
stable under $U$, and $U$ is continuous in the topology of ${\cal P.}$ It
is, therefore an appropriate test space to give a meaning to the spectral
decomposition whose elements belong to $\Phi ^{\times }.$ But other kind of
test functions spaces can be defined and we will obtain different spectra,
e. g.: a continuous set of eigenfunctions can be found, with and adequate
rigging, showing that the Renyi map have continuous spectrum, precisely the
set of complex numbers $z$ such that $|z|<1$ (all the mixing operator have
an spectral decomposition with a continuous spectrum in Hilbert space).

If $t\in Z$ from eqs. (4.5.1.3,6) we can see that the evolution operator is:
\[
U^t=\sum_{n=0}^\infty \frac 1{\beta ^{nt}}|B_n)(\widetilde{B}%
_n|=|1)(1|+\sum_{n=1}^\infty \frac 1{\beta ^{nt}}|B_n)(\widetilde{B}%
_n|....(4.5.1.8) 
\]
If we would like to work in space ${\cal L}$ only we must remember that all
the formulae above are just weak equation, e. g. the last one is just:
\[
(\sigma |U^t\rho )=(\sigma |1)(1|\rho )+(\sigma |\sum_{n=1}^\infty \frac
1{\beta ^{nt}}|B_n)(\widetilde{B}_n|\rho ).............(4.5.1.9) 
\]
for all $\rho ,$$\sigma \in {\cal P,}$ and $\rho $ is a density. Then as $%
\beta >1$ and $\rho $ is normalized we have:
\[
\stackunder{t\rightarrow \infty }{\lim }(\sigma ,U^t\rho )=(\sigma
|1)..................................................(4.5.1.10) 
\]
in perfect agreement with theorem 4.3.1. Eq. (4.5.1.9) is just the weak
version (or coarse graining) of eq. (4.5.1.8) that allows us to work within
space ${\cal L}$ but using always weak limits as (4.5.1.10). But if we work
with functional directly, namely in space $\Phi ^{\times }$, from eq.
(4.5.1.8) we can say that:
\[
\stackunder{t\rightarrow \infty }{\lim }U^t|\rho
)=|1)..........................................................(4.5.1.11) 
\]
which is a strong limit, namely the fine graining version of (4.5.1.10). If
we call $\beta =e^{-\gamma }<1$ , from eq. (4.5.1.9) we can also say:
\[
\rho (t)=U^t\rho =\rho _{*}+\rho _1(t)e^{-\gamma t},...\rho
_{*}=|1).....................(4.5.1.12) 
\]
where $\rho _{*}=|1)$ is the equilibrium distribution function and $\rho
_1(t)e^{-\gamma t}$ is something like a ''fluctuation'' around the
equilibrium state. We write this last equation because we will find a
similar equation in the quantum case.

\subsubsection{The baker's transformation.}

The $\beta $-adic $\beta =2,3,..$ baker's transformation in the unit square $%
Y=[0,1)\times [0,1)$ is a two-step operation:

i.-squeeze the 1$\times 1$ square to a $\beta \times \frac 1\beta $
rectangle, and,

ii.-cut the rectangle vertically into $\beta $ rectangles and pile them up
to form another $1\times 1$ square.

Then:
\[
(x,y)\rightarrow B(x,y)=(\beta x-r,\frac{y+r}\beta ),....for:\frac r\beta
\leq x<\frac{r+1}\beta ,..r=0,...\beta -1.......(4.5.2.1) 
\]
This equation is an obvious generalization of eq. (4.3.2), which is the
particular case of eq. (4.5.2.1) for $\beta =2.$ As we can see we have a
sort of two Renyi maps one in each coordinate. Baker's transformation is a
Bernoulli shift and has Kolmogorov Sinai entropy $\log {}_2\beta $ [25].The
invariant measure is Lebesgue measure. The density function $\rho (x,y)$
evolves according to the Frobenius-Perron operator $U:$%
\[
U\rho (x,y)=\rho (B^{-1}(x,y))=\rho (\frac{x+r}\beta ,\beta y-r),..for:\frac
r\beta \leq y<\frac{r+1}\beta ,..r=0,...\beta -1,..(4.5.2.2) 
\]
This operator is unitary in the Hilbert space ${\cal L}=L^2$ the equilibrium
distribution function is the constant function $\rho _{*}=1,$ and the
Lebesgue spectrum is the unit circle plus the simple eigenvalue 1

As the baker's transformation $B$ is the natural extension of the Renyi map $%
R$ the conclusion that we can obtain are the same and we refer to [29] for
details. $B$ acts on the Liouville-Hilbert space ${\cal L}=L^2=L_x^2\otimes
L_y^2$ and a suitable initial biorthonormal system can be constructed from
the tensor products of the eigenfunctions of the $\beta $-adic Renyi map
(cf. eq. ((4.5.1.4) and (4.5.1.5))
\[
|\varphi _{nm}>=B_n(x)\stackrel{\sim }{B}_m(y)..and...<\stackrel{\sim }{%
\varphi }_{nm}|=\stackrel{\sim }{B_n}%
(x)B_m(y).......................(4.5.2.3) 
\]
Using these bases and the same perturbative method as before the following
spectral decomposition can be obtained:
\[
U=|f_{00}><\stackrel{\sim }{f}_{00}|+\sum_{\nu =1}^\infty \left\{
\sum_{r=0}^\nu \frac 1{\beta ^\upsilon }|f_{\nu ,r}><\stackrel{\sim }{f}%
_{\nu ,r}|+\sum_{r=0}^{\nu -1}|f_{\nu ,r+1}><\stackrel{\sim }{f_{\upsilon
,r}|}\right\} ...(4.5.2.4) 
\]
where the vectors $|f_{\nu ,r}>$ and $<\stackrel{\sim }{f}_{\nu ,r}|$ can be
obtained from the vectors of eqs. (4.5.2.3). As we have said the Liouville
spectrum is the unit circle plus the eigenvalue 1, so in the new spectral
decomposition we have found new eigenvalues $1/\beta ^\nu <1.$

The initial vectors $\varphi _{nm}$ and $\stackrel{\sim }{\varphi }_{nm}$
are linear functionals over the spaces $\Phi _{-}=L_x^2\otimes {\cal P_y}$
and $\Phi _{+}={\cal P}_x\otimes L_y^2.$ Furthermore it can be shown that
the vectors $f_{nm}\in \Phi _{-}^{\times }$ and $f_{nm}\in \Phi _{+}^{\times
}$ are also functional over the same spaces, so the spectral decomposition
(4.5.2.4) can be implemented if we use these functional vector spaces. We
have enlarged the state space with densities that can be distributions in
the $y$ coordinate, in the case of $\Phi _{-}^{\times }$, e. g.: if the $y$
distributions are Dirac's deltas we will have a distribution whose support
is a set of horizontal straight lines, that we shall call a ''horizontal
Dirac's comb''. In the case of $\Phi _{+}^{\times }$ we must change the $y$
by the $x$ and we would have, e. g.:''vertical Dirac's combs''

Now, ''mutatis mutandi'', we can repeat what we have said in eqs. (4.5.1.9)
to (4.5.1.12), and we will find similar equations for the baker's
transformation. The equilibrium distribution, in this case will be:
\[
\rho _{*}=|f_{00}>......................(4.5.2.4) 
\]
where $|f_{00}>\in \Phi _{-}^{\times }$.

\subsection{Appendix 4.A. Rigged Hilbert spaces [10], [29].}

As it is well known all linear spaces of the same dimension are isomorphic
if this dimension is finite. This is not the case if the dimension is
infinite. In fact, let us consider the infinite sequence of the vectors of a
basis of an infinite dimension vector space
\[
\{\Phi _n:n=1,2,...\},.............(4.A.1) 
\]
Let $V$ be the vector space of all finite linear combination of the vectors
of the basis above, namely:$\Psi \in V$ if:
\[
\Psi =\sum_{n=1}^ic_n\Phi _n..................(4.A.2) 
\]
$V$ is a linear space of infinite dimension, but we will see that we can
built other spaces using basis \{$\Phi _n\}$. For instance we can add to $V$
the limit points of all the convergent infinite sequences of vectors of $V$.
But defining different criteria of convergence we will have different set of
limit points and,therefore, different vector spaces. The most use full
convergence is the convergence in the norm. The sequence $\{\Psi _i\}$
converge in the norm to a limit point $\chi $ if:
\[
\stackunder{i\rightarrow \infty }{\lim }\parallel \chi -\Psi _i\parallel
=0...............(4.A.3) 
\]
If the sequences $\{\Psi _i\}$ are sequences of vectors of $V$ and they
converge in the norm and we add the limit points $\chi $ of these sequences
to $V$ we obtain a larger space ${\cal H}$ where we have finite sequences
like (4.A.2) but also limit points of infinite sequences.We will say that $%
{\cal H}$ is the closure of $V$ and also that $V$ is dense in ${\cal H.}$
But we can use other kind of convergencies, namely other topologies, and we
will obtain different spaces.

Let us suppose that we chose the sequences, such that the coefficients $c_n$
satisfy the condition:
\[
\sum_{n=1}^\infty |c_n|^2<\infty ..................(4.A.4) 
\]
Then, adding the corresponding limit points, we obtain a Hilbert space $%
{\cal H}$ which contains all the sequences that converge in norm, and it is
call also the competitions of $V$ respect to the topology of the norm. But
we can also consider a the infinite dimension linear space $\Xi $ of all,
either finite or infinite, linear combinations of the basis $\{\Phi _n\}$,
namely all the linear combinations $\xi =\sum_nc_n\Phi _n$ with limitations
impose over the coefficients $c_n.$ Of course we cannot define a norm in
such a space, but we now we have three infinite dimensional linear spaces
such that:
\[
V\subset {\cal H}\subset \Xi .....................(4.A.5) 
\]
Let us define the inner product:
\[
(f,h)=\sum_nb_n^{*}c_n............(4.A.6) 
\]
Then ${\cal H}$ is the space of the vectors $h=\sum_nc_n\Phi _n$ such that $%
(h,h)=\sum |c_n|^2<\infty .$ Let us now define the conjugated space of $%
{\cal H,\ H^{\times }}\subset \Xi $ of all linear functional over ${\cal H}$
namely vectors $f=\sum_nb_n\Phi _n$ such that the inner product:
\[
f[h]=(f,h)=\sum_nb_n^{*}c_n......(4.A.7) 
\]
is convergent for all $h\in {\cal H.}$ The convergence of this inner product
is a consequence of Schwarz inequality:
\[
|(f,h)|^2\leq (h,h)(f,f)...........(4.A.8) 
\]
so (4.A.7) converge if $(f,f)=\sum_n|b_n|^2$ converge and, therefore, $f\in 
{\cal H}$ so ${\cal H=H^{\times }}$. (We can as well define the space of
antilinear as $f[h]=(h,f)$ i.e. bra are linear functional and ket can be
consider like antilinear functional$)$

Let us now define a new space $\Omega $ as the space of all vectors $\omega
=\sum_nu_n\Phi _n,$ endowed with coefficients $c_n$ such that-they satisfy
the following set of infinite conditions:
\[
\sum_n|u_n|^2n^m<\infty ,...m=1,2,3.....(4.A.9) 
\]
Obviously $\Omega \subset {\cal H.}$ Let us now find the conjugate space of $%
\Omega ,$ $\Omega ^{\times }\subset \Xi ,$ namely the space of convergent
linear continuous functional over $\Omega $. These functional read $\sigma
=\sum_nv_n^{*}\Phi _n$ and they are such that:
\[
\sigma [\omega ]=(\sigma ,\omega )=\sum_nv_n^{*}u_n,.........(4.A.10) 
\]
is convergent for all $\omega \in \Omega .$ Therefore:
\[
\sum_n|v_n|^2n^{-m}<\infty ,...m=1,2,3...(4.A.11) 
\]
In fact: according to Schwarz lemma we have:
\[
|\sum_nv_n^{*}n^{-\frac m2}u_nn^{\frac m2}|^2<\left(
\sum_n|v_n|^2n^{-m}\right) \left( \sum_n|u_n|^2n^m\right) <+\infty
,.....(4.A.12) 
\]
and the r.h.s. is convergent if eqs (4.A.9) and (4.A.11) are fulfilled.

As it is obvious that $V^{\times }=\Xi ,$ we now have the following set of
infinite dimensional spaces:
\[
V\subset \Omega \subset {\cal H=H^{\times }\subset }\Omega ^{\times }\subset
V^{\times }=\Xi .........(4.A.13) 
\]
Any triplet:
\[
\Omega \subset {\cal H}\subset \Omega ^{\times
}............................(4.A.14) 
\]
like those of eq. (4.A.12) and others that can be obtained, e. g.: if limit
the m's of eq. (4.A.9) to be just 1$\leq m\leq M,$ for some $M\in N$, are
called Gel'fand triplets. $\Omega $ is known as the test space and $\Omega
^{\times },$ as the rigged space. Mathematically it is convenient that the
test space would be a nuclear space. Heuristically speaking nuclear spaces
are the infinite dimension spaces endowed with the largest number of
properties of finite dimensional spaces, among then they have discrete
spectral decomposition. Precisely, nuclear spaces are spaces obtained, so to
say, as the union of an infinite sequence of spaces of finite dimension. As
space $\Omega $, from our point of view, is the space of operators
corresponding to real measurement apparatuses and as these devices make only
a finite (so less than discrete) number of measurements; logically $\Omega $
must be a nuclear space. In fact., even if physical devices make a finite
number of measurements, we can conceive that this numbers grows with the
progress of technology. Then an infinite, but discrete, number of
measurements would corresponds to the limit of an infinitely long period of
technological progress. A finite number of measurements will corresponds to
a test space of a finite number of dimensions. Then the test space
corresponding to the limit of technological progress will be a nuclear
space, since this space is the limit of a sequence of finite dimensional
spaces. E. g.: a measurement device make $n$ measurements, that can define $%
n $ points of a curved, that can be interpolated by a polynomial of degree $%
n.$ The space of polynomials of degree $n$ will be the test space that
corresponds to this device. In the limit of technological progress the test
space will be the space ${\cal P}$ of polonomial of any degree, in fact a
nuclear space.Generally speaking choosing different nuclear test functions
spaces we can also chose the physical properties of our measurement devices.

In finite dimensional vector spaces the eigenvalue problem, for every
selfadjoint linear operator $A$ can be solve in a unique way. Namely we can
find a unique spectrum $\{a_i\}$ and an orthonormal basis \{$\Psi _n\}$ such
that $A\Psi _n=a_n\Psi _n$.This is not so for infinite dimensional linear
spaces, since the spectrum depends on the rigging we use, nevertheless it
can be demonstrated the Gel'fand-Maurin

{\bf Theorem 4.A.1.} If $A$ is a self adjoint operator in ${\cal H}$ there
is always a complete set of eigenvectors of $A$ in some rigged Hilbert space 
$\Omega ^{\times }.$

Let us give to very important examples:

i.-Let $\Xi $ be the space of functions $f(x)$ of one real variable $x$ and
let $A=P=-i\frac d{dx}$ be the self adjoint momentum operator in ${\cal H}%
=L^2.$ The eigenvectors of $P$ are the plane waves $e^{ikx}$ with do not
belong to $L^2$ since they have not finite norm. Nevertheless they can be
considered as functionals over a convenient space test function $\Phi $
since:
\[
e^{ikx}[f]=(e^{ikx},f(x))=\int_{-\infty }^{+\infty }e^{-ikx}f(x)dx\approx 
\widehat{f(k)},.........(4.A.15) 
\]
where $\widehat{f(k)}$ is the Fourier-transform of $f(x)$ and $\Phi $ is any
subspace of ${\cal H}$ such that eq. (4.A.15) id convergent. Then we have
the Gel'fand triplet $\Phi \subset {\cal H}\subset \Phi ^{\times }$ and $%
e^{ikx}\in \Phi ^{\times }.$

ii.-Let $\Xi $ be as in the example above and $A=Q=x$ be the position
selfadjoint operator in ${\cal H}=L^2$. The eigenvectors of $Q$ are the
Dirac's deltas $\delta (x-y),$ since $Q\delta (x-y)=y\delta (x-y)$ , these
distributions do not belong to $L^2$ since they are not even functions.
Nevertheless they can be considered as functionals over a convenient space
of test functions $\Phi $, since we can rigorously define these deltas as
the functionals:
\[
\delta _y[f(x)]=f(y)........................(4.A.16) 
\]
where $f(x)$ is any function of $\Phi .$ Usually physicist write this last
equation as:
\[
\delta _y[f(x)]=(\delta (x-y),f(x))=\int_{-\infty }^{+\infty }\delta
(x-y)f(x)dx=f(y),.........(4.A.17) 
\]
even if the integral in this last equation have not a rigorous definition.
Usually $\Phi $ is the set of function with nice properties e.g. they are
continuous, derivable, with compact support, etc. Then $\delta _y\in \Phi
^{\times }.$

these examples show that usual operator don not have their eigenvalues in $%
{\cal H}$ but in properly chosen rigged Hilbert spaces.

\section{The Quantum Evolution.}

As the laws of quantum evolution are well known (cf. [10],[17],[18]), in
this section we will see the use of the no-graining and coarse-graining
technics in quantum mechanics.

\subsection{The case of discrete spectrum.}

Let us begin making just an heuristic calculation. Let ${\cal H}$ {\cal \ }%
be the quantum Hilbert space. Let $\{|i>\}$ be a energy eigen-basis of this
Hilbert space, where $i$ is a discrete index. The quantum Liouville space is 
${\cal L=H\times H}$, and a generic density matrix reads:
\[
\rho =\sum_{i,j}\rho _{ij}|i><j|....................(5.1.1) 
\]
where since $\rho =\rho ^{\dagger },$ it is$\rho _{ij}=\rho _{ji}^{*}$.

Let $O$ be a self adjoin operator, it reads:
\[
O=\sum_{ij}O_{ij}|i><j|..................(5.1.2) 
\]
where $O_{ij}=O_{ji}^{*}.$

The mean value of operator $O$.in the quantum state $\rho $ is:
\[
<O>_\rho =(\rho |O)=tr(\rho |O)=\sum_{ij}\rho _{ij}O_{ji}...........(5.1.3) 
\]
as $|i>$ is an energy eigen state we have:
\[
H|i>=\omega _i|i>.......................(5.1.4) 
\]
where $\omega _i$ is the energy of state $|i>$. The time evolution of this
eigen states reads:
\[
|i(t)>=e^{-i\omega _it}|i>.................(5.1.5) 
\]
Therefore the time evolution of $\rho $ is:
\[
\begin{array}{c}
\rho (t)=\sum_{ij}\rho _{ij}|i(t)><j(t)|=\sum_{ij}\rho _{ij}e^{i(\omega
_i-\omega _j)t}|i><j| \\ 
=\sum_i\rho _{ii}|i><i|+\sum_{i\neq j}\rho _{ij}e^{i(\omega _i-\omega
_j)t}|i><j|...(5.1.6)
\end{array}
\]

Then the time evolution of the mean value of eq.(5.1.3) is:
\[
<O>_{\rho (t)}=(\rho (t)|O)=\sum_i\rho _{ii}O_{ii}+\sum_{i\neq j}\rho
_{ij}e^{i(\omega _i-\omega _j)t}O_{ji}.......(5.1.7) 
\]
Now let us suppose that the steps of the spectrum are so small and the
function under the second summatory of the r.h.s. of the last equation is so
nice that this summatory can be approximated by an integral. Therefore, if
the function is nice enough, from Riemann-Lebesgue theorem we would have:
\[
\stackunder{t\rightarrow \infty }{\lim }(\rho (t)|O)=\sum_i\rho
_{ii}O_{ii}=(\rho _{*}|O)............(5.1.8) 
\]
where we have defined an equilibrium density matrix $\rho _{*ij}=\rho
_{ii}\delta _{ij}$. This equation would be the quantum analog of the
classical equation (4.3.3) for mixing system and it would show that both
system have a similar behavior and opens the possibility to use classical
theorems in the quantum case also. Of course this demonstration is not
rigorous, but it serve to motivate the study of continuous spectra of next
section. Using continuous spectra we will find a rigorous theorem. The role
played by continuous spectra in this case is not strange since evolution
operators of mixing systems have this kind of spectrum.[30]. Anyhow, we can
also say that the most we can get is a weak limit, since Riemann-Lebesgue
theorem cannot be used directly in eq.(5.1.6). Furthermore, de decomposition
of the r.h.s. of this equation is not a decomposition within space ${\cal H}$
{\cal \ }since its second term has null trace (cf. eq.(2.3.2)).

\subsection{The case of a continuous spectrum.[31]}

In the next subsection we will consider the Friedrichs model, with can be
defined in Hilbert space $H${\cal \ }with a energy eigen-basis $%
\{|1>,|\omega >\},0\leq \omega <\infty ,$ with hamiltonian operator:
\[
\begin{array}{c}
H=\omega _1|1><1|+\int_0^\infty d\omega \omega |\omega ><\omega | \\ 
+\lambda \int_0^\infty d\omega g(\omega )[|\omega ><1|+|1><\omega
|].....(5.2.1)
\end{array}
\]

In this section this formula will be only used as an example of an operator
expanded in a continuous spectrum basis to conclude that the expansion of a
generic selfadjoint operator reads:
\[
O=\int_0^\infty d\omega O_\omega |\omega ><\omega |+\int \int_0^\infty
d\omega d\omega ^{\prime }O_{\omega \omega ^{\prime }}|\omega ><\omega
^{\prime }|.......(5.2.2) 
\]
where $O_\omega ,O_{\omega \omega ^{\prime }}$are regular functions such
that $O_\omega \in R,O_{\omega ^{\prime }\omega }^{*}=O_{\omega \omega
^{\prime }}$. Below we will say that $O\in \Phi ,$ a space with some
properties that we will chose for convenience. Thus functions $O_{\omega
,}O_{\omega \omega ^{\prime }}$ will be restricted by this choice.

The first term of the r.h.s. of eq. (5.2.2) will be called the singular
component of $O$ , since it could be written as the second term but with a
singular coefficient $O_{\omega \omega ^{\prime }}=O_\omega \delta (\omega
-\omega ^{\prime }).$ The second term will be called the regular term.

Let us consider density matrix at time $t=0$:
\[
\rho (0)=\int \int_0^\infty d\omega d\omega ^{\prime }\rho _{\omega \omega
^{\prime }}|\omega ><\omega ^{\prime }!.........................(5.2.3) 
\]
at time $t$ this state reads:
\[
\rho (t)=\int \int_0^\infty d\omega d\omega ^{\prime }\rho _{\omega \omega
^{\prime }}|\omega ><\omega ^{\prime }|e^{-i(\omega -\omega ^{\prime
})t}.............(5.2.4) 
\]

If we consider that $O$ can be written as the $\rho $ of eq. (5.2.3) but
with coefficients $O_\omega \delta (\omega -\omega ^{\prime })+O_{\omega
\omega ^{\prime }}$ (as in eq. (5.2.2)) the mean value of operator $O$ in
the state $\rho (t)$ is:
\[
\begin{array}{c}
<O>_{\rho (t)}=(\rho (t)|O)=tr(\rho (t)O)=\int_0^\infty d\omega \rho
_{\omega \omega }O_\omega \\ 
+\int \int_0^\infty d\omega d\omega ^{\prime }\rho _{\omega \omega ^{\prime
}}O_{\omega \omega ^{\prime }}e^{-i(\omega -\omega ^{\prime })t}....(5.2.5)
\end{array}
\]
Now,if space $\Phi $ is chosen in such a way that Riemann-Lebesgue theorem
can be used, namely let the functions of $\Phi $ be $L^1,$ we have:
\[
\stackunder{t\rightarrow \infty }{\lim }<O>_{\rho (t)}=\int_0^\infty d\omega
O_\omega \rho _{\omega \omega }...............................(5.2.6) 
\]
As this equation is valid for any operator $O\in \Phi $ we may try to find a
density matrix $\rho _{*}$ such that:
\[
\stackunder{t\rightarrow \infty }{\lim }<O>_{\rho (t)}=\stackunder{%
t\rightarrow \infty }{\lim }(\rho (t)|O)=(\rho
_{*}|O)......................(5.2.7) 
\]
It is easy to see that the density matrix $\rho _{*}$ cannot be find if $%
\rho _{\omega \omega ^{\prime }}$ is a regular function of variables $\omega
,\omega ^{\prime },$ i.e. from eq. (5.2.5) we see that to obtain this result
it is necessary that $\rho _{\omega \omega ^{\prime }}=0$, $\omega \neq
\omega ^{\prime }$ and $\rho _{\omega \omega }\neq 0,$ but we cannot write $%
\rho _{\omega \omega ^{\prime }}=\rho _\omega \delta (\omega -\omega
^{\prime })$, because in this case the $\rho $ is not regular. [32]. Then we
are forced to consider states with diagonal singularities, namely with the
same operator's pathology. So we are forced to introduce singular components
in the density matrix, namely the$\rho _{\omega ,\omega ^{\prime }}$ of eq
(5.2.3) cannot be regular and it must read something like $\rho _\omega
\delta (\omega -\omega ^{\prime })+\rho _{\omega ,\omega ^{\prime }}$. But
now if we try to find the mean value (5.2.5), the $O_\omega \delta (\omega
-\omega ^{\prime })$ term and the $\rho _\omega \delta (\omega -\omega
^{\prime })$ term produce the result:
\[
\int \int_0^\infty O_\omega \delta (\omega -\omega ^{\prime })\rho _{\omega
^{\prime }}\delta (\omega -\omega ^{\prime })d\omega d\omega ^{\prime
}=\int_0^\infty O_\omega \rho _\omega \delta (0)d\omega \rightarrow \infty 
\]
which is divergent. Therefore to have a formalism free of these problems we
are forced to make a fresh start and to consider that the operators $O$ are
defined by the regular functions $O_\omega ,O_{\omega ,\omega ^{\prime }}$
and the state functions $\rho $ are the matrices of rigged space $\Phi
^{\times }$ defined as the linear operators on space $\Phi $ and therefore
are defined by two regular functions $\rho _\omega ,\rho _{\omega ,\omega
^{\prime }}.$ Then we have:
\[
(\rho |0)=\int_0^\infty d\omega \rho _\omega O_\omega +\int \int_0^\infty
d\omega d\omega ^{\prime }\rho _{\omega \omega ^{\prime }}O_{\omega \omega
^{\prime }}......(5.2.8) 
\]
where $\rho _\omega \in R,\rho _{\omega ^{\prime }\omega }^{*}=\rho _{\omega
\omega ^{\prime }}$. So we are forced to introduce a singular component $%
\rho _\omega $ also in the density matrices. Now $\rho _{*}$ can be found,
it is the functional of space $\Phi ^{\times }$ with $\rho _\omega \neq
O,\rho _{\omega \omega ^{\prime }}=0.$ The consistence of this method is
proved by the logical physical results of paper [31].

Eq. (5.2.7) can now be consider as the rigorous quantum analog of the
classical eq. (4.3.3). We can call the weak limit of this equation the
''quantum mixing'' property and state the following

{\bf Theorem 5.2.1.-}Quantum system with continuous spectrum are endowed
with the quantum mixing property. (provided we use the formalism based in
eq. (5.2.8))

Eq. (5.2.7) can also be considered a prove of a weak decoherence in quantum
systems.This would the no-graining conclusion. But we would like to have a
strong decoherence. Then we can follow two ways: We can use coarse-graining.
This technic is well known, so we refer to papers [33] and [34]. Or we can
use fine-graining. In this case, in order to obtain a strong limit from the
weak limit of eq. (5.2.8) we must gave a precise sense to all terms of the
r.h.s.of eq. (5.2.5), rigging the Hilbert space ${\cal H}$ {\cal \ }in such
a way that all the mathematical characters are well defined. So working with
the functional of space $\Phi ^{\times }$ we can write the strong limit:
\[
\stackunder{t\rightarrow \infty }{\lim }\rho (t)=\rho
_{*}..............(5.2.9) 
\]
that corresponds to the classical strong limit (4.4.3).

Now we would like to obtain, not just a limit, but the time irreversible
evolution of $\rho (t)$ that yields the limit (5.8.9). Unfortunately our
silkiness to work with continuous spectra is very limited [35] so we are
forced to use some mixed technics, as we shall see in the next subsections.

\subsection{Friedrichs model.}

\subsubsection{The general formalism.}

We believe that the, well known, Friedrichs model is the best quantum
example to fix the ideas.\ In this example we have a free (naked) stable
quantum state $|1>$ (which is postulate to be real: $K|1>=|1>$) that becomes
unstable when coupled to a continuous field $|\omega >.$ The stable state
can be considered as a simplified model of an atom in an exited stable
state, that becomes unstable if coupled with an electromagnetic field,
which, in the model, is represented by the continuous field. Thus, let us
consider a Hilber space ${\cal H,}$ with a basis ${\ \{|1>,|\omega >\}}%
,0\leq \omega <\infty $ such that: 
\[
<1|1>=1,<1|\omega >=0,<\omega |\omega ^{\prime }>=\delta (\omega -\omega
^{\prime })....(5.3.1) 
\]
\[
1=|1><1|+\int_0^\infty |\omega ><\omega |d\omega ..........(5.3.2) 
\]
and a system with free hamiltonian:

\[
H_0=\omega _1|1><1|+\int_0^\infty \omega |\omega ><\omega |d\omega
.........(5.3.3) 
\]
and $\omega _1>0.$ Therefore the spectrum of $H_0$ is $R_{+}$ with a
degeneration at $\omega _1.$ Let the interaction hamiltonian be: 
\[
H_I=\lambda \int_0^\infty g(\omega )(|1><\omega |+|\omega ><1|)d\omega
..........(5.3.4) 
\]
where $g(\omega )$ is an interaction function endowed with all sort of nice
properties: it is analytical, well behaved at $\omega \rightarrow +\infty $,
etc., etc. The total hamiltonian is: 
\[
H=H_0+H_I...................(5.3.5) 
\]

This hamiltonian can be diagonalized, using standard technics. Then we
obtain: 
\[
H=\int_0^\infty \omega |\omega ,\stackrel{adv}{ret}><\omega ,\stackrel{adv}{%
ret}|d\omega ..........(5.3.5) 
\]
where $\{|\omega ,\stackrel{adv}{ret}>\}$ are the usual retarded or advanced
bases [38]. We can see, comparing eq. (5.3.3) and eq. (5.36) that the
interaction has erased the discrete component of the spectrum. In fact,
state $|1>$has became unstable and now it is just a pole in the
corresponding S-matrix. Any how using eq. (5.3.6) we can compute the time
evolution of any state, e. g. the state $|1>$ at $t=0$. As we have just said
state $|1>$ of the free system (5.3.2) is transformed in an unstable state
by the interaction (5.3.4), in such a way that the survival probability $%
P(t)=|<1|1(t)>|^2$ vanishes when $t\rightarrow +\infty .$ It is also known
that $P(t)$ has a vanishing derivative when $t=0$ (Zeno effect), then it has
a decreasing exponential behavior, and finally oscillates for big $t$
(Khalfin effect) (fig. 2)[36].

\subsubsection{Hilbert and rigged spaces.}

Let us forget for a moment the problem of the unification of the dynamics
and the thermodynamics and let introduce some equations related with the
problem of the time asymmetry, as it was stated is subsection 1.1. As we
shall see the previous equations are all what we need to define the quantum
arrow of time according to the coarse graining school.. As this school
always work within Hilbert space ${\cal H}$ , the following property hold: 
\[
K:{\cal H\rightarrow H\ }..............(5.3.7) 
\]
Instead for the fine greaning school we need a richer structure. In fact, we
need to define two subspaces $\phi _{\pm }\subset {\cal H}$ , in a
convenient way. To do so let us consider a vector$|\varphi >\in {\cal H}$
and its components $<\omega |\varphi >$ and let us promote the real energy $%
\omega $ to a complex variable $z$, then: 
\[
|\varphi >\in \phi _{\pm }.....iif.......<z|\varphi >\in H_{\pm }^2\cap 
{\cal S\ }..........(5.3.8) 
\]
where $H_{\pm }^2$ are the Hardy classes from above and below respectively
(cf. App. 5.A) and ${\cal S\ }$the Schwarz class of functions. It can be
roved that $\phi _{\pm }$ are nuclear spaces. Then we can then define two
Gel'fand triplets: 
\[
\phi _{-}\subset {\cal H\subset }\phi _{-}^{\times }...............(5.3.9) 
\]
\[
\phi _{+}\subset {\cal H}\subset \phi _{+}^{\times }.............(5.3.10) 
\]

\subsubsection{The rigged Hilbert space formalism.}

Using analytical continuation technics (cf. [15], [36],[37],[38]),
essentially just the Cauchy theorem, it is possible to obtain a new spectral
decomposition of the identity operator $1$ and the hamiltonian operator $H$
as: 
\[
1=|z_1,-><z_1,+|+\int_\Gamma |z,-><z,+|dz.......(5.3.11) 
\]
\[
H=z_1|z_1,-><z_1,+|+\int_\Gamma z|z,-><z,+|dz.......(5.3.12) 
\]
where $|z_1,->\in \phi _{-}^{\times }$, $|z_1,+>\in \phi _{+}^{\times }$ , $%
z_1$ is a complex rot of equation $\alpha (z)=0$ where: 
\[
\alpha (z)=z-\omega _1+\lambda \int_\Gamma \frac{g^{*}(z^{*})g(z)}{z-\omega
_1}dz............(5.3.13) 
\]
and $\Gamma $ is any curve that goes from the origin of the complex plane to
the positive infinity of the real axis and passes below $z_1$ (fig. 3). The
first terms of eqs. (5.3.11) and (5.3.12) are produced by the residues of
the poles corresponding to the roots located at the zeros of equation $%
\alpha (z)=0$ or, what is the same thing, the poles of the S-matrix. We can
see that the discrete component of the spectrum, that we have lost in eq.
(5.3.6), reappears in eq (5.3.12) in the form of a matrix of the rigged
Hilbert space.

Now we have several possibilities to chose the curve $\Gamma ,$ that are
used by different authors:

i.-To use a generic curve $\Gamma .$

ii.-To use curve $\Gamma ^{\prime }$ of fig. 4, in such a way that, as the
vertical paths of the curve are mutually cancel we are mostly integrating on
the real positive axis.

iii.-To take the negative real axis as the integration path.

iv.-To define a tilde operation as:
\[
\int_\Gamma f(z)g(z)dz=\int_0^\infty \widetilde{f}(x)g(x)dx 
\]
for all $g(x)$ in the test function space. In this case the complex integral
formally becomes a real one.

If we use this last method and forget the tilde eq. (5.3.12) reads: 
\[
H=z_1|z_1,-><z_1,+|+\int_0^\infty \omega |\omega ,-><\omega ,+|d\omega
......(5.3.12^{\prime }) 
\]
so we have build a basis $\{|z_1,->,|\omega ,->\}$ for space $\phi _{-}$(cf.
[37] for details). These vectors reads:
\[
|z_1,->=<1|z_1,->(|1>+\lambda \int_0^\infty d\omega \frac{g(\omega )}{%
[z_1-\omega ]_{-}}|\omega >)...........(5.3.12") 
\]
\[
|\omega ,->=|\omega >+\frac{\lambda g(\omega )}{\alpha (\omega )}%
(|1>+\lambda \int_0^\infty d\omega ^{\prime }\frac{g(\omega )}{\omega
-\omega ^{\prime }+i\epsilon }|\omega ^{\prime }>)....(5.3.12^{\prime }") 
\]
where the subindex ``--'' in the denominator of the integral in eq.
(5.3.12'') means that the curve $\Gamma ^{\prime }$ must be used for the
integration.

Now we have two spectra to compare: (5.3.5) and (5.3.12''). The main
difference is that (5.3.5) is structurally unstable when $\lambda
\rightarrow 0$, while (5.3.12'') is stable. In fact, an algorithm is call
structurally stable if it does not change much under small changes of the
coefficients. When $\lambda =0$ the spectral decomposition of H is (5.3.3).
If $\lambda $ is small, a small change of $\lambda $ that makes $\lambda =0,$
produces a big change in the usual decomposition, that goes from (5.3.5),
with no discrete term, to (5.3.3) with the discrete term $\omega _1|1><1|$ .
The sudden vanishing of this term when $\lambda \rightarrow 0$ is a
catastrophe (precisely a Poincar\'e catastrophe) that creates a great number
of problems if we try to perform an expansion around $\lambda =0$ in Hilbert
(or Liouville) space. On the contrary (5.3.12'') is stable, since it has the
term $z_1|z_1,-><z_1,+|$that goes to $\omega _1|1><1|$ when $\lambda
\rightarrow 0$ as we shall see.

From eq. (5.3.12) it can be seen that $|z_1,->$ and $<z_1,+|$ are
respectively the left and right- eigenvector of $H$ corresponding both to
the eigenvalue $z_{1.}$ It can be proved that:
\[
<z_1,+|z_{1,-}>=1.................................... 
\]
\[
<z+|z,->=0....................................... 
\]
\[
<z,+|z_{1,-}>=0...................................... 
\]
\[
<z.+|z,-^{\prime }>=\delta (z-z^{\prime })..........(5.3.14) 
\]
It can also be proved that:
\[
<z_1,-|z_{1,-}>=0..................................... 
\]
\[
<z_{1,}+|z_1,+>=0.....................(5.3.14^{\prime }) 
\]
namely there are non-null vectors of zero norm in spaces $\phi $$%
_{-}^{\times }$ and $\phi $$_{+}^{\times }$ [39].

Let us call $z_1=\beta _1-\frac i2\gamma _1$, where $\gamma _1>0$. Then from
eq.(5.3.12') we can obtain the time evolution of $|z_1(t),->$ and $%
|z_1(t),+> $ precisely:
\[
|z_1(t),->=e^{-iz_1t}|z_1(0),->=e^{-i\beta _1t}e^{-\frac{\gamma _1}%
2t}|z_1(0),->.................. 
\]
\[
|z_1(t),+>=e^{-iz_1^{*}t}|z_1(0),+>=e^{-i\beta _1t}e^{\frac{\gamma _1}%
2t}|z_1(0),+>......(5.3.15) 
\]
These equations show that $|z_1(t),->$is a decaying state and ,in fact all
states of $\phi $$_{-}^{\times }\setminus {\cal H}$ are decaying states,
while $|z_1(t),+>$ is a growing state, and all states of $\phi $$%
_{+}^{\times }\setminus {\cal H}$ are growing states.

It can be proved that [15]:
\[
K|z_{1,-}>=|z_{1,+}>....................................... 
\]
\[
K|z_{1,+}>=|z_{1,-}>..........................(5.3.16) 
\]
which is a natural fact, since growing states must be transformed in
decaying states by the time-inversion operator and vice versa. Therefore we
have:
\[
K:\phi _{-}^{\times }\rightarrow \phi _{+}^{\times
}............................................... 
\]
\[
K:\phi _{+}^{\times }\rightarrow \phi _{-}^{\times
}.................................(5.3.17) 
\]

The following limits are valid (cf. eq. (5.3.12'') :
\[
\stackunder{\lambda \rightarrow \infty }{\lim }|z_1,->=\stackunder{\lambda
\rightarrow \infty }{\lim }|z_1,+>=|1>..................(5.3.18) 
\]
Therefore $|z_1,->$ and $|z_{1,}+>$ can be considered as version of the
unstable state $|1>$ in spaces $\phi $$_{-}^{\times }$ and $\phi $$%
_{+}^{\times }$. In fact, the difference between these vectors and $|1>$ is
a $O(\lambda ),$ since when $\lambda =0$ the interaction disappears. Let us
remember that the survival probability of state $|$$1(t)>$ was:
\[
P(t)=|<1|1(t)>|^2=<1|1(t)><1(t)|1>....(5.3.19) 
\]
$P(t)$ shows the initial Zeno effect behavior, then an exponential behavior
and finally the oscillatory Khalfin effect behavior. If we make the
substitution $|1(t)>\rightarrow |z_1(t),->$ we obtain:
\[
P(t)\rightarrow P^{\prime }(t)=<1|z_1(t),-><z_1(t),-|1>=e^{-\gamma
_1t}..............(5.3.20) 
\]
and only the exponential behavior remains. Thus the physical nature of the
state $|z_1,->$ would be the one of a decaying unstable ideal state, where
we have eliminated the Zeno and Khalfin effects, because these effects are
contained in the last term of the r.h.s. of eq. (5.3.12'') (also called
``the background''). Namely, the three effects are mixed if we use the time
evolution based in eq. (5.3.6), but Zeno and Khalfin effects can be
separate, from the exponential behavior, if we use the evolution based in
eq. (5.3.12). Eq. (5.3.20) show also that $\gamma _1^{-1}$ is the mean life
time of the unstable states. Fine graining can be thought as an
approximation of real states that eliminates the unimportant Zeno and
Khalfin effects. Zeno effects is unimportant because it takes place at $t=0$
while, we are generally interested in the phenomena at $t\rightarrow \infty $%
. Khalfin effect it is uninteresting because essentially it is an
oscillatory effect, around the exponential behavior, while we are interested
in mean values only. But as we have said in the introduction it is not
completely clear if these ideal exponential states are just, mathematical
useful,effective state or real physical states.

Furthermore, fine graining school need to work with rigged Hilbert spaces $%
\phi $$_{-}^{\times }$ and $\phi $$_{+}^{\times }$ to solve the problem of
the arrow of time as we shall see in the next section.

Friedrichs model is just an example, but its rigged Hilbert space structure
can be found in every scattering process [38]. Therefore even if we will
base our reasoning in this model, what we will explain below is rather
general.

\subsubsection{Mixed states.}

Let us now introduce the arguments of the next subsection written the
evolution equations of mixed states in our model. A mixed arbitrary state at
time $t=0$ can be expanded in basis $\{|\omega ,ret>\}$ as:
\[
\rho =\int \int_0^\infty \rho _{\omega \omega ^{\prime }}|\omega
,ret><\omega ^{\prime },ret|d\omega .......................(5.3.21) 
\]
and its time evolution reads.
\[
\rho (t)=\int \int_0^\infty \rho _{\omega \omega ^{\prime }}e^{-i(\omega
-\omega ^{\prime })t}|\omega ,ret><\omega ^{\prime },ret|d\omega
.......(5.3.22) 
\]
We can as well use the advanced basis, but this is all what we can say in
space ${\cal H}$. But in space $\phi _{-}^{\times }$ we can use the basis $%
\{|z_1.->,|\omega ,->\}$ (introduced in eq. (5.3.12')) and expand $\rho $ as:
\[
\begin{array}{c}
\rho =\rho _{11}|z_1,-><z_1,-|+\int_0^\infty (\rho _{1\omega }|z_1,-><\omega
_{,-}| \\ 
+\rho _{\omega 1}|\omega ,-><z_1,-|)d\omega + \\ 
\int \int_0^\infty \stackrel{\sim }{\rho }_{\omega \omega ^{\prime }}|\omega
,-><\omega ,-|d\omega
\end{array}
...(5.3.23) 
\]
and its time evolution reads:
\[
\rho (t)=\rho _{*}(t)+e^{-\frac 12\gamma _1t}\rho _1(t)+e^{-\gamma _1t}\rho
_2(t)............................(5.3.24) 
\]
where:
\[
\rho _{*}(t)=\int \int_0^\infty \stackrel{\sim }{\rho }_{\omega \omega
^{\prime }}e^{-i(\omega -\omega ^{\prime })t}|\omega ,-><\omega ^{\prime
},-|d\omega ............(5.3.25) 
\]
and:
\[
\begin{array}{c}
\rho _1(t)=\int_0^\infty (\rho _{1\omega }e^{-i(\beta _1-\omega
)t}|z_1,-><\omega ,-| \\ 
+\rho _{\omega 1}e^{-i(\omega -\beta _1)t}|\omega ,-><z_1,-|)d\omega
.......(5.3.26)
\end{array}
\]
\[
\rho _2(t)=\rho _{11}|z_1,-><z_1,-|............................(5.3.27) 
\]
Now since $\gamma _1>0,$ $\rho _1(t)$ oscillates, and, $\rho _2(t)$ is time
invariant we have:
\[
\stackunder{t\rightarrow \infty }{\lim }(\rho (t)-\rho
_{*}(t))=0.......................................(5.3.28) 
\]
which seams very close to the strong limit we are looking for. The only
problem is that $\rho _{*}(t)$ it is not a dynamical equilibrium state,
since it oscillates. Nevertheless from the thermodynamical point of view $%
\rho _{*}(t)$ is a thermodynamical equilibrium state, since it has a
constant (and maximum) Gibbs entropy. In fact , it is evident (from the
quantum version of theorem (3.6.2)) that Gibbs entropy is constant for the
time evolution (5.3.22), therefore it is constant for the time evolution
(5.3.25) which it is similar. It is clear that is all what we can ask to the
model, since the field cannot go to dynamical equilibrium because the modes
of the field are uncoupled.

Therefore, from the thermodynamical point of view eq. (5.3,28) reads.
\[
\stackunder{t\rightarrow \infty }{\lim }\rho (t)=\rho
_{*}..............................................(5.3.29) 
\]
and it is the strong limit we are looking for. As in the case of eq.
(4.5.1.11) this limit belongs to the corresponding rigged Hilbert space.

Where is the miracle that allows to pas from the oscillatory
evolution(5.3.22), with no limit to the partially dumped evolution (5.3.24)
with a thermodynamical limit? The miracle is that eq. (5.3.22) is valid in
space ${\cal L}={\cal H}\times {\cal H}$ while eq. (5.3.24) is valid in
space $\Phi _{-}^{\times }=\phi _{-}^{\times }\times \phi _{-}^{\times }$ so
really eq. (5.3.28) is a functional equation that can be interpreted as:
\[
\stackunder{t\rightarrow \infty }{\lim }(\rho (t)|O_{-})=(\rho
_{*}|O_{-}).............................(5.3.30) 
\]
where $O_{-}$ is an operator of the test operator space $\Phi _{-}=\phi
_{-}\times \phi _{-}$, the space of measurement operators we have chosen.
Therefore the miracle happens just because we have chosen a convenient test
space for our physical measurement apparatuses. Eq. (5.3.30) is a weak limit
that is similar to the weak limit of mixing classical states and a
consequence of theorem 5.2.1 since our model has a continuous spectrum. We
will continue this line of reasonings once the more complete example of the
next subsection would be introduced.

In the practical case we will study in section 8 the field $|\omega >$ will
be the thermic radiation field within the universe that we can considered as
thermalized., from its beginning, by other interactions than those of eq.
(5.3,4), then it can be classically chosen as a Boltzmann thermic
distribution function:
\[
\rho _{*}=ZT^{-\frac 32}e^{-\frac \omega
T}...................................(5.3.31) 
\]
where $T$ is the temperature and $Z$ a normalization function, and the
dumping terms are produced by nuclear reaction phenomena within the stars
and $\gamma _1^{-1}$ is the characteristic time of these nuclear reactions.
We will use this model in section 8.

\subsection{Friedrichs model for many oscillators.}

We will now introduce a not very realistic physical model, that
nevertheless, is the simplest one for our purpose. Let us consider a set an
infinite (or a great number) of uncoupled harmonic oscillator, labelled by $%
\omega $ , with hamiltonian:
\[
H_\omega =\omega (a_\omega ^{\dagger }a_\omega +\frac
12)..................(5.4.1) 
\]
where $a_\omega ^{\dagger }$ and $a_\omega $ are the creation and
annihilation operators of the harmonic oscillator. The total hamiltonian
reads:
\[
H=\int_0^\infty H_\omega d\omega .........................(5.4.2) 
\]
$H_\omega $ can also be written:
\[
H_\omega =\sum_nH_\omega ^{(n)}...........................(5.4.3) 
\]
where:
\[
H_\omega ^{(n)}=\omega (n|n,\omega ><n,\omega |+\frac 12).......(5.4.4) 
\]

$|n,\omega >$ is the $\omega $-oscillator in the $n$ exited state ($n=0$
corresponds to the ground state)

Let us suppose that each of these states is coupled with a field represented
by a set of infinite states $|n,\omega ,w>$ in such a way that now the
coupled $H_\omega ^{(n)}$ reads:
\[
\begin{array}{c}
H_\omega ^{(n)}=\omega (n|n,\omega ><n,\omega |+\frac 12)+\int_0^\infty
dw|n,\omega ,w><n,\omega ,w|+ \\ 
\lambda \int_0^\infty dwg_{n,\omega }(w)(|n,\omega ><n,\omega ,w|+|n,\omega
,w><n,\omega |).....(5.4.5)
\end{array}
\]
where $\lambda $ is a coupling constant and $g_{n,\omega }(w)$ a coupling
function which necessarily has the property $g_{0,\omega }(w)=0$, since the
ground state of each oscillator is stable and therefore it is not coupled
with the corresponding field that would produce its instability. Therefore
we have constructed a model which can be considered as a infinite repetition
of the Friedrichs model of last subsection. In this non-realistic model the
instability of all the states, with the exception of the ground states, is
obtained by coupling a field to each oscillation mode. It is, in fact, a
non-economical way, but it serve to our purpose, which it is now only to
find the laws of unstable evolutions.

Now, using the procedure of the previous subsection in each Friedrichs model
of each mode we can diagonalize each operator $H_\omega ^{(n)}$ and we will
obtain:
\[
H_\omega ^{(n)}=z_\omega ^{(n)}n|n,\omega ,+><n,\omega ,-|+\frac 12\omega
.................(5.4.6) 
\]
where, for simplicity, we have omitted the field term, where $z_\omega
^{(0)}=\omega ,$ since the ground states are not perturbed,and, $Imz_\omega
^{(n)}<0$ for $n\neq 0.$ If we renormalize and eliminate the $\frac 12$%
-terms we obtain:
\[
H=\int_0^\infty \sum_nz_\omega ^{(n)}(|n,\omega ,+><n,\omega ,-|d\omega
...............(5.4.7) 
\]
where we have put the factor $n$ inside $z_\omega ^{(n)}.$

Let us now consider a density matrix $\rho =\rho (0)\in {\cal L=H\times H,}$
that can be expanded in basis $\{|n,\omega ,->\}$ as:
\[
\rho =\int_0^\infty \sum_n\rho _{n,n^{\prime };\omega }|n,\omega
,-><n,\omega ,-|d\omega ..............(5.4.8) 
\]

We will always work with these density matrices below. And these is the
essential fact. Since $|n,\omega ,->\in \phi _{-}^{\times }$ and, therefore $%
\rho \in \phi _{-}^{\times }\times \phi _{-}^{\times },$ what we have done,
in choosing the expansion (5.4.8), is to assume that our operator space is $%
\Phi _{-}=\phi _{-}\times \phi _{-}$ in such a way that now we have the
Gel'fand triplet:
\[
\Phi _{-}\subset {\cal L}\subset \Phi _{-}^{\times
}....................................(5.4.9) 
\]
and,therefore, $\rho \in \Phi _{-}^{\times }=\phi _{-}^{\times }\times \phi
_{-}^{\times }$. On physical ground what we are doing is to postulate that
our measurement apparatuses correspond to operators in $\Phi _{-}$. We will
discuss this postulate below, but we can see immediately that this is the
price to pay to get the strong limit we are looking for and the
corresponding unstable time evolution.

In fact, from eq. (5.3.12) we can obtain the time evolution:
\[
\begin{array}{c}
\rho (t)=e^{-iLt}\rho (0)= \\ 
\int_0^\infty \sum_n\rho _{n,n^{\prime };\omega }e^{-i(z_\omega
^{(n)}-z_\omega ^{(n^{\prime })*})t}|n,\omega ,-><n^{\prime },\omega
,-|d\omega .......(5.4.10)
\end{array}
\]
but since

\[
-i(z_\omega ^{(n)}-z_\omega ^{(n^{\prime })*})=-i(\beta _\omega ^{(n)}-\beta
_\omega ^{(n^{\prime })})-\frac i2(\gamma _\omega ^{(n)}-\gamma _\omega
^{(n^{\prime })})...........(5.4.11) 
\]
and $\gamma _\omega ^{(n)},\gamma _\omega ^{(n^{\prime })}\geq 0,$ (only $%
\gamma _\omega ^{(0)}=0)$, thus:
\[
\stackunder{t\rightarrow \infty }{\lim }\rho (t)=\int_0^\infty \sum_n\rho
_{0,0;\omega }|0,\omega ,-><0,\omega ,-|=\rho _{*}..................(5.4.12) 
\]
and we have obtained our strong limit, equivalent to (5.2.9). Furthermore,
now we have the time evolution to this limit, eq. (5.4.10). To obtain this
result we have used an infinite set of continuous field that we neglect in
all the formulas above. Some how we have ''traced away'' these fields. But
the result will not change, from the physical point or view, if we write all
these fields. The result we have obtaining regarding the states of the
harmonic oscillator will be the same, these oscillators reach to the
equilibrium showed in the last equation, but the fields will continue to
oscillate and they will be always far from the equilibrium (like in the last
part of the last subsection). This is not surprising, since these fields
have no self-interaction or mutual interaction, therefore they cannot reach
to equilibrium. Then, to neglect these field was only a useful shorthand
with no physical consequences-(provided we take into account all the
warnings we made in the last part of the last subsection). We must also
remember that the quantities of eq. (5.4.12) are just functionals over the
space $\Phi _{-}$ thus if we contract this equation with any vector of this
space we will find the weak version of limit (5.4.12) showing that in this
example theorem 5.2.1 is fulfilled.

If we collectively call $2\gamma $ to all the $\gamma ^{\prime }s$ or $%
2\gamma $ is the inverse of the characteristic life time of the system or if
we call $2\gamma $ to the smaller of them to maintain the leading term only,
eq. (5.4.11) reads:

\[
\rho (t)=\rho _{*}+\rho _1e^{-\gamma t}..................(5.4.13) 
\]
as usual we have:
\[
tr\rho =tr\rho _{*}=1........................(5.4.14) 
\]
.since matrix $\rho $ is the usual one and matrix $\rho _{*}$ is an
expansion of stable states (5.4.12) and norm must be conserved (cf. Appendix
5.B). But:
\[
tr\rho _1=0..............................(5.4.15) 
\]
as a consequence of eq. (5.3.14'), showing that this matrix is something
like a fluctuation around the equilibrium state.

Let us finally observe that in this model we can not pretend that $\rho _{*}$
would be the equilibrium state of the canonical ensemble. To obtain that
result obviously we must coupled the oscillators-among themselves and the
model will be much more complicated. To mimic a canonical ensemble at
temperature $T$ in this model the best we can do is to make the following
choice:
\[
\rho _{0,0;\omega }=\frac Z{T^{\frac 32}}e^{-\frac \omega
T}....................(5.4.16) 
\]
with this choice we have the correct equilibrium distribution and the field
produce the irreversible evolution toward this equilibrium. The evolution of 
$\rho (t)$ then reads:
\[
\rho (t)=\int_0^\infty \left[ \frac Z{T^{\frac 32}}e^{-\frac \omega T}\rho
_{*}^{(\omega )}+e^{-2\gamma t}f(\omega )\rho _1^{(\omega )}\right] d\omega
........(5.4.17) 
\]
where $f(\omega )$-is an arbitrary function, so the conclusions are
essentially the same as in the last subsection.

\subsection{Appendix 5.A.-Hardy class functions [29].}

A complex function $f(\omega )$ on $R^{+}$ is a Hardy class function from
above (below) if:

i.-$f(\omega )$ is the boundary value of a function $f(z)$ of the complex
variable $z=x+iy$ that is analytic in the half plane $y>0$ $(y<0).$

ii.-$\int_{-\infty }^{+\infty }|f(x+iy|^2dx<k<\infty $, for all $y$ with $%
0<y<\infty $ $(-\infty <y<0).$

\subsection{Appendix 5.B. Computation and conservation of norm, trace, and
energy.}

The trace of a density matrix $\rho $ is:
\[
tr\rho =\int_0^\infty <\omega ,ret|\rho |\omega ,ret>d\omega
..............(5.B.1) 
\]
and it is invariant under changes of basis. Making the same procedure that
we used to go from eq (5.3.6) to eq. (5.3.12) we can obtain:
\[
tr\rho =<z_1,+|\rho |z_1,->+\int_0^\infty <\omega ,+|\rho |\omega ,->d\omega
.........(5.B.2) 
\]
As the fine graining theory deals with states that vanishes when $%
t\rightarrow \infty $, we can be worried guessing if the norms, the traces,
or the energy are conserved in this theory. There is no problem since we can
state the following results.

i.-From eqs. (5.3.14,14') we see that unstable density matrices like $%
|z_1,-><z_1-|$ or $|z_1,+><z_1,+|$ have null trace this fact is possible
since we are not working in Hilbert space..

ii.-Using eqs. (5.3.15) and (5.3.14') we can see that the trace of eq.
(5.B.2) is conserved as the usual trace of eq. (5.B.1)

iii.-If the trace is conserved also the norm of pure states is conserved.

iv.-The mean value of the energy in one of these unstable states like $%
|z_1,-><z_1,-|$ reads:
\[
<H>=<z_1,-|H|z_1,->=z_1<z_1,-|z_1,->=0......(5.B.3) 
\]
and therefore the energy of the states that vanish when $t\rightarrow \infty 
$ it is zero, creating no problems with energy conservation..

\section{Coarse-Graining and Trace. Time Asymmetry.}

\subsection{Coarse graining.}

Let us go back, for a while, to the classical regime. Usually
coarse-graining is based in the fact that the dynamical variables cannot be
measured with infinite precision, i.e., there is always an error and also we
cannot compute with an infinite number of digits. Perhaps there is a
fundamental graininess in nature but this graininess it is not yet neither
theoretically nor experimentally found.

Coarse graining can be introduced by partitioning the space $X$ into finite
(or discrete) number of cells $A_i$ that satisfy:
\[
\bigcup_iA_i=X,.........A_i\cap A_j=\emptyset ,.if..i\neq j..........(6.1.1) 
\]
This partition is arbitrary but it must be non-trivial with respect to some
measure $\mu $ namely:
\[
0<\mu (A_i)<\mu (X).............................................(6.1.2) 
\]
for all values of $i.$ For every density $\rho $ within each cell $A_i$ of
the partition, we can compute the average of $\rho $ as:
\[
<\rho >_i=\frac 1{\mu (A_i)}\int_{A_i}\rho (x)\mu
(dx)........................(6.1.3) 
\]
and the coarse grained density respect to the partition is given by:
\[
\stackrel{\sim }{\rho }(x)=\sum_i<\rho >_i1_{A_i}(x)=\{\sum_i\frac 1{\mu
(A_i)}|1_{A_i})(1_{A_i}|\}|\rho )=\Pi \rho (x).....(6.1.4) 
\]
where $1_{A_i}$ is the characteristic function of the cell $A_i$ and $\Pi $
is the projector defined by the partition. $\Pi $ is a projector since:
\[
\Pi ^2=\left[ \sum_i\frac 1{\mu (A_i)}|1_{A_i})(1_{A_i}|\right] \left[
\sum_j\frac 1{\mu (A_j)}|1_{A_j})(1_{A_j}|\right] = 
\]
\[
\sum_{ij}\frac 1{\mu (A_i)}\frac 1{\mu (A_j)}|1_{A_i})(1_{A_j}|\mu
(A_i)\delta _{ij}=\sum_i\frac 1{\mu (A_i)}|1_{A_i})(1_{A_i}|=\Pi 
\]
From the reasonings of the introduction or from theorem 4.3.1 we can deduce,
in the case of finite number partition (for discrete number see [7]) the

{\bf Theorem 6.1.1.-}If $P_t$ is a $\rho _{*}$-mixing Markov operator with a
unique stationary density $\rho _{*}$ and $\{A_i\}$ is a non-trivial
partition of the phase space $X$, then:
\[
\stackunder{t\rightarrow \infty }{\lim }(P_t\rho )^{\sim }=\stackunder{%
t\rightarrow \infty }{\lim }\stackrel{\sim }{\rho }%
_{*}..........................................(6.1.5) 
\]
for all initial densities.

Thus we have obtained our coarse graining strong limit. Now we can consider
the transformation:
\[
\stackrel{\sim }{P_t}\stackrel{\sim }{\rho (x)}=(P_t\rho (x))^{\sim
}..........................................(6.1.6) 
\]
from eq. (6.1.5) we see that this transformation has a strong limit (i.e. in
the norm) and therefore, according to theorem 4.4.1 it is exact.

Then using theorem 4.4.2-we can say about entropy that:

{\bf Theorem 6.1.2.-}If $P_t$ is a reversible $\rho _{*}$ mixing Markov
operator with a unique stationary density $\rho _{*}$ and $\{A_i\}$ is a non
trivial partition of the phase space $X$, then: $.$%
\[
\stackunder{t\rightarrow \infty }{\lim }H_C((P_t\rho (t))^{\sim }|\stackrel{%
\sim }{\rho }_{*})=0.............................(6.1.7) 
\]
for all initial densities $\rho $.

But we must realize that the way the conditional entropy converge to zero
depends on the way in which the coarse graining is carried out [6]. It can
be proved that the rate of convergency of entropy to equilibrium becomes
slower as the measurement technics improve and the coarse graining becomes
finer (!!!). Such phenomena have not been observed. Thus is most unlikely
that trivial coarse graining would play a role in determining
thermodynamical behavior, if a natural graininess it is not found.
Candidates for this natural and universal graininess would be:

i.- The graininess produced by operators $\stackrel{\sim }{\Pi },\stackrel{%
\wedge }{\Pi }$ which are introduced using fine graining methods [15],[40].
But, really, this is only the coarse-graining old version of the
fine-graining method.

ii.- The graininess introduced by the universe event horizon [41].

iii.-The graininess introduced by Plank's quantities, e.g. it seems that it
is absolutely impossible to measure length smaller than Plank's length.

But the physic related with these graininess is still under research.

More general projectors than those defined in eq. (6.1.4) can be used, as we
have seen in the introduction, since any projector will do the job done in
eqs. (1.2.1.2) and (1.2.1.3). A theory that uses one of these generalized
projectors will be called, by extension, also a ''coarse-graining'' theory.

Coarse-graining can be used also in the quantum case. Then $\Pi $ is a
projector over the quantum Liouville space ${\cal L=H\otimes H.}$ Using the
quantum theorem (5.2.1) we can obtain the same results as in the classical
case, where we use the theorem (4.3.1).

\subsection{Time asymmetry in coarse graining theories.}

All the limit of the previous section were computed as $t\rightarrow +\infty 
$ but it is clear that all these limits are also valid when $t\rightarrow
-\infty $. Therefore in coarse graining theories there is equilibrium both
in the far past and the far future. This fact can be easily verifier with
the baker transformation, where we have a set of infinite lines, horizontal
for the far future and vertical for the far past, that will be considered as
a uniform equilibrium distribution function for any coarse graining
partition. It is also evident that if the initial distribution function has
an adequate symmetry, e.g. the one of the characteristic function of a
square domain, the evolutions, toward the past and toward the future will be
strictly symmetric, but this will not be the case if the initial
distribution function is not symmetric (other calculations about the baker
transformation behavior can be found in ref. [6])

Going now to the quantum case we see a quite similar phenomena, even with no
coarse graining. Let us consider the state $|1>$ of Friedrichs model (which
can be considered as a symmetric initial condition, as the characteristic
function of an square domain, in the case of the baker transformation). The
behavior of the survival probability $P(t)$ , as shown in fig.2, is
completely symmetric with respect to $t=0$. Thus classically if we use
coarse graining technics, or quantum mechanically if we use only states of
the Hilbert space we will find that past is only conventionally different
than future. What is, then, the way to distinguish past from future? It is
the method that we explained in the introduction: Take the time $t=0$.
Consider the set of evolutions of the system for $t>0$ (for all possible
initial condition) and let as call it ${\cal H_{-}}$ in the quantum case $%
{\cal (}$or ${\cal L_{-}}$ in the classical case${\cal )}$. It is identical
to the set of evolutions for $t<0$ (for all possible-initial conditions),
that we shall call ${\cal H_{+}(}$or ${\cal L_{+})}$. The existence $T$ or $%
K,$ the mathematical transformation, that relates the future evolutions with
the past evolutions shows that these sets of evolutions are identical. In
fact.

{\bf Theorem 6.2.1.} For every evolution $\rho (t)\in {\cal L_{-\text{ t }}}%
(t>0)$ there is a time symmetric evolution $\rho (-t)\in {\cal L_{+}\ }(t<0)$
if the evolution equation are reversible.

Proof: From the definition of a reversible evolution (2.3.21) for every $%
\rho (t)\in {\cal L_{-\text{ }}}$ there is a physical evolution $\rho
(-t)\in {\cal L_{+}}$ defined by 
\[
\rho (-t)={\cal K}\rho (t),....t>0,..................\Box 
\]
Also we have:
\[
T:{\cal L_{-}\rightarrow L_{+},..........}T:{\cal L_{+}\rightarrow L_{-}} 
\]
These two sets of evolutions ${\cal L_{-,}L_{+}(}$or ${\cal H_{-,}H_{+})}$
are the two mathematical structures that we introduced in subsection 1.1. As
they are identical (cf. eq. (6.2.2)it is irrelevant to chose one or the
other. So let us chose one of these structures to build our theory, let us
say ${\cal L_{-},}$ and forget the other. Now we can say that the theory
begins at $t=0$ and goes toward the future for $t>0$ (or toward the past
since the choice of one word or the other as just conventional as the choice
between ${\cal L_{-\text{ }}}$ and ${\cal L_{+}}$). It is quite evident that
this theory developed in the lapse $0\leq t<\infty $ will fulfill all our
requirements,{\it \ provided we forget all about the lapse }$-\infty <t\leq
0.$ These are the characteristics of the resulting theory if we use coarse
graining and usual Hilbert space quantum states. Even if successful in many
respect it produces, should we say, a certain uneasiness.

\subsection{Traces.}

Let us now consider the classical case only.

Let $X$ and $Y$ be to topological Hausdorff phase spaces, $\varphi
:Y\rightarrow X$ a given continuous function on $X$ , and $S_t:Y\rightarrow
Y $ a dynamical system operating in phase space $Y$. A function $%
h:R\rightarrow X$ is a trace of the dynamical system if there is a point $y$
in space $Y$ such that $h(t)=\varphi (S_t(y)),$ for all times $t$ (this
meaning of the word trace must not be confuse with the one we use when we
speak about the trace of a matrix).

It can be proved that every continuous function in a space $X$ is the trace
of a single dynamical system operating in a phase space $Y$, therefore we
have the quite surprising

{\bf Theorem 6.3.1.-}[6] Let the phase space $X$ be an arbitrary but
topological Hausdorff space. Then there is a second phase space $Y$ also
topological and Hausdorff, a dynamical system $S_t$ operating in $Y$ and, a
continuous function $\varphi :Y\rightarrow X$ such that every continuous
function $h:R\rightarrow X$ is the trace of $S_t$. ( A topological space is
Hausdorff (or separable) if any two distinct points possess disjoint
neighborhoods)

That is, for every $h$ there is a point $y$ in phase space $Y$ such that $%
h(t)=\varphi (S_t(y))$, for all times $t.$

Let us now consider the trajectories of a dynamical system: If we have a
dynamical system $S_t$ operating in a phase space $Y$ , then only three
possible types of trajectories may be observed:

i.- The trajectory is a fixed point $x_{*}$ such that $S_tx_{*}=x_{*}$, for
all $t$.

ii.-The trajectory is a non intersecting curve, with the property $%
S_t(x)\neq S_{t^{\prime }}(x)$ if $t\neq t^{\prime }$.

iii.-A periodic trajectory such that $S_t(x)=S_{t+T}(x)$, for all times $t$,
being $T$ the period.

But nothing prevents the existence or non periodic intersecting trajectories 
$h(t)$, in space $X$ if $\varphi :Y\rightarrow X.$ Thus we can demonstrate
the following

{\bf Theorem 6.3.2.-}Let the phase spaces $X$ and $Y$ be topological
Hausdorff spaces and $h:R\rightarrow X$ an intersecting and non periodic
trace of a dynamical system $S_t:Y\rightarrow Y$. Then the entropy of
densities evolving under the action of $h$ is either constant or increasing.

Proof. The proof is based on the trivial observation that if $h$ is
intersecting and non periodic, then at every intersection point $x$ on the
trajectory $h$ the inverse $h^{-1}(x)$ is not unique. Therefore the trace $h$
is the trajectory of a semidynamical system, and since semidynamical systems
are irreversible, from theorems 3.6.1 and 3.6.2 the entropy is either
constant or increasing. $\Box $

Thus the simple act of taking a trace of a dynamical system (with
time-constant entropy) may be sufficient to generate a system in which the
entropy is increasing. But for certain class of traces much more can be said.

Let $X$ and $Y$ be two different phase spaces with normalizable measures $%
\mu _{*}$ and $\nu _{*}$ and associated densities $\rho _{*}$ and $\sigma
_{*}$,respectively, and $T_t:X\rightarrow X$ and $S_t:Y\rightarrow Y$ be two
measure preserving transformations. If there is a transformation $\varphi
:Y\rightarrow X$ that is also measuring preserving, i. e. if
\[
\nu _{*}(\varphi ^{-1}(A))=\mu _{*}(A)....................(6.3.1) 
\]
for all subsets $A$ of the phase space $Y$, and such that $T_t\circ \varphi
=\varphi \circ S_t$, then $T_t$ is called a factor of $S_t$. From this
definition the trajectory of the factor $T_t$ is a trace of the system $S_t$%
. Then we have the following

{\bf Theorem 6.3.3.}-[42]. Every $\rho _{*}$-exact transformation is the
factor of a Kolmogorov automorphism.

This theorem precise the things we must do if we want to find an exact
transformation with all its nice properties:

i.-We must show that the system we are working with is a Kolmogorov system.
This can be difficult from the mathematical point of view, but as chaos is
very frequent in nature it is not a very restricting physical condition.

ii.-Then, according to theorem 6.3.3., every measure preserving factor will
produce an exact transformation. The problem is just to find the most
convenient one.

As an example let us consider again the baker transformation. It can be
prove that this transformation is a Kolmogorov automorphism, endowed with a
constant entropy. However the system corresponding to coordinate $x$ is a
factor of baker transformation. Also it is identical to the dyadic Renyi
transformation:
\[
T(x)=2x..(mod1)............(6.3.2) 
\]
which is uniformly exact and whose entropy smoothly increases to zero by
theorem 4.4.2

We have show that coarse graining produces no substantial difference between
past and future. This is not the case with traces, as we can see from baker
transformation where that $x$-side of a parallelogram will always increase
toward the future and decrease toward the past, Thus coarse-graining do not
produces time asymmetry while traces do produce this phenomenon.

Let us now give all the whole panorama :

-We have projectors $\Pi $ like those introduced in section 6.1, namely such
that:
\[
\Pi :{\cal L}_Y\rightarrow {\cal L}_X,.....\Pi ^2=\Pi ,.....(6.3.3) 
\]
where ${\cal L}_Y={\cal L}$ is the state space and ${\cal L}_X$ is the space
of relevant states. $\Pi $ has not inverse $\Pi ^{-1}$ since $\Pi ^2\Pi
^{-1}=\Pi \Pi ^{-1}$ yields $\Pi =1.$

-We have traces:
\[
\varphi :Y\rightarrow X,.............................(6.3.4) 
\]
namely mapping between phase spaces. $\varphi $ can have an inverse, and in
this case $\varphi (Y)$ is dense in $X$, or it do not have an inverse when
the dimension of $X$ is smaller than the dimension of $Y,$ like in the case
of eq. (6.3.2)

Finally let us remark that traces define a mapping in the corresponding
Liouville spaces. Let ${\cal L}_X$ and ${\cal L_Y}$ be the corresponding
Liouville spaces to the phase spaces $X$ and $Y$. Then to the mapping $%
\varphi :Y\rightarrow X$ corresponds the mapping:
\[
\Lambda ^{-1}:{\cal L}_Y\rightarrow {\cal L}_X..................(6.3.5) 
\]
(the -1 is just a matter of convention) defined by:
\[
\Lambda ^{-1}\rho (y)=\rho (\varphi ^{-1}(x))...........(6.3.6) 
\]
In the next subsection we will study even more general mappings.

\subsection{Generalized traces.}

In section 6.3 we are forced to work in the classical case only, since we
have used phase space. Now we would like to generalize the trace notion in
order to work also in the quantum case.

A generalized trace is given by eq. (6.3.3) if eq. (6.3.4) is not fulfilled,
i. e. it is a mapping between Liouville spaces not originated by a mapping
between phase spaces. Being a mapping like (6.3.3) it is something like a
''projector with an inverse''.But now spaces ${\cal L}_Y$ and ${\cal L}_X$
can be classical or quantum Liouville spaces. These generalized traces are
typical of the fine-graining formalism, and try to show it as a kind of
generalization of the coarse-graining one.

Let us consider the particular case ${\cal L_{Y=}}\Phi _{-}^{\times }{\cal ,L%
}_X={\cal L}.$ Let us also consider the basis $\{|1>,|\omega >\}$ of eq.
(5.3.1) that we shall call $\{|i>\}$ and such that $H|i>=z_i|i>,z_i\in C.$
let us define the basis $\{|ij)\},|ij)=|i><j|.$Let us also consider the
basis \{$|z_1,->,|\omega .->\}$ of eq. (5.3.23), that we shall call \{$%
|i,->\}$ and in the same fashion let us define the basis \{$%
|ij,-)\},|ij,-)=|i,-><j,-|.$Using the basis $\{|z_1,+>,|\omega ,+>\}$ we
can, as well define a basis $\{|ij,+)\}.$ Then we can define a generalized
trace as:
\[
\Lambda ^{-1}:\Phi _{-}^{\times }\rightarrow {\cal L}%
........................................................(6.4.1) 
\]
\[
\Lambda =\sum_{ij}|ij,-)(ij|,...,\Lambda
^{-1}=\sum_{ij}|ij)(ij,+|............(6.4.2) 
\]

Namely $\Lambda $ is the transformation that make correspond each state $%
\rho $ of space ${\cal L}$ to a functional in space $\Phi _{-}^{\times }.$ $%
\Lambda $ looks like just a ''change of basis''. But really $\Lambda $ is
much more than a change of coordinates since it takes vectors of one space
to vectors in another space. Therefore to weak limits in ${\cal L}$
corresponds strong limits in $\Phi _{-.}^{\times }$ and the generalized
trace $\Lambda $ embodied the solution of our problem: to go from weak
limits to strong limits and can be considered as the symbol that synthesize
the fine graining technic.

Some observations are in order:

i.- Since $\Lambda $ is a generalized trace, therefore as a trace it
contains time asymmetry. In fact $\Lambda $ defined in eq. (6.4.1) is
related with dumping phenomena that produces equilibrium toward the future
and should be called $\Lambda _{-}.$ We can, as well, define a $\Lambda _{+}$
related with creation phenomena that implies equilibrium in the far past,
namely:
\[
\Lambda _{+}:\Phi _{+}^{\times }\rightarrow {\cal L}%
...................(6.4.3) 
\]
\[
\Lambda _{+}=\sum_{ij}|ij,+)(ij|,.....\Lambda
_{+}^{-1}=\sum_{ij}|ij)(ij,+|,.........(6.4.4) 
\]
if we chose $\Lambda _{-}$rather than $\Lambda _{+}$ we are creating a
time-asymmetry.

In order to see the relation of the two $\Lambda ^{\prime }s$ let us
introduce the star-conjugation:
\[
A^{\star }={\cal K}A^{\dagger }{\cal K}^{\dagger
}....................(6.4.5) 
\]
then it is easy to see that:
\[
\Lambda _{+}=\Lambda _{-}^{\star },....\Lambda _{-}=\Lambda _{+}^{\star
}........(6.4.6) 
\]
\[
\Lambda _{-}\Lambda _{-}^{\star }=\Lambda _{+}\Lambda _{+}^{\star
}=1,\Lambda _{-}^{-1}=\Lambda _{-}^{\star }=\Lambda _{+},\Lambda
_{+}^{-1}=\Lambda _{+}^{\star }=\Lambda _{-,}...(6.4.7) 
\]
for these last equation we can say that the $\Lambda ^{\prime }s$ are
star-unitary.

ii.- Using generalized trace $\Lambda $ we do not lose any information. (in
the case of usual traces we lose information if $\dim X<\dim Y,$ as in the
example of the baker transformation before eq (6.3.2) but in the case of the 
$\Lambda $-trace the dimension of the two spaces is the same and ${\cal L}$
is dense in $\Phi _{-}^{\times }$$).$ But it can be demonstrated that this
generalized trace $\Lambda ,$ some how, renormalize the infinite amount of
information contained in ${\cal L}$ [43]

iii.-A generalized trace is not a trace, so there is not mapping in the
corresponding phase spaces that make trajectories correspond to
trajectories. In this sense using fine-graining technics trajectories loose
al their importance and even have no meaning. [44],[45].

iv. $\Lambda $-trace allows to define a Hilbert space where the
time-evolution are irreversible..

In fact: using the bases we have introduce we can deduce that:
\[
1=\sum_{ij}|ij,-)(ij,+|...................(6.4.8) 
\]
\[
L=\sum_{ij}(z_i-z_j^{*})|ij,-)(ij,+|.......(6.4.9) 
\]
where as $Imz_i=-\frac{\gamma _i}2<0,$ it is $Im(z_i-z_j^{*})\leq 0,$ The
time evolution operator is $U(t)=e^{-iLt}$and $UU^{\dagger }=1,$ i. e. $U$
is unitary. Let us now define a modify liuovillian:
\[
G=\Lambda ^{\star }L\Lambda =\sum_{ij}(z_i-z_j^{*})|ij)(ij|....(6.4.10) 
\]
which induce a evolution $W(t)=e^{-iGt}$ such that $WW^{\dagger }\neq 1$,
and, therefore it is not unitary but star-unitary $WW^{\star }=1.$ the two
evolution are related by:
\[
W(t)=\Lambda ^{\star }U(t)\Lambda ........................(6.4.11) 
\]
We can also define $\Lambda $-density matrices as, related by the $\Lambda $%
-trace (6.4.1), as:
\[
\rho _\Lambda (t)=\Lambda ^{\star }\rho (t),...\rho (t)=\Lambda \rho
_\Lambda (t)..(6.4.12) 
\]
where $\rho (t)\in \Phi _{-}^{\times },\rho _\Lambda (t)\in {\cal L}$ that
evolve as:
\[
\rho (t)=U(t)\rho (0),..\rho _\Lambda (t)=W(t)\rho _\Lambda
(0)...............(6.4.13) 
\]
Eq. (5.3.24), translated to the $\rho _\Lambda $ language reads:
\[
\rho _\Lambda (t)=\rho _{\Lambda *}+e^{-\frac 12\gamma _1t}\rho _{\Lambda
1}(t)+e^{-\gamma _1t}\rho _{\Lambda 2}(t),......(6.4.13^{\prime }) 
\]
that can be also obtain from eq. (6.4.13), since the $\rho _\Lambda (t)$
evolve under the action of the operator $e^{-iGt}$ and $G$ has complex
eigenvalues (cf. (6.4.11).

Then the space $\Phi _{-}^{\times }$ of the $\rho ^{\prime }s$ can be
considered as an ideal reversible world of reversible equations, namely the
ideal world of Newton, endowed with unitary evolutions,(David Bohm would say
that this is the space of implicate order [43]), while the space ${\cal L}$
of the $\rho _\Delta $ is the real, physical, irreversible world of
Boltzmann, endowed with non-unitary evolution (just star-unitary, David Bohm
would say that this is the space of explicate order [43]). Between these two
worlds $\Lambda $ establishes a canonical mapping (David Bohm would say a
''metamorphosis'' [43]). Even if the $\rho _\Delta ^{\prime }s$ live in the
ordinary Liouville space they evolve with a non-unitary law (cf. (6.4.13)),
so $\Lambda -$trace achieve the dream of physicist: it creates an ordinary
Hilbert space where the evolutions are non-unitary and irreversible.
Precisely, the essence of the fine-graining formalism was to maintain the
time symmetric primitive equations (with operator evolution $U(t))$ and to
obtain time-asymmetry by choosing a typical time-asymmetric space ${\cal L}%
_Y=\Phi _{-}^{\times }$. $\Lambda $-trace change these roles We get a
time-asymmetric equation (with evolution $W(t)$) in a time-symmetric space $%
{\cal L}$ as in the coarse-graining case. But, of course, the physics remain
the same.

v.- Using the $\Lambda $ Lyapunov variables can be find very easy since:
\[
(\rho (t)|\rho (t))=(\rho (0)|U^{\dagger }U|\rho (0))=(\rho (0)|\rho
(0))=const.,.....(6.4.14) 
\]
therefore it is not a Lyapunov variable, but:
\[
Y=(\rho _\Lambda (t)|\rho _\Lambda (t))=(\rho _\Lambda (t)|W^{\dagger
}W|\rho _\Lambda (0))=var.,............(6.4.15) 
\]
or
\[
Y=(\rho (t)|(\Lambda ^{\star })^{\dagger }\Lambda ^{\star }|\rho (t))=(\rho
(t)|M|\rho (t))=var.,...M=(\Lambda ^{\star })^{\dagger }\Lambda ^{\star
},...(6.4.15^{\prime }) 
\]
Precisely: if
\[
|\rho _\Lambda (0))=\sum_{ij}\rho _{i\text{ }%
j}|ij)...............................(6.4.16) 
\]
the corresponding time evolution is:
\[
|\rho _\Lambda (t))=\sum_{ij}\rho
_{ij}e^{-i(z_i-z_j^{*})}|ij)...................(6.4.17) 
\]
but $Im(z_i-z_j^{\star })=-\Gamma _{ij}\leq 0,$ so:
\[
(\rho _\Lambda (t)|\rho _\Lambda (t))=\sum_{ij}|\rho _{ij}|^2e^{-\Gamma
_{ij}t}...................(6.4.18) 
\]
is always decreasing and it is, therefore, a Lyapunov variable.

In more general cases, than the one of eq. (6.4.1)-(6.4.2), it can be proved
that every rigging corresponds to a $\Lambda $-trace and vice versa [43]

\subsection{Time asymmetry in fine graining theories.}

Let us begin computing the conditional entropy $H_C(\rho |\rho _{*})$ (cf.
eq. (3.6.1,2)) in the case of the classical evolution (4.5.1.12) (in the
quantum case we have the time evolution (5.4.13)). If we want to use the
classical equation for $H_{C\text{ }}$and we have a quantum density matrix
we must first transform this quantum matrix to the corresponding classical
distribution function, using eq. (6.A.1). But we can use directly the
definition of $H_C$ if we define the logarithm of a quantum density matrix
as the operator whose eigenvalues are equal to the logarithms of the
eigenvalues of the primitive operator [3]. As Wigner integral is linear the
quantum analog of eq. (5.4.13) is the same equation). Then:
\[
H_C(\rho |\rho _{*})=-\int_X(\rho _{*}+e^{-\gamma t}\rho _1)\log
(1+e^{-\gamma t}\frac{\rho _1}{\rho _{*}})dx........(6.5.1) 
\]
considering that $|\rho _1|<<\rho _{*}$ or $t>>\gamma ^{-1}$ we can expand
the logarithm and since $tr\rho _1=0,$ taking into account eq. (6.A.3) we
have:
\[
H_C(\rho |\rho _{*})=-e^{-2\gamma t}\int_X\frac{\rho _1^2}{\rho _{*}}%
dx................................(6.5.2) 
\]
which is negative, growing, and, with a vanishing limit when $t\rightarrow
\infty $, so it accomplish all the properties to formulate the second law of
thermodynamics in its third order form. But we have obtained this
satisfactory conclusion because we have work with the operator test function
space $\Phi _{-}$ and the quantum states belong to space $\Phi _{-}^{\times
} $ (albeit some mathematical problems, since we are computing the $\log $
of a vector of a rigged Hilbert space; these problems can be solved, in
principle, if we use the generalized trace $\Lambda ,$ of the previous
section, and if we substitute the $\rho ^{\prime }s$ by $\rho _\Lambda
^{\prime }s,$ since these last density matrices belong to ${\cal L}$ ,so
they can be used with no problem, but they keep the evolution properties of
the $\rho ^{\prime }s,$ namely the dumping factors of eqs. (6.5.1) and
(6.5.2) as in eq. (6.4.13'); namely we define
\[
H_C(\rho |\rho _{\star })=-\int_X\rho _\Lambda \log \frac{\rho _\Lambda }{%
\rho _{\Lambda *}}dx) 
\]
. So we can go now to the central problem of the origin of time asymmetry in
fine graining theories.

Let us now consider an isolated system which is all our universe, there is
nothing we can know about the exterior of the system and the system cannot
interact with something out side the system. If the time evolution equations
of a theory are time symmetric it is quite impossible to brake this symmetry
by rigorous mathematical manipulations, symmetry will always appear, some
way or another. Nevertheless, the examples we gave show that normally in
these theories we can find two extensions of Liouville space ${\cal L.}$%
{\cal .} They are the rigged Hilbert spaces $\Phi _{-}^{\times }$ and $\Phi
_{+}^{\times }$ which are defining using the test spaces $\Phi _{-}$ and $%
\Phi _{+}$(usually these test spaces are nuclear spaces that can be consider
the spaces corresponding to the operators of the measurement devices, as
explained in appendix 4.A). Time symmetry makes that these spaces-are
related by:
\[
{\cal K:}\Phi _{-}^{\times }\rightarrow \Phi _{+}^{\times },.....{\cal K:}%
\Phi _{+}^{\times }\rightarrow \Phi _{-}^{\times
}..........................(6.5.3) 
\]
and, therefore, they are identical. To chose one of the other is an
irrelevant choice. As irrelevant as to trough a dice with the same number in
all its faces. So if we chose one space or the other, physics do not change.
Both spaces are only conventionally different. Any possible difference could
came only for the exterior of the system and there is nothing there to
interact with. Nevertheless in each space $\Phi _{-}^{\times }$ or $\Phi
_{+}^{\times }$ future is substantially different than past, since there is
equilibrium toward only one of these directions and we can call this
direction the future. So let us chose one of the spaces, we then establish a
time asymmetry and we can formulate the second law of thermodynamics, as we
have done, and our problem is solved (compare the solution with the coarse
graining case, it is not so different)

We can say the same talking about the choice of the generalized traces $%
\Lambda _{-}$ or $\Lambda _{+}$ and work within the Liouville space ${\cal L}
$, the space of physical states.

The same trick can be done in various different ways e. g.:

i.-In papers [26],[27],[28], and [32] two semigroups are defined, each
related to a rigged Hilbert space, and one of these semigroups is
arbitrarily chosen. One semigroup is obtained expanding the solution of the
evolution equation in a basis of $\Phi _{-}^{\times }$ and the evolution
turns out to be well defined for $t\in (-\infty ,+\infty ]$, namely it is
not well defined for $t\rightarrow \infty .$ The other semigroup has the
inverse properties.

ii.-In the book [40] a projector $\Pi _{-}=\sum_i|ii,-)(ii,+|$ is defined
and consider as the projector on the really relevant space. But $\Pi _{+}=$$%
{\cal K}\Pi $ is identical to $\Pi _{-}$ , so we must chose one or the other
as in the previous cases.

So in all these cases we must do a conventional choice to find a
mathematical structure: a space, a semigroup, a projector,... such that
using this structure the future exhibit substantially different properties
than the past.

Someone may say that we have not explain time asymmetry, since we have just
introduce it by an arbitrary choice. To answer this criticism we must
remember that physics really never explains. It merely find the mathematical
structure more adequate to foresee the physical phenomena: e. g.: the more
adequate mathematical space, the more adequate mathematical equations, etc.
The curvature of space-time do not explains gravity, it happens that a
riemannian manifold is the best mathematical structure to deal with gravity.
Analogously, it simply turns out that the most adequate mathematical space
to explain time asymmetry and the second law of thermodynamics is a rigged
Hilber space not the usual Hilbert space:{\it \ So the relevant important
choice is between the mathematical structures }${\cal L\ }${\it or $\Phi
_{-}^{\times }$} (or which is the same think $\Phi _{+}^{\times })$ {\it The
choice between these last two rigged spaces, $\Phi _{-}^{\times }$} {\it or $%
\Phi _{+}^{\times }$, is on the contrary, irrelevant and physically
unimportant}

\subsection{Comparison between Fine-Graining and Coarse-Graining.}

As we can see coarse-graining and fine-graining are very similar.

Both are based in theorem 4.3.1 about the weak limit of mixing evolutions.
Coarse-graining obtains a strong limit via a projection, fine-graining
obtains the ''strong limit'' using functionals.

Both obtains their arrow of time defining a pair of time.symmetric
structures. The pair ${\cal L_{-},L_{+}}$, of $t>0,$ and $t<0$ evolutions in
the case of coarse-graining. The pair of rigged Hilbert spaces $\Phi
_{-}^{\times },\Phi _{+}^{\times }$ in the case of fine-graining. In both
methods one of these structures is conventionally chosen.

The main weakness of coarse-graining is that the projector is not defined in
a canonical way.

The main weakness of fine-graining is that we are force to enlarge the space
and we do not know the exact nature of the objects we must add. Are these
ideal unstable states just mathematical usefull tools (like Fadeev-Popov
ghost) or real physical objects? The answer to this question depend on the
point of view that we would take studying the problem. In fact:

i. Any decaying state was always created by a creation process. The quantum
state that corresponds to the creation process followed by the decaying
belongs to ${\cal H}$ (like the vector $|1>$ of the Friedrichs model with
the survival probability (5.3.19), the one of fig. 2). Nevertheless, if the
lifetime of the decaying state is very large, we use to neglect the creation
process and to consider the state just like a decaying state with
exponential decaying survival probability (as in eq. (5.3.20)). This is the
state $|z_1,->$ that belongs to $\phi _{-}^{\times }$. So the quantum
theory, that uses these states, could be considered as an effective theory,
where creation process are neglected. We can say the same for classical
theories. In Bakers transformation a regular density, with a regular
support, will have a creation process and a symmetric decaying process,
towards equilibrium, much in the same way as state $|1>.$ But if we study
the time evolution of a ''horizontal Dirac comb'' states we will find that
these ideal states have no creation process, as the state $|z_1,->.$

ii. Nevertheless, what could be just an usefull simplification, when using
states with large lifetime, can be a rigorous fact in the case of the
universe, where, as we do not know its creation process, this process must
be necessarily neglected.

iii. So the new unstable states added to physical space are similar to
planes waves, they are eternal objects with no creation process and, in
fact, if we define plane waves in a rigorous way we need a rigged Hilbert
space to do it. From this point of view, coarse-graining physicist would be
like stubborn persons that only work with waves packet and refuse to use
plane waves because they ''are not physical objects''.

iv. If we allows time go to infinity and we would like to consider the
rigorous equilibrium state at infinite time, this state belongs to $\Phi
_{-}^{\times }$ as in the case of Baker's transformation, so we are forced
to work with a fine-graining theory. But if we content ourselves with
approximate equilibrium states at finite time, arguing that $t\rightarrow
\infty $ is physically impossible, we do not need these states.

So the real nature of the new states is open to discussion. However a
fine-graining physicist can take a conservative attitude and consider the
new states just as ideal states, namely just as useful mathematical devices
as plane waves are.

Is it the choice of the fine-graining or the coarse-graining just a matter
of taste or there are physical or mathematical reasons to chose one or
another? The reader must decide by himself.

\subsection{Appendix 6.A : Wigner function integral ([46],[47]).}

We have continuously jumped from the classical to the quantum case and back.
Therefore it is interesting to present a theory to formalize these jumps and
to make some applications of it.

Let $\rho $ be a density matrix of Liouville space ${\cal L}={\cal H\times H}
$ and let $\{|q>\}$ be the configuration or position basis of the Hilbert
space ${\cal H}.$ The corresponding Wigner function reads:
\[
\rho _W(q,p)=\pi ^{-1}\int <q+\lambda |\rho |q-\lambda >e^{2i\lambda
p}d\lambda ................(6.A.1) 
\]
It can be proved that:
\[
L\rho _W(q,p)=\pi ^{-1}\int <q+\lambda |L\rho |q-\lambda >e^{2i\lambda
p}d\lambda +O(\hbar )......(6.A.2) 
\]
where $L$ is respectively the classical and quantum Liouville operator. In
the classical limit $\hbar \rightarrow 0$ therefore $\rho _W$ can be
considered as the classical distribution function corresponding to $\rho $.
As in the classical regime we practically works in this limit we will
consider that eq.(6.A.1) is the relation between the quantum density matrix
and the classical distribution function. In fact, even if $\rho _W$ is not
generally positive definite, using the Wigner integral from classical
equation we can pas to quantum equation and vice versa, as a few examples
will show. E.g., let us observe that:
\[
\begin{array}{c}
\parallel \rho _W\parallel =\int \int \rho _W(q,p)dqdp= \\ 
=\int dq\int <q+\lambda |\rho |q-\lambda >\delta (\lambda )d\lambda =tr\rho
\end{array}
.......(6.A.3) 
\]
so to the classical norm corresponds the quantum trace. Also:
\[
\begin{array}{c}
(\rho _W|O_W)=\int \int \rho _W(q,p)O_W(q,p)dqdp= \\ 
\pi ^{-2}\int dq\int \int \int <q+\lambda |\rho |q-\lambda ><q+\mu |O|q-\mu >
\\ 
\times e^{2ip(\lambda +\mu )}dpd\lambda d\mu = \\ 
\pi ^{-1}\int dq\int <q+\lambda |\rho |q-\lambda ><q-\lambda |O|q-\lambda
>d\lambda \cong tr(\rho O)
\end{array}
..(6.A.4) 
\]
Therefore to the inner product in classical Liouville space corresponds the
inner product in the quantum Liouville space. This fact complete the analogy
between classical and quantum spaces implemented by the Wigner integral.

As an exercise we can compute the classical distribution function
corresponding to density matrices $\rho _1(t)$ and $\rho _2(t)$ of
eqs.(5.3.26) and (5.3.27). As these equations will be used in section 8
where we will use eq (6.5.2) to compute the entropy neglecting $O(\lambda )$
we will do so in this exercise:
\[
\begin{array}{c}
\rho _{W1}(t)=\pi ^{-1}\int <q+\lambda |\int_0^\infty (\rho _{1\omega
}e^{-1(\omega _1-\omega )t}|1><\omega | \\ 
+h.c.)d\omega |q-\lambda >e^{2ip\lambda }d\lambda
\end{array}
......(6.A.5) 
\]
where:
\[
<q|\omega >=\frac 1{\sqrt{2\omega }}e^{-i\sqrt{\omega }%
q}.............................................(6.A.6) 
\]
where $2m=1,$ $\hbar =1,$ etc. After an easy calculation we obtain:
\[
\rho _{W1}(q,p,t)\approx \rho _{1,(2p-\sqrt{\omega _1})^2}e^{4ip(\sqrt{%
\omega _1}-p)t}e^{2i(p+\sqrt{\omega _1})q}+h.c......(6.A.7) 
\]
We can see that the main values of this distribution function are obtained
when $p=\sqrt{\omega _1}$ since for other values there are rapid
oscillations. Analogously:
\[
\rho _{W2}(t)=\pi ^{-1}\rho _{11}\int <q+\lambda |1><1|q-\lambda
>e^{2ip\lambda }d\lambda \approx \delta (p-\sqrt{\omega _1})....(6.A.8) 
\]
therefore also in this case all the effect is concentrated around the energy 
$\omega _1$ that will correspond, in the application of section 8, to the
characteristic energy of nuclear reactions.

\section{Entropy in Curved Space Time.}

We have mention cosmology twice:

i.-The cosmological event horizon could be a way to explain a universal
graininess of nature.

ii.-Fine graining time asymmetry is explained using a system with no
exterior, namely the universe.

Furthermore there is a cosmological arrow to investigate, so we cannot avoid
cosmology in a complete discussion of our subject.

Many years ago Mach thought us that most of the basic physical facts can
only be explained only if we consider the universe as a whole, e. g.: if we
want to explain why a system is a inertial one or not we must consider the
whole universe, the system will be inertial if it is in uniform translatory
motion with respect to the matter of the whole universe. The arrows of time
are not exceptions, since they have a global nature. In fact, lumps are
solved by the coffee in all places in the same time direction, here and in
the Andromeda nebula. We must explain why it is so, and we will find the
explanation only if we define the arrow of time in global cosmological
models. Thus let us begin studying the notion of entropy in curved
space-time because cosmological model are presented in this kind of spaces,
[2].

\subsection{Thermodynamics in special relativity.}

For phenomenological reasons we can assume that the laws of thermodynamics
are valid in the special relativity proper system of coordinates $S^0$. From
the relativity principle we then know that this laws are also valid in every
inertial system $S$ in translatory uniform motion with respect to $S^0$,
provided the quantities involved in these laws would be transformed in a
convenient way. In other word we would like to obtain the ''Lorentz
transformation'' that makes invariant the following laws

i.-The first law::
\[
\Delta E=\Delta Q-\Delta W.........................(7.1.1) 
\]
where $E$ is the energy, $Q$ the heath and $W$ the work.

ii.- The second law:
\[
\Delta S\geq \frac{\Delta Q}T.....................................(7.1.2) 
\]
where $S$ is the entropy and $T$ the temperature and the equality holds only
for reversible evolution. To do this let us suppose that:

i.- The pressure is isotropic namely it is equal in all directions and

ii.- Let us temporarily use for simplicity axes chosen in such a way that
the velocity $u$ of the system $S$ with respect to the system $S^0$ is
parallel to the $x$-axis.

Then from ordinary special relativity we know the coordinate transformation
equations for the following mechanical quantities.:

i.-For the volume $v$:
\[
v=v_0\sqrt{1-u^2}......................(7.1.3) 
\]

ii.-For the pressure $p$:
\[
p=p_0................................(7.1.4) 
\]

iii.-For the energy $E$:
\[
E=\frac{E_{0+}p_0v_0}{\sqrt{1-u^2}}.....................(7.1.5) 
\]

iv.-For the work $W:$%
\[
dW=\sqrt{1-u^2}dW_0+\frac{u^2}{\sqrt{1-u^2}}d(E_0+p_0v_0).......(7.1.6) 
\]
where the quantities with subscript ''0'' refers to system $S^0.$

Then for the covariance of the first law eq. (7.1.1) it is necessary and
sufficient that:
\[
Q=\sqrt{1-u^2}Q_{0......................}(7.1.7) 
\]
so we have obtained the transformation law of the heath.

Let us now consider a thermic system at $S^0$. We can accelerate this
thermic system up to the velocity $u$ in a reversible and adiabatic way, so
the entropy of the system is not modified and we obtain:
\[
S=S_0..............................(7.1.8) 
\]

Finally, from eqs. (7.1.7,8) it is evident that the second law, eq. (7.1.2)
will be covariant iif:
\[
T=\sqrt{1-u^2}T_0.................(7.1.9) 
\]
so we have obtained the change of coordinate equation of all the basic
thermodynamical quantities.

Let us now find the corresponding equations in four-dimensional language.

The first law is just a form of the conservation of energy, therefore its
four-dimensional version will be:
\[
\partial _\mu T^{\mu \nu }=0.....................(7.1.10) 
\]
where $T^{\mu \nu }$ is a convenient energy momentum tensor-($\mu ,\nu
,...=0,1,2,3)$.

To deduce the four-dimensional form of the second law let us consider a
small volume of a thermodynamical fluid $v$ and let us call $\phi $ the
entropy density at the point where this element of volume is located in such
a way that $\phi v$ is the entropy of the element. If $\delta t$ is a small
period of time the second law reads:
\[
\frac d{dt}(\phi v)\delta t\geq \frac{\delta Q}T.............(7.1.11) 
\]
or ($i,j...=1,2,3)$:
\[
\left( \frac{d\phi }{dt}v+\phi \frac{dv}{dt}\right) \delta t=\left(
u_i\partial _i\phi +\frac{\partial \phi }{\partial t}+\phi \partial
_iu_i\right) v\delta t\geq \frac{\delta Q}T.........(7.1.12) 
\]
where $u^i=\frac{dx^i}{dt}.$ Combining terms we have:
\[
\left( \partial _i(\phi \frac{dx_i}{dt})+\frac{\partial \phi }{\partial t}%
\right) v\delta t\geq \frac{\delta Q}T......(7.1.13) 
\]
but:
\[
\frac{ds}{dt}=\sqrt{1-u^2}........(7.1.14) 
\]
and from eqs. (7.1.3) and (7.1.8):
\[
\phi =\frac{\phi _0}{\sqrt{1-u^2}}..........(7.1.15) 
\]
\[
\frac{\delta Q}T=\frac{\delta Q_0}{T_0}.............(7.1.16) 
\]

Thus we obtain:
\[
\partial _\mu \left( \phi _0\frac{dx^\mu }{ds}\right) \delta v\geq \frac{%
\delta Q_0}{T_0}.....(7.1.17) 
\]
where $\delta v=v\delta t$ is the coordinate four-dimensional volume
element, which is equal to $\delta v_0=\frac v{\sqrt{1-u^2}}\sqrt{1-u^2}%
\delta t=\delta v$ (cf. eqs. (7.1.3), (7.1.14)), the proper four dimensional
volume element, so we can use either one or the other. Thus if we define the
flow of proper entropy or ''entropy vector'' as:
\[
S^\mu =\phi _0\frac{dx^\mu }{ds}..................(7.1.18) 
\]
we obtain the four-dimensional version of the second law:
\[
\partial _\mu S^\mu \delta v_0\geq \frac{\delta Q_0}{T_0}%
..............(7.1.19) 
\]
which is valid for all inertial systems and we can put $\delta v$ instead of 
$\delta v_{0.}$.

\subsection{Thermodynamics in general relativity.}

Using the transcription rules to go from special relativity to general
relativity, namely:
\[
\eta _{\mu \nu }\rightarrow g_{\mu \nu },.....\partial _\mu \rightarrow
\nabla _{\mu ,........}\delta v\rightarrow \sqrt{-g}\delta v,......(7.2.1) 
\]
the first law, eq. (7.1.3) reads:
\[
\nabla _\mu T^{\mu \nu }=0.....................(7.2.2) 
\]
or introducing the tensor density ${\sf T}^{\mu \nu }=\sqrt{-g}T^{\mu \nu }$
we have the ordinary divergence:
\[
\partial _\mu ({\sf T}^{\mu \nu }+t^{\mu \nu })=0...........(7.2.3) 
\]
where $t^{\mu \nu }$ is the pseudo tensor density of potential
energy-momentum. This would be the general relativity covariant equation
that shows the closest resemblance to eq. (7.1.3).

Using the transcription rules on the second law, eq. (7.1.19) we obtain:
\[
\nabla _\mu S^\mu \sqrt{-g}\delta v\geq \frac{\delta Q_0}{T_0}%
........(7.2.4) 
\]
where, being all the factors scalars, we have, in fact, obtained a equation
which is valid for all coordinate systems. Introducing the density ${\sf S}%
^\mu =\sqrt{-g}S^\mu $, since $\nabla _\mu S^\mu =\frac 1{\sqrt{-g}}\partial
_\mu \sqrt{-g}S^\mu =\frac 1{\sqrt{-g}}\partial {\sf _\mu S}^\mu $, this
last equation reads:
\[
\partial _\mu {\sf S}^\mu \delta v\geq \frac{\delta Q_0}{T_0}%
..............(7.2.5) 
\]
which,again, is the general relativity covariant equation that shows the
closet resemblance with the special relativity second law (7.1.19).

Of course these are not the unique covariant generalization of the
thermodynamical laws of general relativity but they are the simplest and
they lead to successful applications.

\subsection{Thermodynamics in cosmology.}

Let us consider a Robertson-Walker metric:
\[
ds^2=dt^2+a^2d\sigma ^2........(7.3.1) 
\]
where $d\sigma $ is the comoving arc length and $a$ the scale factor or the
radius of the universe.

If the energy momentum tensor corresponds to a isotropic fluid with density $%
\rho _{00}$ and pressure $p_0$ the first law reads:
\[
\frac d{dt}(\rho _{00}a^3\delta \sigma )+p_0\frac d{dt}(a^3\delta \sigma
)=0.......(7.3.2) 
\]
where $\delta \sigma $ is a comoving-coordinate three-dimensional volume.

If we consider a comoving thermic fluid there will be not exchange of heath
among the comoving volumes, and $u^\mu =\frac{dx^\mu }{ds}=(1,0,0,0),$ so
the second law, as expressed by eq. (7.2.5) reads:
\[
\partial _\mu (\phi _0u^\mu \sqrt{-g})\geq 0=\frac d{dt}(\phi _0a^3)\geq
0....(7.3.3) 
\]
where $\phi _0$ is the proper entropy density and $a$ the scale factor or
radius of the universe. If we multiplies this equation by the constant
coordinate comoving volume $\delta \sigma $ we obtain:
\[
\frac d{dt}(\phi _0a^3\delta \sigma )\geq
0.............................(7.3.4) 
\]
This equation gives the recipe to compute the entropy in that comoving frame
of a Robertson-Walker metric:{\it \ multiplied the local proper entropy
density by the proper volume.}, which is, of course a very reasonable and
natural result, that perhaps it is so natural it that can simply be assumed
from the beginning, but now it is rigorously proved.

Let us check this result with just one calculation: We now that in a
radiation dominated universe temperature follows the law:
\[
T=T_0\frac{a_0}a.....................................(7.3.5) 
\]
that can be obtained integrating eq. (7.3.2) if we take $p_0=\frac 13\rho
_{00}\sim T^4$, namely the radiation state equation, and that entropy of a
black-body radiation in given by the formula:
\[
S=\frac 43C_ST^3V...............................(7.3.6) 
\]
where $C_S$ is the Stefan coefficient, $T$ the temperature and $V$ the
volume. If we substitute this two last equation into eq.(7.3.4) we see that
the evolution of a radiation dominated universe is reversible, as can be
expected.

From these consideration we can deduce that the only effect produced by the
expansion of the universe, in isotropic models, is the temperature
decreasing. This is the only effect we must take into account below.

\section{ The Cosmological Problem.}

\subsection{The problem of the time asymmetry.}

The problem of the existence of the arrows of time or, what is the same
think, the time asymmetry of the universe, can be stated, as we explained in
the introduction, in the following question:

{\it i.-How can it be that there is time asymmetry in the universe if all
the relevant physical laws are time symmetric?}

In fact, universe has several time asymmetries, namely the various arrows of
time: the thermodynamical one, the electromagnetical one, the psicological
one, etc., while its main laws are time-symmetric (because, as usual in this
kind of discussions, we will neglect the time-asymmetric laws of weak
interactions [1], since it is very difficult to imagine a reason that
explains macroscopic time asymmetry based only in the asymmetry of these
laws)

A second question is to explain the fact that all the arrows of time point
in the same direction.

In this section we would like to answer these questions, giving an adequate
mathematical framework to the problem and using several, old and new, well
known ideas ([2],[3],[4])

Let us first review the main equation of section 2. If the state of a
physical system is described by $\rho $ (being $\rho $ classically the
distribution function or quantum mechanically the density matrix) we will
call $\rho ^{rev}=$${\cal K}\rho $ the state with reversed initial
conditions (e.g.: if $K$ is the Wigner operator of quantum mechanics then: $%
{\cal K}\rho =K\rho K^{\dagger }.$,[17],[15],[48]). We will say that the
conditions at $t=0$ are time symmetric if $\rho ^{rev}(0)={\cal K}\rho
(0)=\rho (0)$ and time asymmetric otherwise, If $\rho (t)$ is the state of
the universe at time $t$ , the universe would have a time symmetric
evolution with respect to $t=0$ if (cf. eq. (2.2.21)):
\[
{\cal K}\rho (t)=\rho (-t).......................................(8.1.1) 
\]
But the universe has, in fact, a time-asymmetric evolution, at least with
respect with some instant of time, that we call $t=0,$ such that:
\[
{\cal K}\rho (t)\neq \rho (-t).......................................(8.1.2) 
\]
If the evolution equations, embodied in the universe liouvillian operator $L$%
, are time-symmetric, namely (cf. eq. (2.2.11)):
\[
{\cal K}L{\cal K}^{\dagger
}=L............................................(8.1.3) 
\]
to time symmetric conditions at $t=0$ will corresponds a time symmetric
evolution (8.1.1) and to time asymmetric conditions will correspond time
asymmetric evolutions like (8.1.2). In fact,:
\[
\rho (t)=e^{-iLt}\rho (0)......................................(8.1.4) 
\]
therefore if the $t=0$ condition is time-symmetric we have:
\[
{\cal K}\rho (t)=e^{i{\cal K}L{\cal K}^{\dagger }t}{\cal K}\rho (0)=e^{iLt}%
{\cal K}\rho (0)=\rho (-t)...........(8.1.5) 
\]
since ${\cal K\ }$is an antilinear operator (namely ${\cal K}i=-i).$ In the
same way the time asymmetric case can be demonstrated. Then the observed
time asymmetry of the universe evolution which obey eq, (8.1.2) can be
explained only in two alternative ways:

i.-Really eq (8.1.3) is not exact and there is a small, but relevant,
time-asymmetric term in the liouvillian (e.g. cause perhaps by the weak
interactions) or:

ii.-
\[
{\cal K}\rho (0)\neq \rho
(0)..........................................(8.1.6) 
\]
namely the initial state of the universe is not time symmetric.

So, if we reject weak interactions, or any clever manipulation of the,
otherwise time-symmetric physical laws, as the origin of time-asymmetry, we
must necessarily consider eq. (8.1.6) as the only possible cause of this
phenomenon. As,-in principle, asymmetry is a more generic property than
symmetry (as complex numbers are more frequent than real ones) eq, (8.1.6)
seems very natural and, therefore, this will be the idea that we will adopt
in this section. If eq. (8.1.6) is valid, from eq. (8.1.5) we have:
\[
{\cal K}\rho (t)\neq e^{iLt}\rho (0)=\rho
(-t)........................(8.1.7) 
\]
i.e. eq. (8.1.2), the equation we must prove.

Finally, let us remark that the same explanation can be used to explain the
other two fundamental asymmetries of nature ${\cal P\ }$and ${\cal C}${\cal .%
}In fact, if:
\[
{\cal P}\rho (0)\neq \rho (0),.........{\cal C}\rho (0)\neq \rho
...............(8.1.8) 
\]
we will have:
\[
{\cal P}\rho (t)\neq \rho (t),........{\cal C}\rho (t)\neq \rho
(t)...............(8.1.9) 
\]
even if:
\[
{\cal P}L{\cal P}^{\dagger }=L,............{\cal C}L{\cal C}^{\dagger
}=L...............(8.1.10) 
\]
e.g.: Eq. (1.9,2) can be demonstrated if we postulate the existence of a
small fluctuation between the amounts of matter and antimatter at the
beginning of the universe.

\subsection{Entropy, Fluctuations, and Irreversibility.}

Let us first study the thermodynamical arrow of time. So, let us consider
the entropy $S$ as the state function that represent more eloquently the
thermodynamical state of the universe ($S$ can be computed using
coarse-graining entropy, or fine-graining entropy). We know that the vast
proportion of possible states of the universe will be near the equilibrium
state $\rho _{*}$ and will have the equilibrium entropy $S_{*}$.
Nevertheless we know that fluctuations around the equilibrium state, namely
less probable unstable states near the equilibrium, will spontaneously
appear and we also know that, in these fluctuations states, entropy will be
smaller than $S_{*}$.

Anyhow steady equilibrium state satisfy Liouville equation:
\[
L\rho _{*}=0..............................................(8.2.1) 
\]
For simplicity let us consider that there is just one equilibrium state in
the universe, as it is very likely since the universe looks chaotic
and,therefore,it is at least ergodic, therefore from eq. (8.1.3) we have:
\[
L{\cal K}\rho _{*}={\cal K}L{\cal K}^{\dagger }{\cal K}\rho
_{*}=0........................(8.2.2) 
\]
Therefore:
\[
{\cal K}\rho _{*}=\rho
_{*}..............................................(8.2.3) 
\]
Thus if we have $\rho (0)=\rho _{*}$ we will have a time-symmetric evolution
and no thermodynamical arrow of time (in fact, the universe will always
remain in state $\rho _{*}$). But for an unstable
non-equilibrium-fluctuation state $\rho $ ,in general, we will have- that $%
{\cal K}\rho \neq \rho $. Therefore it is enough to assume that the universe
began (al $t=0$) in one of these states and we will have a time-asymmetric
evolution and a thermodynamical arrow of time, because the initial entropy
is $S<S_{\star }$, and therefore there will be growing of entropy, both to
the past and to the future of $t=0$, since entropy will try to reach the
equilibrium entropy in both directions. (in the exceptional case that the
initial non-equilibrium unstable state would be such that ${\cal K}\rho
(0)=\rho (0)$ at $t=0$ it would be ${\cal K}\rho \neq \rho $ at a different
time, close to $t=0$, that can, as well, be taken as the origin of time in
eq, (8.1.7)). Then it is enough to suppose that the universe began in a
non-equilibrium unstable state to obtain the thermodynamical arrow of time
and the second law of thermodynamics, if we conventionally-consider only
times $t\geq 0$ (and conventionally call to this period the ''future'' of $%
t=0$).

This low entropy initial state of the universe could be consider as a
fluctuation. In fact, irregular fluctuation of the equilibrium entropy are
present in systems with a finite number of particles [3], but vanish if this
number goes to infinity. Then, these fluctuations cannot be consider if we
work with a distribution $\rho $ in Liouville space, as we have done in
these lectures, because these distributions are probabilities computed
assuming an infinite number of particles (or an infinite number of copies of
the system). Fluctuations can be introduced in several way, e. g.:

i.- Using Boltzmann entropy as in [7].

ii.- Working in a rigged space where distribution, corresponding to a finite
number of particles, namely $\rho ^{\prime }s,$ built using a finite number
of Dirac's deltas, can be consider, etc.

We will not discuss this subject further here.

About this solution, to the problem of the initial low entropy state of the
universe, it can be argue that this initial fluctuation is very unlikely
[8], since the universe is very big, perhaps even infinite. Nevertheless, we
shall prove, in the next section, that this conjecture is unnecessary, since
the initial instability is naturally produced by the universe expansion, so
really no fluctuations are needed, that is why we do not discuss
fluctuations in these lectures.

For isolated subsystems within the universe time asymmetry can be obtained
in a similar way. In fact, we use to imagine that these subsystem (Gibbs ink
drop spreading in a glass of water, or the perfume spreading into the room,
etc.,etc.) began in an unstable initial state with low entropy (a
concentrated ink drop, all the perfume inside a bottle, etc., etc.). But
these initial states are always produced, not by unlikely fluctuations but
by external agencies (the ink or the perfume factories), that use energy to
produce these concentrations that they obtain from other subsystems in
unstable initial states (chemical-unstable coal or nuclear-unstable
isotopes, etc. etc.) that, in turn, obtain their energy, via a chain of
unstable states (like those of the stars), from the universe unstable
initial state. Therefore,we conclude that all time-asymmetric processes have
a cosmological origin. The only difference is that in the case of a
subsystem we have a reason to consider only times $t\geq t_0$, being the
time $t_0=0$ the time of creation of the initial unstable state of the
subsystem, since time $t<t_0$ corresponds to a period before the creation of
the unstable states by the external agency (the concentration period of the
ink drop or the bottle of perfume), where the subsystem is not isolated. The
subsequent diffusion of the ink drop, the perfume, etc. will produce the
growth of the thermodynamical entropy.

The quantum analog of these reasonings can be found in papers [15] [49],
[50] and [39].

The creation of a low entropy state is, therefore produce either by a
initial fluctuation, in the case of the universe (but we will see that the
initial fluctuation is not necessary in the next subsection), or by an
external agency, in the case of subsystems within the universe. Thus,
neglecting for a moment, the fluctuations, we will call a ''conspiracy'' to
the appearance of a low entropy state not produced by an external agency.
Then we can conclude that conspiracies do not exists in nature.In fact, let
us consider a system in a low energy unstable state produced by external
agencies (e.g. a glass store and an elephant). Any process, within the
system, will produced a growth of entropy (e.g. the elephant moving by the
store and braking all the glasses). This is an irreversible process. In
fact, its time reverse process (a film of the motion of the elephant played
backward) is full of conspiracies and therefore do not exists in nature.
Irreversibility, therefore, can also be explained in this way in our
formalism.

\subsection{The problem of the coordination of the arrows of time.}

Now we must solve the second problem that can be stated in the following
question:

{\it ii.-Why all the arrows of time point in the same direction?}

Also we would like to show that the initial fluctuation is not strictly
necessary.

To solve these problems we will consider that the cosmological arrow of
time, namely the growth of the radius or scale factor of the universe $a$,
is the master arrow of time, that defines the direction of all the others.
First we will show that the thermodynamical arrow of time, namely the
tendency to obtain a final equilibrium, points in the same direction than
the master arrow.

Let $S_{*}$ be the equilibrium entropy and $S(t)$ the actual entropy of the
matter and radiation within the universe at time $t$ (therefore now we will
work with an open system since we exclude the entropy of the gravitational
field). The entropy gap:
\[
\Delta S=S_{*}-S(t)..........................................(8.3.1) 
\]
would be minus the conditional entropy $-H_c(\rho |\rho _{*})$ .according to
eq. (3.6.2), in full agreement with general relativity, if we take into
account the change of the universe temperature, as explained in section 7,
namely:

\[
\Delta S=\int_X\rho (x)\log \frac{\rho (x)}{\rho _{*}(x)}%
dx................(8.3.2) 
\]
where $\rho (t,x)$ and $\rho _{\star }(x)$ are the corresponding local
distribution functions, $X$ the phase space $x\in X$ a point of this space.
The distribution functions are normalized as:
\[
\int_X\rho dx=1,.....\int_X\rho _{\star }dx=1,...............(8.3.3) 
\]
Now we can consider that as in eq, (5.3.24):
\[
\rho (t)=\rho _{*}+(\rho _1+\rho _2e^{-\frac \gamma 2t})e^{-\frac \gamma
2t}=\rho _{*}+\rho _\Delta e^{-\frac \gamma 2t}.....(8.3.4) 
\]
where the second term of the r.h.s. is some kind of correction around the
equilibrium term, with a dumping factor with a characteristic time $\approx $%
$\gamma ^{-1}.$ We will only consider the universe evolution after
decoupling time, the universe will be matter dominated and $\gamma
^{-1}=t_{NR}$, will be the characteristic time of nuclear reactions, that
make the matter within the star evolve toward thermal equilibrium with the
cosmic microwave background. Eq. (8.3.4) can be considered only as a
phenomenological equation, that can be obtained if we use coarse-graining
technics and we neglect the Zeno and Khalfin effects; but we know there is a
rigorous way to eliminate these effects, if we use the rigged Hilbert space
formalism (fine-graining technics) as in eq. (5.3.24).

$\rho _\Delta =\rho _1+\rho _2e^{-\frac \gamma 2t}$ is normalized as:
\[
e^{-\frac \gamma 2t}\int_X\rho _\Delta dx=\int_X\rho dx-\int_X\rho
_{*}dx=0....(8.3.5) 
\]
This normalization is also a consequence of eq. (5.3.14').

We will consider that $|\rho _\Delta |\ll \rho _{\star }$, or $t\gg \gamma
^{-1}$, namely that the fluctuation is small compared with the equilibrium
distribution function. Then the entropy gap $\Delta S,$ expanding the
logarithm and neglecting unimportant terms, as in eq. (6.5.2),reads:

\[
\Delta S\approx e^{-\gamma t}\int_X\frac{\rho _\Delta ^2}{\rho _{*}}%
dx>0.............................(8.3.6) 
\]
Thus, when $\gamma =0$ the growing of entropy variation disappear. To
compute the time derivative of $\Delta S$ let us use the model of eqs.
(5.3.31) and. (5.4.17), described at the end of subsection 5.3, then the
last equation reads:
\[
\Delta S\approx e^{-\gamma t}\int_X\frac{T^{\frac 32}}Ze^{\frac \omega
T}\rho _\Delta ^2dx.........(8.3.7) 
\]
where we have explicitated the time variation in the first exponential
function and in $T(t)$ being the rest of the quantities time constant, since
we can neglect the second time variable term of $\rho _\Delta $ with respect
to the first constant one (or we could keep both terms with an small
modification of the formulae). $\rho _1$ and $\rho _2$ are independent of
the temperature because they are related with the nuclear reaction processes
only. From paper [31] (or eqs. (6.A.7) and (6.A.8)) we can introduce a
reasonable simplification and suppose that the only important values of the
last integral are those around $\omega _1$, the characteristic energy of the
nuclear processes,then:
\[
\Delta S=Ce^{-\gamma t}T^{\frac 32}e^{\frac{\omega _1}%
T}..............................(8.3.7) 
\]
where $C$ is a time independent constant. The temperature evolution will be
dominated by the radiation within the universe and,therefore,will follow eq.
(7.3.5) so:
\[
\Delta S=C^{\prime }e^{-\gamma t}a^{-\frac 32}e^{\frac{\omega a}{T_0a_0}%
}........................(8.3.8) 
\]
where $C^{\prime }$ is another time independent constant. Now we can compute
the time derivative that reads:
\[
\stackrel{\bullet }{\Delta S(t)}=(-\gamma -\frac{3\stackrel{\bullet }{a}}{2a}%
+\frac{\omega \stackrel{\bullet }{_1a}}{T_0a_0})\Delta S...........(8.3.9) 
\]
where $\frac{\stackrel{\bullet }{a}}a=H(t)\approx t_U^{-1}$ is the Hubble
coefficient. Since we are in the matter dominated period we have:
\[
a=a_0\left( \frac t{t_0}\right) ^{\frac
23}..................................(8.3.10) 
\]
thus:
\[
\Delta \stackrel{\bullet }{S}=(-\gamma -t^{-1}+\frac{2\omega _1}{3T_0t_0}%
\left( \frac{t_0}t\right) ^{\frac 13})\Delta S..(8.3.11) 
\]
Eq. (8.3.8) shows two antagonic. effects (fig 5). The universe gravitational
field, embodied in the positive coefficient (and in the term $t^{-1}$), is
the external agency that mostly try to take the system away from
equilibrium, while, on the other hand, the nuclear reaction, embodied in $%
\gamma $ try to convey the system toward equilibrium (but the gravitational
term $t^{-1}$ try to establish equilibrium). This to effects are equal at a
critical times $t_{cr}$such that:
\[
\gamma t_0+\left( \frac{t_0}{t_{cr}}\right) =\frac{2\omega _1}{3T_0}\left( 
\frac{t_0}{t_{cr}}\right) ^{\frac 13}............(8.3.12) 
\]
Usually this equation will have two positive roots $t_{cr1}<t_{_{cr2.}}$
(fig. 6)

It is premature to give physical numerical values to the parameters of the
model. In fact, this model is extremely simplified, since it is based in an
homogeneous space geometry while the decaying processes are produced within
the stars, so what we really need is an inhomogeneous geometry to properly
describe the phenomenon. However, with reasonable numerical values
(essentially taking $\omega _1>>T_{0,}$ $\gamma ^{-1}\approx t_0)$ we can
obtain the following conclusions;

a.-The first root is in the region $t<<t_0$ so the first term of the l.h.s.
of the last equation can be neglected to obtain $t_{cr1}=t_0\left( \frac{%
3T_o }{2\omega _1}\right) ^{\frac 32}.$ (This quantity, with minus sign,
gives the third negative root) At this time the entropy gap has a minimum.

b.-The second root is in the region $t>>t_0$ so the second term of the
l.h.s. can be neglected to obtain $t_{cr2}=\left( \frac{2\omega _1t_{NR}}{%
3T_0t_0}\right) ^3t_0$. At this time the entropy gap has a maximum.

Then we can state the following conclusions:

i.-If $t<t_{cr1}$ then the second term of the l.h.s.of eq. (8.3.12)
dominates $\Delta \stackrel{\bullet }{S}<0,$ and there is a big value for
the entropy gap that is rapidly thermalized..

ii.-If: $t_{cr1<}t<t_{cr2}$, then $\stackrel{\bullet }{\Delta S}>0$ , the
r.h.s. of eq. (8.3.12) dominates and there will be a growing of the entropy
gap, produced by the universe expansion, that is driving the universe away
from equilibrium. There will be a growing of complexity in this period, them
particles, atoms, molecules, galaxies stars, planets, and living beings
appear..

iii.-On the contrary, if $t>t_{cr2}$ then $\stackrel{\bullet }{\Delta S}<0$
, the first term of the l.h.s. of eq. (8.3.12) dominates ,the entropy gap
will diminish and the universe goes toward its final equilibrium state,
driven by the nuclear reaction processes, in agreement with paper [4]. All
the structures within the universe decay and disappears.

Therefore:
\[
\stackunder{t\rightarrow \infty }{\lim \Delta S}%
=0.........................................(8.3.13) 
\]
Numerical estimations show that $t_{cr1}<<t_0<<t_{cr2}$, in such a way that
the first period can be, some how, neglected since probably this period take
place before decoupling time. (Also $t_{cr2}\gg t_0$ as in paper [51])

iv.-Eq, (8.3.8) shows how the universe expansion creates, in a continuous
way, the universe instability and complexity. This fact make the initial
fluctuation hypothesis unnecessary. This instability is created toward the
future, defined as the direction of the universe expansion. Eq. (8.3.8)
shows also how the local nuclear reaction try to restore equilibrium{\it ,
in the same time direction. }The thermodynamical arrow of time is the local
tendency to thermodynamical equilibrium (and not the total entropy
growth,since our system is not isolated because the entropy of the
gravitational field was not considered). Therefore the thermodynamical arrow
coincide with the cosmological arrow.

v.- All this reasonings are also valid before recombination time, where we
must use a much bigger $\gamma ,$ because in that period we must consider
reaction with much smaller characteristic time, in fact, smaller than
recombination time. Since the period $t<t_{cr1}$ probably lays in this
period perhaps the universe reach also a thermodynamical equilibrium, and we
can use the arguments of reference [52] to show that the electromagnetical
arrow of time coincide with the cosmological one. Also the dumping factor $%
e^{-\gamma t}$ can be obtained if we consider a pole in the lower half-plane
of the unphysical sheet of the energy complex plane; thus we must use the
upper rim of the positive real axis cut and, therefore, retarded solutions
[35].

vi.- Finally ourselves are just subsystem with unstable initial state,
produced by external agencies, like the ink drop or the bottle of perfume,
therefore our thermodynamical arrow, that can be identify with our
psicological arrow, points also as the cosmological arrow. So all the arrows
of time points in the same direction.

vii.- Therefore we have given a mathematical formalism to the answers of the
to main questions about the universe time asymmetry. We believe that the
presented solution is quite satisfactory, only much more physical examples
must be studied with the fine-graining method and some mathematical
refinement are missing (like those of paper [48]). When these examples would
be studied and this refinements will be implemented we will have a
definitive and rigorous answer to these, long standing, fundamental
questions.

\section{Conclusions.}

After all these explanation and discussions we believe that we can drawn the
following conclusions:

i.-There are not compelling local-physical motivation to choose one technic
or the other. Therefore it is not easy to see how to find a local
cross-experiment to settle the matter. Probably this cross-experiment not
even exists, so really both technics are physically equivalent.

ii.-Coarse-graining is more ''physical'', since it works directly in usual
Hilbert space. The price to pay is the introduction of an object, which is
really alien to the theory, the projector. This projector is essentially
arbitrary, so coarse grainig will not have a deep physical meaning until a
natural graininess will be not find.

iii.-Fine-graining is more ''mathematic'', since it works in Rigged Hilbert
Space. But after paying this price, we are not force to introduce any object
alien to the theory. In this sense fine-grainig is more pure, and really it
cannot be distinguish from no-graining. Therefore it seems that
fine-graining is conceptually superior even if, from the operational point
of view, coarse-graining could be more convenient. Anyhow fine-graining has
also its ambiguities, e.g.: the choice of the test function space, even if
it seems more probable that we would find a canonical choice of this space,
in the future, than a canonical choice of the coarse-graining projector.

iv. For conceptually difficult chapters of physic, like cosmology or quantum
measurement theory, it is advisable to use fine-graining, since it is
conceptually superior to coarse-graining. Thus, perhaps a global
cross-experiment that shows the convenience of use one technic or the other
could be find using cosmological reasonings.

\section{Bibliography.}

[1]R.G. Sachs, The Physics of time reversal, Univ. Chicago Press, Chicago,
1987.

[2]R.C. Tolman, Relativity, Thermodynamics, and Cosmology, Dover Pub., New
York, 1987.

[3]L.D. Landau and E.M.Lifshtz, Statistical Physics, Pergamon Press, Oxford,
1958

[4]P.C. Davies, Stirring up trouble. Adelaide Univ., Preprint, 1994

[5]M.C. Mackey, Rev. Mod. Phys. {\bf 61}, 981, 1989.

[6]A. Lasota and M.C. Mackey, Probabilistic properties of deterministic
systems, Cambridge University, Cambridge, 1985.

[7]J.L. Lebowitz, Time's arrow and Boltzmann's entropy, Rutgers Univ.,
Preprint, 1994.

[9]I. Prigogine, From being to becoming: time and complexity in physical
sciences , Freeman, San Francisco, 1980.

[10]L.E. Ballentine, Quantum Mechanics, Prentice Hall, New Jersey, 1990.

[11]I. Gel'fand and G. Shilov, Generalized Functions, Academic Press,New
York, 1968.

[12]B.L. Hu, J.P. Paz, and Y. Zhang, Phys. Rev. D 4{\bf 5}, 2843, 1992.

[13]R.W. Zwanzig, Statistical Mechanics of Irreversibility, in Lectures in
theoretical physics III, eds. W.E. Britten et al., Interscience, New York,
1961.

[14]R.W. Zwanzig, I. Chem. Phys. {\bf 33},1388, 1960.

[15]M. Castagnino, F. Gaioli, and E. Gunzig, Cosmological features of time
asymmetry, submitted to Foundations of Cosmic Physics, 1995.

[16]M. Castagnino, E. Gunzig, and F. Lombardo, Gen. Rel. Grav., 1995.

[17]A. Messiah, Quantum mechanics, North-Holland Pub., Amsterdam, 1962.

[18]P. Roman, Advanced quantum theory, Addison Wesley, New York, 1965.

[19]I. Prigogine, C. George, F. Henin, and L. Rosenfeld, Chem. Scripta {\bf 4%
}, 5, 1980.

[20]J. Voigt, Commun. Math. Phys. {\bf 81}, 31, 1981.

[21]G. Tabor, Chaos and integrability in non-linear dynamics, J. Wiley \&
sons, New York, 1980.

[22]V.I. Arnold and A. Avez, Ergodic problems of classical mechanics,
Benjamin Inc., New York, 1968.

[23]P. Walter, An introduction to ergodic theory, Graduate texts in
mathematics, Vol.79, Springer Verlag, New York,1982.

[24]D.V. Anosov, Sov. Math. Dokl. {\bf 4,} 1153, 1963.

[25]P. Schild, ''The theory of Bernoulli shift'' Univ. Chicago Press,
Chicago,1979

[26]I. Antoniu and S. Tasaki, Physica A {\bf 190}, 303, 1991.

[27]I. Antoniu and S. Tasaki, Int. J. Quantum Chem. {\bf 46}, 427, 1993.

[28]I. Antoniu and S. Tasaki, U.L.B. Preprint 1993.

[29]A. Bohm, M. Gadella, Dirac Kets, Gamow Vectors, and Gel'fand Triplets,
Springer-Verlag, Berlin, 1989.

[30]P. R. Halmos, ''Lectures on ergodic theory'', Publ. Math. Soc. of Japan,
1956, and Chelsea Pub. Co., New York, 1956.

[31]R. Laura, Unified description of equilibrium and non-equilibrium
systems. The Friedrichs model, Preprint IFIR 1995.

[32]I. Antoniou, R. Laura,S. Tasaki, and N. Suchanecki, U.L.B. Preprint 1995.

[33]B.L. Hu, J.P. Paz, and Y. Zhang, Phys. Rev. D {\bf 47}, 1776, 1993.

[34]A. Caldeira and A. Leggett, Phys. Rev. {\bf 31}, 1059, 1995.

[35]M. Gadella and G. Rudin, U.L.B. Preprint 1995.

[36]E.C.G. Sudarsham, C.B. Chiu, and V. Gorini, Phys. Rev. D {\bf 18,} 2914,
1978.

[37]I. Antoniou and I. Prigogine, Physica A {\bf 192, }443, 1993.

[38]A. Bohm, Quantum Mechanics: foundations and applications,
Springer-Verlag, Berlin, 1979.

[39]M. Castagnino, M. Gadella, F. Gaioli, and R. Laura, IAFE Preprint 1995.

[40]R. Balescu, Equilibrium and non-equilibrium Statistical Mechanics, J.
Wiley \& Sons, New York, 1975.

[41]B.L. Hu, J.P. Paz, and Y. Zhang, ``Quantum origin of noise on
fluctuation in Cosmology'', Proc. Conference on the origin of structure in
the Universe, Chateaux du Pont d'Oye 1992, World Scientific, 1992.

[42]V.A. Rochlin, Am. Math. Soc. Transl. (2),{\bf \ 39,} 1, 1969.

[43]A. Ordo\~nez,''Prigogine's $\Lambda $ and rigged Hilbert spaces'',
Preprint IFIR 1995.

[44]I. Prigogine, T. Petrosky, Phys. Lett. A, {\bf 182, 5, 1993.}

[45]I. Prigogine,''Time, dynamics, and chaos'',XXVI Nobel conference,
Gustavus Adolphus College , preprint 1990.

[46]N. Balazs and A. Voros, Ann. Phys. {\bf 199}, 123, 1990.

[47]M. Hillery, R. F. O'Conell, M. D. Scully, E. P. Wigner, Phys. Rep. {\bf %
106}, 1984.

[48]M. Castagnino, R. Laura, A. Ordo\~nez, and S. Sonego, ``When time
reversal can be defined?'', submitted to J. Math. Phys., 1995.

[49]M. Castagnino and R. Laura, ``The cosmological essence of time
asymmetry'', Proc. SILARG VIII, Ed. W. Rodrigues, World Scientific,
Singapore, 1983.

[50]M. Castagnino, R. Laura, and M. Gonzalez Eiras, IAFE Preprint 1995.

[51]H. Reeves,''The growth of complexity in expanding universes'' in ''The
Anthropic Principle'',Proceedings Second Venice Conference on Cosmology, Ed.
F. Bertolo, U. Cino., Cambridge Univ. Press, Cambridge, 1993.

[52]H. D. Zeh, ''The physical bases of the direction of time'', Springer
Verlag, Berlin, 1989.

\section{Figures.}

0.-$q(t)$ and $p(t)$ functions for time-symmetric solutions, with respect to 
$t=0.$

1.-The baker transformation.

2.-The $P(t)$ graphic, showing Zeno effect, the exponential behavior, and,
Khalfin effect.

3.-The $\Gamma $ curve-

4.-The $\Gamma ^{\prime }$ curve.

5.-$\Delta S$ showing the minimum and the maximum.

6.-$\Delta S^{\prime }$ showing the two roots.

\end{document}